\documentclass[12pt]{article}

\catcode`\@=11
\@addtoreset{equation}{section}

\global\arraycolsep=2pt
\usepackage{amsbsy,amssymb,latexsym,amsfonts,amsmath,cite}

\newcommand\CL{{\mathcal L}}

\newcommand\CP{{\mathcal P}}
\newcommand\CF{{\mathcal F}}
\newcommand\CM{{\mathcal M}}
\newcommand\CH{{\mathcal H}}
\newcommand\CI{{\mathcal I}}
\newcommand\CJ{{\mathcal J}}
\newcommand\CC{{\mathcal C}}
\newcommand\CB{{\mathcal B}}

\newcommand\CN{{\mathcal N}}

\newcommand\bR{{\mathbb R}}
\newcommand\bE{{\mathbb E}}
\newcommand\g{{\mathfrak g}}

\newcommand\BL{{\bf{L}}}

\renewcommand\mod{~\text{mod}~}

\renewcommand\NG{\mathrm{NG}}
\newcommand\DBI{\mathrm{DBI}}
\newcommand\PST{\mathrm{PST}}
\newcommand\WZ{\mathrm{WZ}}
\newcommand\NR{\mathrm{NR}}
\newcommand\fin{\mathit{fin}}
\renewcommand\div{\mathit{div}}

\newcommand\e{\mathrm{e}}
\newcommand\vol{\mathrm{vol}}

\newcommand\ih{{\hat i}}
\newcommand\jh{{\hat j}}
\newcommand\kh{{\hat k}}
\newcommand\lh{{\hat l}}

\newcommand\pp{{\mathit{pp}}}

\begin{document}

\begin{flushright}
\parbox{4.2cm}
{KEK-TH-1084 \hfill \\
OIQP-06-06 \\
May 2006 \\ 
}
\end{flushright}

\begin{center}
{\Large \bf
Non-Relativistic AdS Branes \\
and Newton-Hooke Superalgebra
}
\end{center}
\vspace{10mm}

\centerline{\large Makoto Sakaguchi$^{a}$
 and Kentaroh Yoshida$^{b}$
}

\vspace{8mm}

\begin{center}
$^a$ 
{\it Okayama Institute for Quantum Physics\\
Kyoyama 1-9-1, Okayama 700-0015, Japan} \\
{\tt makoto$\_$sakaguchi@pref.okayama.jp}
\vspace{5mm}

$^b${\it Theory Division, Institute of Particle and Nuclear Studies, \\ 
High Energy Accelerator Research 
Organization (KEK),\\
Tsukuba, Ibaraki 305-0801, Japan.} 
\\
{\tt kyoshida@post.kek.jp}
\end{center}

\vfill

\begin{abstract}
We examine a non-relativistic limit of D-branes in AdS$_5\times$S$^5$
 and M-branes in AdS$_{4/7}\times$S$^{7/4}$.
First, 
Newton-Hooke superalgebras for the AdS branes are derived from
AdS$\times$S superalgebras as In\"on\"u-Wigner contractions. It is shown
that the directions along which the AdS-brane worldvolume extends are
restricted by requiring that the isometry on the AdS-brane worldvolume
and the Lorentz symmetry in the transverse space naturally extend to the
super-isometry. 
We also derive Newton-Hooke superalgebras
for pp-wave branes and show that the directions along which a
brane worldvolume extends are restricted.
Then the Wess-Zumino terms of the AdS branes are derived
by using the Chevalley-Eilenberg cohomology on the super-AdS$\times$S
algebra, and the non-relativistic limit of the AdS-brane actions is
considered. We show that the consistent limit is possible for the
following branes: D$p$ (even,even) for $p=1\mod 4$ and D$p$ (odd,odd) for
$p=3\mod 4$ in AdS$_5\times$S$^5$, and M2 (0,3), M2 (2,1), M5 (1,5) and
M5 (3,3) in AdS$_{4}\times$S$^{7}$ and S$^{4}\times$AdS$_{7}$. We
furthermore present non-relativistic actions for the AdS branes.

\end{abstract}

\thispagestyle{empty}
\setcounter{page}{0}

\section{Introduction}

The AdS/CFT conjecture \cite{AdS} predicts 
that type IIB superstring theory in AdS$_5\times$S$^5$ is dual to
the four-dimensional $\CN=4$ SU($N$) super Yang-Mills theory in large
$N$ limit. Though it is too hard to analyze the full AdS superstring,
Berenstein-Maldacena-Nastase (BMN) 
found a nice way to extract a solvable subsector (referred to as BMN
sector) \cite{Berenstein:2002jq}.
Taking this subsector corresponds to the so-called Penrose limit for the
AdS geometry \cite{Penrose}, and the relevant symmetry to the BMN sector
is the pp-wave superalgebra, which is obtained as an In\"on\"u-Wigner
(IW) contraction \cite{IW} of the super-AdS$_5\times$S$^5$ algebra
\cite{Hatsuda:2002xp} (see \cite{Hatsuda:2002kx} for the
eleven-dimensional cases). 

A non-relativistic limit of strings in flat spacetime provides another
solvable sector \cite{Gomis:2000bd} (see also
\cite{Danielsson:2000gi}). This limit is a truncation of 
the full theory in the sense that light states satisfying a 
Galilean invariant dispersion relation are kept and the rest decouples. 
The relevant symmetry is the Galilean limit of the
Poincar\'e algebra. The non-relativistic flat branes are examined in
\cite{Garcia:2002fa, Brugues:2004an, Gomis:2004pw, Gomis:2005bj,
Kamimura:2005rz}.  In \cite{Brugues:2006yd, Gomis:2005pg} 
these studies have been extended to branes in AdS spaces. In particular
a Lorentzian F-string in AdS$_5\times$S$^5$, i.e. AdS$_2$ brane, was
examined in \cite{Gomis:2005pg}.  They showed that the F-string theory
in AdS$_5\times$S$^5$ is reduced to a free theory in the
non-relativistic limit, and so the resulting theory is exactly solvable.
In the non-relativistic limit, the super-AdS$_5\times$S$^5$ algebra is
also contracted to the Newton-Hooke (NH) superalgebra for the F-string.
Then the isometry of the AdS$_2$-brane worldvolume, the AdS$_2$ algebra
so(1,2), and the Lorentz symmetry in the transverse space,
so(3)$\times$so(5), extend to a super-isometry algebra. 

In this paper we consider
D-branes in AdS$_5\times$S$^5$ and M-branes in
AdS$_{4/7}\times$S$^{7/4}$. First we examine D-branes in
AdS$_5\times$S$^5$.  In addition to AdS$_2$ brane, there exist various
AdS branes in AdS$_5\times$S$^5$, ($m,n$) branes of which worldvolume
extends along $m$ directions in AdS$_5$ and $n$-directions in S$^5$.  In
our previous works \cite{Sakaguchi:2003py, Sakaguchi:2004md,
Sakaguchi:2004mj, SY:NCD(AdS)}, we have classified some possible
configurations of the D-branes in AdS$_5\times$S$^5$ by examining the
$\kappa$-variation surface terms of an open superstring. Here we will
classify possible configurations of D-branes by requiring that the
isometry of the AdS brane worldvolume AdS$_m\times$S$^n$
(H$^m\times$S$^n$) and the Lorentz symmetry in the transverse space
$\bE^{5-m}\times\bE^{5-n}$ ($\bE^{4-m,1}\times\bE^{5-n}$), i.e.,
so$(m-1,2)\times$so$(n+1)\times$so$(5-m)\times$so$(5-n)$ for a
Lorentzian brane and
so$(m,1)\times$so$(n+1)\times$so$(4-m,1)\times$so$(5-n)$ for a Euclidean
brane, naturally extend to the super-isometry. The result  
surely contains our previous result, but some new configurations are
allowed to exist. 
We furthermore derive the NH superalgebras for these branes as IW
contractions of the super-AdS$_5\times$S$^5$ algebra.
The similar analyses are applied to branes in IIB pp-wave,
and derive the NH superalgebras for these branes as IW
contractions of the IIB pp-wave superalgebra.

The Wess-Zumino (WZ) terms for $p$-branes in flat spacetime
can be classified \cite{DeAzcarraga:1989vh} as non-trivial elements of the
Chevalley-Eilenberg (CE) cohomology \cite{CE}.
This is generalized to D-branes in 
\cite{Chryssomalakos:1999xd,Sakaguchi:1999fm} 
by introducing an additional two form which corresponds to a modified
field strength of the background $B$ field.  Here we examine the WZ
terms for AdS branes by using the CE cohomology on $\g$ of the
supergroup 
\[
G= {\rm PSU(2,2|4)/(SO(4,1)\times SO(5))}\,, \quad 
\mbox{i.e.~
``super-AdS$_5\times$S$^5$''/``Lorentz''}\,. 
\]
We show that the WZ terms of
AdS branes can be classified as non-trivial elements of the CE
cohomology, except for the WZ term of a string which is a trivial element
\cite{Hatsuda:2002hz,Hatsuda:2002iu}.

Expanding the supercurrents with respect to the scaling used in the IW
contraction, we obtain the non-relativistic limit of the brane action.
In comparison to the Penrose limit in which the leading terms in the
expansion contribute to the pp-wave brane actions (see Appendix C), 
in the
non-relativistic limit the leading order terms of the Dirac-Born-Infeld
(Nambu-Goto) part and the WZ part cancel out each other, and the
next-to-leading order terms contribute to the non-relativistic action.
We find that the consistent non-relativistic limit exists only for
D$p$ (even, even) for $p=1\mod 4$ and D$p$ (odd, odd) for $p=3\mod 4$ in
AdS$_5\times$S$^5$. We derive the non-relativistic AdS D-brane action
and find that it is reduced to a simple action by fixing the
$\kappa$-gauge symmetry and the worldvolume reparametrization.
While the non-relativistic AdS D-string action is a free field action,
the non-relativistic AdS D$p$-brane action ($p>1$)
contains
an additional term which originates from the flux contribution in the
WZ term. The non-relativistic flat D-brane actions obtained in
\cite{Gomis:2005bj} are reproduced as a flat limit of the
non-relativistic AdS D-brane actions.

Next we examine a non-relativistic limit of M-branes in
AdS$_{4/7}\times$S$^{7/4}$.  The NH superalgebra for M-branes are
derived as IW contractions of the super-AdS$_{4/7}\times$S$^{7/4}$
algebras.  To achieve this, we show that the directions along which a
brane worldvolume extends are restricted by requiring that the isometry
of the AdS brane worldvolume and the Lorentz symmetry in the transverse
space naturally extend to the super-isometry, and that possible M-branes
are classified.  As expected, the configurations obtained in
\cite{Sakaguchi:2003hk,Sakaguchi:2004bu} by examining the
$\kappa$-variation surface term of an open supermembrane are contained
in the above classification.  
The similar analyses are applied to branes in M pp-wave,
and derive the NH superalgebras for these branes as IW
contractions of the M pp-wave superalgebra.
We obtain the WZ terms of AdS branes as
non-trivial elements of the CE cohomology on $\g$ of the supergroup
\begin{eqnarray*}
\text{
$G=$OSp(8$|$4)/(SO(3,1)$\times$SO(7)) ~or~
OSp(8$^*|$4)/(SO(4)$\times$SO(6,1)).
}
\end{eqnarray*}
We find that the non-relativistic
limit exists for M2 (0,3), M2 (2,1), M5 (1,5) and M5 (3,3) in
AdS$_{4}\times$S$^{7}$ and S$^{4}\times$AdS$_{7}$. By taking the
non-relativistic limit of these AdS brane actions, we derive the
non-relativistic M-brane actions in AdS$_{4/7}\times$S$^{7/4}$. It is
shown that by fixing the $\kappa$-gauge symmetry and the
reparametrization the non-relativistic action for AdS M2- and AdS
M5-branes is reduced to a simple action which contains an additional
term originating from the flux contribution of the WZ term. The
non-relativistic flat M2-brane action given in \cite{Gomis:2004pw} is
reproduced as a flat limit of the non-relativistic AdS M2-brane action.

\medskip

This paper is divided into the two parts. Sections 2--5 are devoted to 
studies of AdS branes in ten-dimensions, and those in eleven-dimensions
are examined in sections 6--9.  In section 2, NH superalgebras for
branes in AdS$_5\times$S$^5$ are derived as IW contractions of the
super-AdS$_5\times$S$^5$ algebra. It is shown that the directions along
which the AdS brane worldvolume extends are restricted by requiring that
the isometry on the AdS brane worldvolume and the Lorentz symmetry in
the transverse space naturally extend to the super-isometry. 
The similar analyses are applied to branes in IIB pp-wave
in section 3.
WZ terms of
AdS branes are derived by using the CE cohomology on the AdS$\times$S
superalgebra in section 4. Examining a non-relativistic limit of AdS
brane actions, we obtain non-relativistic AdS brane actions in section
5. From section 6, M-theory in AdS$_{4/7}\times$S$^{7/4}$ is
examined. We derive NH superalgebras for M-branes as IW contractions of
the super-AdS$_{4/7}\times$S$^{7/4}$ algebras in section 6. 
The similar analyses are applied to branes in M pp-wave
in section 7.
After
deriving WZ terms of AdS M-branes by using the CE cohomology on the
AdS$_{4/7}\times$S$^{7/4}$ superalgebras in section 8, we examine the
non-relativistic limit of AdS M-brane actions in section 9. The last
section is devoted to a summary and discussions.

The supervielbeins and the super spin-connections are given in Appendix A.
In Appendix \ref{appendix:kappa}, the $\kappa$-symmetry of
Euclidean/Lorentzian brane actions is derived. Our construction of brane
actions is applicable to branes in a pp-wave by taking the Penrose limit
instead of non-relativistic limit. In fact, we derive brane actions in
the pp-wave in Appendix C.

\section{NH Superalgebra of Branes in AdS$_5\times$S$^5$}

The super-AdS$_5\times$S$^5$ algebra, psu(2,2$|$4), is generated 
by translation $P_A=(P_a,P_{a'})$, Lorentz rotation $J_{AB}=(J_{ab},J_{a'b'})$
and Majorana-Weyl supercharges $Q_I$($I=1,2$) as
\begin{eqnarray}
&&
{[}P_a,P_b]=
\lambda^2J_{ab}~,~~~
{[}P_{a'},P_{b'}]=
-\lambda^2J_{a'b'}~,
\nonumber\\&&
{[}P_a,J_{bc}]=\eta_{ab}P_c-\eta_{ac}P_b~,~~~
{[}P_{a'},J_{b'c'}]=\eta_{a'b'}P_{c'}-\eta_{a'c'}P_{b'}~,
\nonumber\\&&
{[}J_{ab},J_{cd}]=\eta_{bc}J_{ad}+\text{3-terms}~,~~~
{[}J_{a'b'},J_{c'd'}]=\eta_{b'c'}J_{a'd'}+\text{3-terms}~,
\nonumber\\&&
{[}Q_I,P_A]=
-\frac{\lambda}{2}Q_J(i\sigma_2)_{JI}\widehat\Gamma_{A}~,~~~
{[}Q_I,J_{AB}]=
-\frac{1}{2}Q_I\Gamma_{AB}~,
\nonumber\\&&
\{Q_I,Q_J\}=
2i\CC\Gamma^A\delta_{IJ}h_+P_A
-i\lambda\CC\widehat\Gamma^{AB}(i\sigma_2)_{IJ}h_+J_{AB}~,
\label{AdS5xS5 algebra}
\end{eqnarray}
where $a=0,\cdots,4$ and $a'=5,\cdots,9$
are vector indices of AdS$_5$ and S$^5$ respectively.
The gamma matrix $\Gamma^A\in$ Spin(1,9) satisfies
\begin{eqnarray}
\{\Gamma^A,\Gamma^B\}=2\eta^{AB}~,~~~
(\Gamma^A)^T=-\CC\Gamma^A\CC^{-1}~,~~~
\CC^T=-\CC
\label{Clifford}
\end{eqnarray}
where $\CC$ is the charge conjugation matrix.
We use almost positive Minkowski metric $\eta_{AB}$ and define
\begin{eqnarray}
&&\widehat\Gamma_A=(-\Gamma_a\CI,\Gamma_{a'}\CJ)~,~~~
\widehat\Gamma_{AB}=(-\Gamma_{ab}\CI,\Gamma_{a'b'}\CJ)~,~~~
\CI=\Gamma^{01234}~,~~~\CJ=\Gamma^{56789}~,
\nonumber\\&&
Q_Ih_+=Q_I~,~~~
h_+=\frac{1}{2}(1+\Gamma_{11})~,~~~
\Gamma_{11}=\Gamma_{01\cdots9}~,
\end{eqnarray}
and $\lambda=1/R$ where $R$ is the radii of AdS$_5$ and S$^5$\,.

By using an element $g\in$ PSU(2,2$|$4), 
a left-invariant (LI) Cartan one-form is defined as
\begin{eqnarray}
\Omega=g^{-1}dg\equiv
\BL^AP_A
+\frac{1}{2}\BL^{AB}J_{AB}
+Q_IL^I~.
\end{eqnarray}
Then the Maurer-Cartan (MC) equation, which is satisfied by LI Cartan one-forms
\begin{eqnarray}
dL^{\hat A}=\frac{1}{2}L^{\hat B}L^{\hat C}f_{\hat C\hat B}{}^{\hat A}~,~~~
\Omega=L^{\hat A}T_{\hat A}\,,
\end{eqnarray}
is equivalent to the superalgebra 
${[}T_{\hat A},T_{\hat B}\}=f_{\hat A\hat B}{}^{\hat C}T_{\hat C}$\,. 
The Jacobi identities
$f_{[\hat A\hat B}{}^{\hat D}f_{|{\hat D}|{\hat C})}{}^{\hat E}=0$
 of the commutation relation of the superalgebra
is stated as the nilpotency of the differential, $d^2=0$\,. 
Thus (\ref{AdS5xS5 algebra}) is equivalent to
\begin{eqnarray}
d\BL^A&=&
-\eta_{BC}\BL^{AB}\BL^C
+i\bar L\Gamma^AL~,\nonumber\\
d\BL^{ab}&=&
-\lambda^2\BL^a\BL^b
-\eta_{cd}\BL^{ca}\BL^{bd}
+i\lambda\bar L\Gamma^{ab}\CI i\sigma_2 L~,\nonumber\\
d\BL^{a'b'}&=&
+\lambda^2\BL^{a'}\BL^{b'}
-\eta_{c'd'}\BL^{c'a'}\BL^{b'd'}
-i\lambda\bar L\Gamma^{a'b'}\CJ i\sigma_2 L~,\nonumber\\
dL^\alpha&=&
-\frac{\lambda}{2}\BL^A\widehat\Gamma_A i\sigma_2L
-\frac{1}{4}\BL^{AB}\Gamma_{AB}L~.
\label{MC AdSxS in 10-dim}
\end{eqnarray}

We derive NH superalgebras for AdS branes 
as IW contractions of the super-AdS$_5\times$S$^5$ algebra. 

First we consider the bosonic subalgebra.
Let us introduce the following coordinates: 
\begin{eqnarray}
\bar A=A_0,\cdots,A_p~,~~~
\underline{A}=A_{p+1},\cdots,A_9~,
\end{eqnarray}
where $\bar A=(\bar a,\bar a')$ represent the worldvolume directions of
the AdS brane. When the worldvolume extends along $m$ directions in
AdS$_5$ and $n$ directions in S$^5$, we call it an $(m,n)$-brane. We
rescale the generators as follow: 
\begin{eqnarray}
P_{\underline A}\to\frac{1}{\Omega}  P_{\underline A}~,~~~
J_{\bar A\underline B}\to\frac{1}{\Omega}  J_{\bar A\underline B}~.
\label{Omega bosonic}
\end{eqnarray}
The limit $\Omega\to 0$ leads to the NH algebra for the AdS brane
\begin{eqnarray}
&&
[P_{\bar a},P_{\bar b}]=\lambda^2J_{\bar a\bar b}~,~~~
[P_{\bar a'},P_{\bar b'}]=-\lambda^2J_{\bar a'\bar b'}~,~~~
\nonumber\\&&
{[}P_{\bar a},P_{\underline b}]=\lambda^2J_{\bar a\underline b}~,~~~
{[}P_{\bar a'},P_{\underline b'}]=-\lambda^2J_{\bar a'\underline b'}~,
\nonumber\\&&
{[}P_{\bar A},J_{\bar B\bar C}]=
\eta_{\bar A \bar B}P_{\bar C}-\eta_{\bar A \bar C}P_{\bar B}~,~~~
{[}P_{\underline A},J_{\underline B\underline C}]=
\eta_{\underline A \underline B}P_{\underline C}
 -\eta_{\underline A \underline C}P_{\underline B}~,
\nonumber\\&&
{[}P_{\bar A},J_{\bar B\underline C}]=
\eta_{\bar A \bar B}P_{\underline C}~,~~~ 
\nonumber\\&&
{[}J_{\bar A\bar B},J_{\bar C\bar D}]=
\eta_{\bar A\bar D}J_{\bar B\bar C}+\text{3-terms}~,~~~
{[}J_{\underline A\underline B},J_{\underline C\underline D}]=
\eta_{\underline A\underline D}J_{\underline B\underline C}+\text{3-terms}~,
\nonumber\\&&
{[}J_{\bar A\bar B},J_{\bar C\underline D}]=
\eta_{\bar B\bar C}J_{\bar A\underline D}
- \eta_{\bar A\bar C}J_{\bar B\underline D}
~,~~~
{[}J_{\underline A\underline B},J_{\bar C\underline D}]=
\eta_{\underline B\underline D}J_{\bar C\underline A}
- \eta_{\underline A\underline D}J_{\bar C\underline B}~.
\label{IIB NH bosonic}
\end{eqnarray}
This is the NH algebra of a brane
given in \cite{Brugues:2006yd} (see also \cite{GP}). 
The NH algebra
contains two subalgebras. One is the isometry of $(m,n)$-brane
worldvolume generated by $\{P_{\bar A},J_{\bar A\bar B}\}$, the
AdS$_m\times$S$^n$ algebra so$(m-1,2)\times$so$(n+1)$ for a Lorentzian
brane and the $H^m\times$S$^n$ algebra so$(m,1)\times$so$(n+1)$ for an
Euclidean brane.  The other is the Poincar\'e algebra,
iso$(5-m)\times$iso$(5-n)$ for a Lorentzian brane and
iso$(4-m,1)\times$iso$(5-n)$ for a Euclidean brane, generated by
$\{P_{\underline A},J_{\underline A\underline B}\}$ which is the
isometry of the transverse space $\bE^{5-m}\times\bE^{5-n}$ and
$\bE^{4-m,1}\times\bE^{5-n}$ respectively.

Next, we consider the fermionic part. Let us introduce a condition
\begin{eqnarray}
\theta= M\theta \qquad \mbox{with} \quad 
M=\ell\Gamma^{\bar A_0\cdots \bar A_p}\otimes\rho~
\end{eqnarray}
where $\ell^2(-1)^{[\frac{p+1}{2}]}\rho^2=1~$
for  $M^2=1$.
The $2\times 2$ matrix $\rho$ is determined below.
As $\theta=h_+\theta$,
 $[M,h_+]=0$ is required so that $p=$ odd.
We demand  that $M$ satisfies following relations
\begin{eqnarray}
M'\Gamma^{\bar A}&=&\Gamma^{\bar A}M~,
\label{condition 1 10}
\\
M'\widehat\Gamma^{\bar A\bar B} i\sigma_2&=&\widehat\Gamma^{\bar A\bar B} i\sigma_2M~,
\label{condition 2 10}
\end{eqnarray}
where $M'=C^{-1}M^TC$.
If these are satisfied,
the isometry of the AdS brane worldvolume
and the Lorentz symmetry in the transverse space,
so$(m-1,2)\times$so$(n+1)\times$so$(5-m)\times$so$(5-n)$
 for a Lorentzian brane
and 
so$(m,1)\times$so$(n+1)\times$so$(4-m,1)\times$so$(5-n)$
for a Euclidean brane,
naturally extend to the super-isometry as will be seen below.  It is
straightforward to see that the first condition is satisfied by
$\rho^T=\rho$ for $p=1\mod 4$ and by $\rho^T=-\rho$ for $p=3\mod 4$.
The second condition restricts the direction along which branes extend.
Since, for $\rho=1 (p=1\mod 4) $ and $\rho=i\sigma_2 (p=3\mod 4) $, we
derive
\begin{eqnarray}
M'\widehat\Gamma^{\bar A\bar B} i\sigma_2=
(-1)^{\mathrm{d}}\widehat\Gamma^{\bar A\bar B} i\sigma_2M~,
\end{eqnarray}
we have (odd,odd)-branes.
$\mathrm{d}$ denotes the number of Dirichlet directions contained in 
AdS$_5$.
On the other hand, for $\rho=\sigma_1,\sigma_3$ ($p=1\mod 4$),
since
\begin{eqnarray}
M'\widehat\Gamma^{\bar A\bar B} i\sigma_2=
-(-1)^{\mathrm{d}}\widehat\Gamma^{\bar A\bar B} i\sigma_2M~,
\end{eqnarray}
 (even,even)-branes
 are allowed.
In both cases, we have $\ell=\sqrt{-s}$ and
\begin{eqnarray}
M'=-M~.
\end{eqnarray}
We summarize branes in Table \ref{IIB-branes}.
\begin{table}[tb]
 \begin{center}
  \begin{tabular}{|c|c|c|c|c|c|}
    \hline
 $\rho$ &   1-brane  &3-brane    &5-brane    &7-brane    &9-brane    \\
    \hline
$\sigma_1$, $\sigma_3$  &  (2,0),~(0,2)   &  &(4,2),~(2,4)    &   &    \\
$i\sigma_2$     &    &(3,1),~(1,3)    &   &(5,3),~(3,5)    &    \\
1 &(1,1)      &    & (5,1),~(3,3),~(1,5)   &    &(5,5)    \\
    \hline
  \end{tabular}
 \end{center}
  \caption{Branes in AdS$_5\times$S$^5$}
\label{IIB-branes}
\end{table}
The 9-brane is nothing but AdS$_5\times$S$^5$ itself as $M=h_+$ in this
case. The (even,even)-branes ($p=1\mod 4$) and (odd,odd)-branes
($p=3\mod 4$) are 1/2 BPS Dirichlet branes of F- and D-strings in
AdS$_5\times$S$^5$ derived in
\cite{Sakaguchi:2003py,Sakaguchi:2004md,Sakaguchi:2004mj}\footnote{ The
brane probe analysis \cite{SMT} is also consistent with this result.}.
In the presence of gauge field condensates, see \cite{SY:NCD(AdS)}.  As
will be seen in section 4, we find consistent non-relativistic limits
for these AdS branes.

Let us decompose $Q^\alpha$ with the projection operator
\begin{eqnarray}
\CP_\pm=\frac{1}{2}(1\pm M) \qquad \mbox{as} \quad 
Q=Q_++Q_-~,~~~Q_\pm \CP_\pm=Q_\pm\,, 
\end{eqnarray}
and rescale fermionic generators as
\begin{eqnarray}
Q_+\to Q_+~,~~~
Q_-\to\frac{1}{\Omega }Q_-~.
\label{Omega fermionic}
\end{eqnarray}
Taking $\Omega\to 0$ leads to (anti-)commutation relations
\begin{eqnarray}
&&{[}P_{\bar A},Q_+]=\frac{\lambda}{2}Q_+\widehat\Gamma_{\bar A}i\sigma_2~,~~~
{[}P_{\underline A},Q_+]=\frac{\lambda}{2}Q_-\widehat\Gamma_{\underline A}i\sigma_2~,~~~
{[}P_{\bar A},Q_-]=\frac{\lambda}{2}Q_-\widehat\Gamma_{\bar A}i\sigma_2~,~~~
\nonumber\\&&
{[}J_{\bar A\bar B},Q_\pm]=\frac{1}{2}Q_\pm\Gamma_{\bar A\bar B}~,~~~
{[}J_{\underline A\underline B},Q_\pm]=\frac{1}{2}Q_\pm\Gamma_{\underline A\underline B}~,~~~
{[}J_{\bar A\underline B},Q_+]=\frac{1}{2}Q_-\Gamma_{\bar A\underline B}~,~~~
\nonumber\\&&
\{Q_+,Q_+\}=2i\CC\Gamma^{\bar A}h_+\CP_+P_{\bar A}
-i\lambda\CC\widehat\Gamma^{\bar A\bar B}i\sigma_2h_+\CP_+J_{\bar A\bar B}
-i\lambda\CC\widehat\Gamma^{\underline A\underline B}i\sigma_2h_+\CP_+
 J_{\underline A\underline B}~,
\nonumber\\&&
\{Q_+,Q_-\}=
2i\CC\Gamma^{\underline  A}h_+\CP_-P_{\underline  A}
-2i\lambda\CC\widehat\Gamma^{\bar A\underline  B}i\sigma_2h_+\CP_-
J_{\bar A\underline  B}~.
\label{IIB NH fermionic}
\end{eqnarray}
In summary, we have derived the NH superalgebra for AdS brane,
(\ref{IIB NH bosonic}) and (\ref{IIB NH fermionic}), as an IW
contraction of psu(2,2$|4$). The NH superalgebra for an F-string
\cite{Gomis:2005pg} is contained as the $p=1$ case.

We note that generators $P_{\bar A}$, $J_{\bar A\bar B}$, $J_{\underline
A\underline B}$ and $Q_+$ form a super-subalgebra
\begin{eqnarray}
&&
[P_{\bar a},P_{\bar b}]=\lambda^2J_{\bar a\bar b}~,~~~
[P_{\bar a'},P_{\bar b'}]=-\lambda^2J_{\bar a'\bar b'}~,~~~
{[}P_{\bar A},J_{\bar B\bar C}]=
\eta_{\bar A \bar B}P_{\bar C}-\eta_{\bar A \bar C}P_{\bar B}~,~~~
\nonumber\\&&
{[}J_{\bar A\bar B},J_{\bar C\bar D}]=
\eta_{\bar A\bar D}J_{\bar B\bar C}+\text{3-terms}~,~~~
{[}J_{\underline A\underline B},J_{\underline C\underline D}]=
\eta_{\underline A\underline D}J_{\underline B\underline C}+\text{3-terms}~,
\nonumber\\&&
{[}P_{\bar A},Q_+]=\frac{\lambda}{2}Q_+\widehat\Gamma_{\bar A}i\sigma_2~,~~~
{[}J_{\bar A\bar B},Q_+]=\frac{1}{2}Q_+\Gamma_{\bar A\bar B}~,~~~
{[}J_{\underline A\underline B},Q_+]=\frac{1}{2}Q_+\Gamma_{\underline A\underline B}~,~~~
\nonumber\\&&
\{Q_+,Q_+\}=2i\CC\Gamma^{\bar A}h_+\CP_+P_{\bar A}
-i\lambda\CC\widehat\Gamma^{\bar A\bar B}i\sigma_2h_+\CP_+J_{\bar A\bar B}
-i\lambda\CC\widehat\Gamma^{\underline A\underline B}i\sigma_2h_+\CP_+
 J_{\underline A\underline B}\,,~
\label{superalgebra}
\end{eqnarray}
which is a supersymmetrization of
so$(m-1,2)\times$so$(n+1)\times$so$(5-m)\times$so$(5-n)$ for a
Lorentzian brane and
so$(m,1)\times$so$(n+1)\times$so$(4-m,1)\times$so$(5-n)$ for a Euclidean
brane. The superalgebra for the (5,5)-brane is psu(2,2$|$4). Since the
dimension of the bosonic subalgebra is 14 for (1,1)-, (3,1)-, (1,3)- and
(3,3)-branes, 16 for (2,0)-, (0,2)-, (4,2)-, (2,4)-branes, and 22 for
(5,1)-, (1,5)-, (5,3)- and (3,5)-branes, one may guess the corresponding
superalgebras as those including variants of su(2$|$2)$\times$su(2$|$2),
osp(4$|$4) and osp(6$|$2)$\times$psu(2$|$1), respectively. The existence
of these superalgebras is ensured by (\ref{condition 1 10}) and
(\ref{condition 2 10}).

It is straightforward to derive MC equations for the AdS brane NH
superalgebra (\ref{IIB NH bosonic}) and (\ref{IIB NH fermionic})
\begin{eqnarray}
d\BL^{\bar A}&=&
-\eta_{\bar B\bar C}\BL^{\bar A\bar B}\BL^{\bar C}
+ i\bar L_+\Gamma^{\bar A}L_+
~,
\label{MC NH 10}\\
d\BL^{\underline A}&=&
-\eta_{\bar B\bar C}\BL^{\underline A\bar B}\BL^{\bar C}
-\eta_{\underline{B}\underline C}\BL^{\underline A\underline B}
 \BL^{\underline C}
+i\bar L_+\Gamma^{\underline  A}L_-
+i\bar L_-\Gamma^{\underline  A}L_+
~,\\
d\BL^{\bar a\bar b}&=&
-\lambda^2\BL^{\bar a}\BL^{\bar b}
-\eta_{\bar c\bar d}\BL^{\bar c \bar a}\BL^{\bar b \bar d}
-i\lambda \bar L_+\widehat \Gamma^{\bar a\bar b}i\sigma_2 L_+
~,\label{MC 10 J N}\\
d\BL^{\bar a'\bar b'}&=&
+\lambda^2\BL^{\bar a'}\BL^{\bar b'}
-\eta_{\bar c'\bar d'}\BL^{\bar c' \bar a'}\BL^{\bar b' \bar d'}
-i\lambda \bar L_+\widehat \Gamma^{\bar a'\bar b'}i\sigma_2 L_+
~,\label{MC 10 J N'}\\
d\BL^{\underline A\underline B}&=&
-\eta_{\underline C\underline D}\BL^{\underline C \underline A}
\BL^{\underline B \underline D}
- i\lambda \bar L_+\widehat \Gamma^{\underline A\underline B}i\sigma_2 L_+
~,\label{MC 10 J D}\\
d\BL^{\bar a\underline b}&=&
-\lambda^2\BL^{\bar a}\BL^{\underline b}
-\eta_{\bar c\bar d}\BL^{\bar c \bar a}\BL^{\underline b \bar d}
-\eta_{\underline c\underline d}\BL^{\underline c \bar a}\BL^{\underline b \underline d}
\nonumber\\&&
- i\lambda \bar L_+\widehat \Gamma^{\bar a\underline b}i\sigma_2 L_-
-  i\lambda \bar L_-\widehat \Gamma^{\bar a\underline b}i\sigma_2 L_+
~,\\
d\BL^{\bar a'\underline b'}&=&
+\lambda^2\BL^{\bar a'}\BL^{\underline b'}
-\eta_{\bar c'\bar d'}\BL^{\bar c' \bar a'}\BL^{\underline b' \bar d'}
-\eta_{\underline c'\underline d'}\BL^{\underline c' \bar a'}\BL^{\underline b' \underline d'}
\nonumber\\&&
- i\lambda \bar L_+\widehat \Gamma^{\bar a'\underline b'}i\sigma_2 L_-
-  i\lambda \bar L_-\widehat \Gamma^{\bar a'\underline b'}i\sigma_2 L_+
~,\\
dL_+&=&
-\frac{\lambda}{2}\BL^{\bar A}\widehat\Gamma_{\bar A}i\sigma_2L_+
-\frac{1}{4}\BL^{\bar A\bar B}\Gamma_{\bar A\bar B}L_+
-\frac{1}{4}\BL^{\underline A\underline B}
 \Gamma_{\underline A\underline B}L_+
 ~,\label{MC 10 Q +}\\
dL_-&=&
-
 \frac{\lambda}{2}\BL^{\bar A}\widehat\Gamma_{\bar A}i\sigma_2L_-
-\frac{\lambda}{2}\BL^{\underline A}
 \widehat\Gamma_{\underline A}i\sigma_2L_+
\nonumber\\&&
-\frac{1}{4}\BL^{\bar A\bar B}\Gamma_{\bar A\bar B}L_-
-\frac{1}{4}\BL^{\underline A\underline B}
 \Gamma_{\underline A\underline B}L_-
- \frac{1}{2}\BL^{\bar A\underline B}
 \Gamma_{\bar A\underline B}L_+
 ~.\label{MC NH 10 last}
\end{eqnarray}
An alternative way to derive these MC equations is to rescale the Cartan
one-forms in the MC equation (\ref{MC AdSxS in 10-dim}) as
\begin{eqnarray}
\BL^{\underline A}\to\Omega\BL^{\underline A}~,~~
\BL^{\bar A\underline B}\to\Omega\BL^{\bar A\underline B}~,~~
L_-\to\Omega L_-
\end{eqnarray}
and take the limit $\Omega\to 0$\,. This provides the leading order
terms of the expansion considered in the non-relativistic limit in
section 4.

Finally, let us consider an alternative scaling
\begin{eqnarray}
\lambda\to\frac{1}{\omega}\lambda~,~~
P_{\bar A}\to\frac{1}{\omega} P_{\bar A}~,~~
J_{\bar A\underline B}\to{\omega}  J_{\bar A\underline B}~,~~
Q_+\to\frac{1}{\sqrt{\omega}}Q_+~,~~
Q_-\to{\sqrt{\omega}}Q_-~.
\label{alternative scaling}
\end{eqnarray}
Since $\lambda$ is absorbed as  
\begin{eqnarray}
P_{\bar A}\to \frac{1}{\lambda}P_{\bar A}~,~~
P_{\underline A}\to \frac{1}{\lambda}P_{\underline A}~,~~
Q_\pm\to\frac{1}{\sqrt{\lambda}}Q_\pm\,, 
\label{alternative lambda}
\end{eqnarray}
this is equivalent to (\ref{Omega bosonic}) and (\ref{Omega fermionic})
with $\Omega=1/\omega$\,. In this paper, we use (\ref{Omega bosonic})
and (\ref{Omega fermionic}) instead of (\ref{alternative scaling}), 
though both limits lead to the same results.

\section{NH Superalgebra of Branes in IIB PP-Wave}

Type IIB PP-wave superalgebra is obtained as an IW contraction of the
super-AdS$_5\times$S$^5$ algebra. First of all, let us introduce the
following quantities for later convenience, 
\begin{eqnarray}
&& P_\pm=\frac{1}{\sqrt{2}}(P_9\pm P_0)~,~~~
P_{\hat i}^*=(P^*_i=J_{0i}, P^*_{i'}=J_{9i'})~, \\ &&
Q=Q^{(+)}+Q^{(-)}~,~~~
Q^{(\pm)}=Q^{(\pm)}\ell_{\pm}~,~~~
\ell_\pm=\frac{1}{2}\Gamma_\pm\Gamma_{\mp}~,~~~
\Gamma_\pm=\frac{1}{\sqrt{2}}(\Gamma_9\pm\Gamma_0)~ \nonumber 
\end{eqnarray}
where $i=1,2,3,4$ and $i'=5,6,7,8$.
The IW contraction is performed in \cite{Hatsuda:2002xp}
by scaling generators in the super-AdS$_5\times$S$^5$ algebra as
\begin{eqnarray}
P_+\to\frac{1}{\Lambda^2}P_+~,~~~
P_{\hat i}\to\frac{1}{\Lambda}P_{\hat i}~,~~~
P_{\hat i}^*\to\frac{1}{\Lambda}P_{\hat i}^*~,~~~
Q^{(+)}\to\frac{1}{\Lambda}Q^{(+)}~,
\end{eqnarray}
and then taking the limit $\Lambda\to 0$\,. 
After the contraction, the super-AdS$_5\times$S$^5$ algebra
is reduced to the IIB pp-wave superalgebra
\begin{eqnarray}
&&
[P_-,P_{\hat i}]=-\frac{\lambda^2}{\sqrt{2}}P^*_{\hat i}~,~~~
[P_-,P_{\hat i}^*]=\frac{1}{\sqrt{2}}P_{\hat i}~,~~~
[P_{\hat i},P_{\hat j}^*]=-\frac{1}{\sqrt{2}}\eta_{\hat i\hat j}P_+~,
\nonumber\\&&
[P_{\hat i},J_{\hat j\hat k}]=\eta_{\hat i\hat j}P_{\hat k}-\eta_{\hat i\hat k}P_{\hat j}
~,~~~
[P_{\hat i}^*,J_{\hat j\hat k}]=\eta_{\hat i\hat j}P^*_{\hat k}-\eta_{\hat i\hat k}P^*_{\hat j}~,
~~~
[J_{\hat i\hat j},J_{\hat k\hat l}]=\eta_{\hat j\hat k}J_{\hat i\hat l}+\text{3-terms}~,
\nonumber\\&&
[Q^{(-)},P_{\hat i}]=\frac{1}{2}Q^{(+)}\Gamma_{\hat i}\CI i\sigma_2~,~~~
[Q^{(+)},P_-]=-\frac{1}{2}Q^{(+)}\Gamma_+\CI i\sigma_2~,
\nonumber\\&&
[Q^{(-)},P^*_{\hat i}]=\frac{1}{2\sqrt{2}}Q^{(+)}\Gamma_+\Gamma_{\hat i}~,~~~
[Q^{(\pm)},J_{\hat i\hat j}]=-\frac{1}{2}Q^{(\pm)}\Gamma_{\hat i\hat j}~,~~~
\nonumber\\&&
\{Q^{(+)},Q^{(+)}\}=
2i\CC\Gamma_-P_+~,~~~
\nonumber\\&&
\{Q^{(-)},Q^{(-)}\}=
2i\CC\Gamma_+P_-
-i\frac{\lambda}{\sqrt{2}}\CC\widehat\Gamma^{\hat i\hat j}i\sigma_2J_{\hat i\hat j}~,
\nonumber\\&&
\{Q^{(\pm)},Q^{(\mp)}\}=
2i\CC\Gamma^{\hat i}\ell_{\mp}P_{\hat i}
+i\lambda\CC\widehat\Gamma^{\hat i}\ell_\mp i\sigma_2P^*_{\hat i}~,
\end{eqnarray}
where $\widehat\Gamma^{\hat i\hat
 j}=(-\Gamma^{ij}\Gamma_+f,\Gamma^{i'j'}\Gamma_+g)$,
 $\widehat\Gamma^{\hat i}=(\Gamma^{i}f,\Gamma^{i'}g)$, $f=\Gamma^{1234}$
 and $g=\Gamma^{5678}$\,. The bosonic subalgebra, the pp-wave algebra,
 is the semi-direct product of the Heisenberg algebra generated by
 $\{P_\ih, P^*_\ih\}$ with an outer automorphism $P_-$ and the Lorentz
 algebra generated by $J_{\ih\jh}$\,.

\subsection{Lorentzian branes}

Here we consider the case that $(+,-)$ are contained in the Neumann
directions. Let us denote the Neumann and the Dirichlet directions,
respectively, as 
\begin{eqnarray}
\bar A=(+,-,\bar{\hat i})~,~~~
\underline A=\underline{\hat i}\,. 
\end{eqnarray}
We derive the NH superalgebra of a Lorentzian pp-wave brane
as an IW contraction of the pp-wave superalgebra.

Let us first consider the bosonic subalgebra. We rescale generators in
the pp-wave algebra as
\begin{eqnarray}
P_{\underline A}\to\frac{1}{\Omega}P_{\underline A}~,~~~
J_{\bar{\hat i}\underline{\hat j}}\to\frac{1}{\Omega}J_{\bar{\hat i}\underline{\hat j}}~,~~~
P^*_{\underline{\hat i}}\to\frac{1}{\Omega}P^*_{\underline{\hat i}}~,~~~
\end{eqnarray}
and then take the limit $\Omega\to 0$\,. The resulting algebra is the NH
algebra of a pp-wave brane
\begin{eqnarray}
&&
[P_-,P_{\bar{\hat i}}]=-\frac{\lambda^2}{\sqrt{2}}P^*_{\bar{\hat i}}~,~~~
[P_-,P_{\underline{\hat i}}]=-\frac{\lambda^2}{\sqrt{2}}P^*_{\underline{\hat i}}~,~~~
[P_-,P^*_{\bar{\hat i}}]=\frac{1}{\sqrt{2}}P_{\bar{\hat i}}~,~~~
[P_-,P^*_{\underline{\hat i}}]=\frac{1}{\sqrt{2}}P_{\underline{\hat i}}~,~~~
\nonumber\\&&
[P_{\bar{\hat i}},P^*_{\bar{\hat i}}]=
-\frac{1}{\sqrt{2}}\eta_{\bar{\hat i}\bar{\hat j}}P_+~,~~~
[P_{\bar{\hat i}},J_{\bar{\hat j}\underline{\hat k}}]=
\eta_{\bar{\hat i}\bar{\hat j}}P_{\underline{\hat k}}~,~~~
[J_{\bar{\hat i}\underline{\hat j}},P^*_{\bar{\hat k}}]=
-\eta_{\bar{\hat i}\bar{\hat k}}P^*_{\underline{\hat j}}
~,
\label{NH bosonic lo pp 10}
\end{eqnarray}
and
\begin{eqnarray}
&&
[P_{\bar{\hat i}},J_{\bar{\hat j}\bar{\hat k}}]=
\eta_{\bar{\hat i}\bar{\hat j}}P_{\bar{\hat k}}
-\eta_{\bar{\hat i}\bar{\hat k}}P_{\bar{\hat j}}~,~~~
[P_{\underline{\hat i}},J_{\underline{\hat j}\underline{\hat k}}]=
\eta_{\underline{\hat i}\underline{\hat j}}P_{\underline{\hat k}}
-\eta_{\underline{\hat i}\underline{\hat k}}P_{\underline{\hat j}}~,~~~
\nonumber\\&&
[J_{\bar{\hat i}\bar{\hat j}},P^*_{\bar{\hat k}}]=
\eta_{\bar{\hat j}\bar{\hat k}}P_{\bar{\hat i}}^*
-\eta_{\bar{\hat i}\bar{\hat k}}P_{\bar{\hat j}}^*~,~~~
[J_{\underline{\hat i}\underline{\hat j}},P^*_{\underline{\hat k}}]=
\eta_{\underline{\hat j}\underline{\hat k}}P_{\underline{\hat i}}^*
-\eta_{\underline{\hat i}\underline{\hat k}}P_{\underline{\hat j}}^*~,~~~
\nonumber\\&&
[J_{\bar{\hat i}\bar{\hat j}},J_{\bar{\hat k}\bar{\hat l}}]=
\eta_{\bar{\hat j}\bar{\hat k}}J_{\bar{\hat i}\bar{\hat l}}+\text{3-terms}~,~~~
[J_{\underline{\hat i}\underline{\hat j}},J_{\underline{\hat k}\underline{\hat l}}]=
\eta_{\underline{\hat j}\underline{\hat k}}J_{\underline{\hat i}\underline{\hat l}}+\text{3-terms}~,
\nonumber\\&&
[J_{\bar{\hat i}\bar{\hat j}},J_{\bar{\hat k}\underline{\hat l}}]=
\eta_{\bar{\hat j}\bar{\hat k}}J_{\bar{\hat i}\underline{\hat l}}
-\eta_{\bar{\hat i}\bar{\hat k}}J_{\bar{\hat j}\underline{\hat l}}
~,~~~
[J_{\underline{\hat i}\underline{\hat j}},J_{\bar{\hat k}\underline{\hat l}}]=
\eta_{\underline{\hat i}\underline{\hat l}}J_{\underline{\hat j}\bar{\hat k}}
-\eta_{\underline{\hat j}\underline{\hat l}}J_{\underline{\hat i}\bar{\hat k}}
~.
\label{NH bosonic Lorentz pp 10}
\end{eqnarray}

Next we consider the fermionic part.
We introduce a matrix $M$
\begin{eqnarray}
M=\ell\Gamma^{+-\bar A_1\cdots\bar A_{p-1}}\rho
\end{eqnarray}
where $\rho$ is a $2\times 2$ matrix. 
Then $Q^{(\pm)}$ are decomposed into the two parts as follows: 
\begin{eqnarray}
Q^{(\bullet )}_\pm=\pm Q^{(\bullet )}_\pm M~.
\end{eqnarray}
The chirality of $Q^{(\bullet )}$ is preserved only when $p=$odd.
In addition, requiring that $M^2=1$\,, we obtain the following condition,  
\begin{eqnarray}
\ell^2(-1)^{[\frac{p-1}{2}]}\rho^2=1~.
\end{eqnarray} 
Then we demand that
\begin{eqnarray}
M'\Gamma^{\bar A}&=&\Gamma^{\bar A}M~,
\label{condition 1 pp 10}\\
M'\widehat\Gamma^{\bar{\hat i}\bar{\hat j}}i\sigma_2&=&
\widehat\Gamma^{\bar{\hat i}\bar{\hat j}}i\sigma_2M~
\label{condition 2 pp 10}
\end{eqnarray}
where
\begin{eqnarray}
M'=\CC^{-1}M^T\CC=\pm (-1)^{p+[\frac{p-1}{2}]}M~,~~~
\rho^T=\pm\rho~.
\end{eqnarray}
Since
\begin{eqnarray}
M'\Gamma^{\bar A}=\pm(-1)^{[\frac{p-1}{2}]}\Gamma^{\bar A}M~,~~~\rho^T=\pm\rho,
\end{eqnarray}
the first condition is satisfied by
\begin{eqnarray}
\pm(-1)^{[\frac{p-1}{2}]}=1~,~~~\rho^T=
\pm\rho
~.
\end{eqnarray}
This implies that $\rho^T=\rho$ for $p=1\mod 4$ and $\rho^T=-\rho$ for
$p=3\mod 4$\,, and that $M'=-M$ and $\ell=1$\,. The second condition is
rewritten as
\begin{eqnarray}
\pm(-1)^{\mathrm{n}}=1~,~~~\rho^T=\left\{
  \begin{array}{l}
1,i\sigma_2       \\
\sigma_1, \sigma_3      \\
  \end{array}
\right.
~
\end{eqnarray}
where $\mathrm{n}$ is the number of the Neumann directions contained in
$\{1,2,3,4\}$ and $\{5,6,7,8\}$ so that the directions along which a
pp-wave brane worldvolume extends are restricted.  We summarize the
results in Table \ref{lorentzian pp brane 10}.
\begin{table}[htbp]
 \begin{center}
  \begin{tabular}{|c|c|c|c|c|c|}
    \hline
  $\rho$     &1-brane    &3-brane    &5-brane    &7-brane    &9-brane    \\
    \hline
 $\sigma_1,\sigma_3$      &   &    &$(+,-;1,3)$   &    &    \\
                          &   &    &$(+,-;3,1)$   &    &    \\
    \hline
 $i\sigma_2$      &    &$(+,-;0,2)$    &    &$(+,-;4,2)$    &    \\
      &    &$(+,-;2,0)$    &    &$(+,-;2,4)$    &    \\
    \hline
 $1$      &$(+,-)$     &    &$(+,-;0,4)$     &    &$(+,-;4,4)$    \\
       &    &    &$(+,-;2,2)$     &    &    \\
       &     &    &$(+,-;4,0)$     &    &    \\
    \hline
  \end{tabular}
 \end{center}
  \caption{Lorentzian pp-wave branes.}
\label{lorentzian pp brane 10}
\end{table}
($+,-;$odd,odd)-branes with $p=1\mod 4$
and ($+,-$;even,even)-branes with $p=3\mod 4$
are 1/2 BPS D-branes of an open pp-wave superstring 
\cite{Bain:2002tq,Sakaguchi:2003py}.
Our results are consistent with those obtained in the brane probe analysis
\cite{SMT},
the supergravity analysis \cite{Bain}
and the CFT analysis
in the light-cone gauge \cite{DP,BP,BGG,SMT2}.

Scaling $Q^{(\bullet)}_\pm$ as
\begin{eqnarray}
Q^{(\bullet)}_+\to Q^{(\bullet)}_+~,~~~
Q^{(\bullet)}_-\to \frac{1}{\Omega}Q^{(\bullet)}_-
\label{Q scale ppNH 10}
\end{eqnarray}
and taking the limit $\Omega\to 0$,
we obtain the fermionic part of the NH superalgebra
\begin{eqnarray}
&&
[Q^{(-)}_\pm,P_{\bar{\hat i}}]=
-\frac{1}{2\sqrt{2}}Q^{(+)}_\pm\Gamma_{\bar{\hat i}}\Gamma_+fi\sigma_2~,~~~
[Q^{(-)}_+,P_{\underline{\hat i}}]=
-\frac{1}{2\sqrt{2}}Q^{(+)}_-\Gamma_{\underline{\hat i}}\Gamma_+fi\sigma_2~,~~~
\nonumber\\&&
[Q^{(+)}_\pm,P_-]=
-\frac{1}{\sqrt{2}}Q^{(+)}_\pm fi\sigma_2~,~~~
[Q^{(-)}_\pm,P^*_{\bar{\hat i}}]=
\frac{1}{2\sqrt{2}}Q^{(+)}_\pm\Gamma_+\Gamma_{\bar{\hat i}}~,~~~
\nonumber\\&&
[Q^{(-)}_+,P^*_{\underline{\hat i}}]=
\frac{1}{2\sqrt{2}}Q^{(+)}_-\Gamma_+\Gamma_{\underline{\hat i}}~,~~~
[Q^{(\bullet)}_\pm,J_{\bar{\hat i}\bar{\hat j}}]=
-\frac{1}{2}Q^{(\bullet)}_\pm\Gamma_{\bar{\hat i}\bar{\hat j}}~,~~~
[Q^{(\bullet)}_\pm,J_{\underline{\hat i}\underline{\hat j}}]=
-\frac{1}{2}Q^{(\bullet)}_\pm\Gamma_{\underline{\hat i}\underline{\hat j}}~,~~~
\nonumber\\&&
[Q^{(\pm)}_+,J_{\bar{\hat i}\underline{\hat j}}]=
-\frac{1}{2}Q^{(\pm)}_-\Gamma_{\bar{\hat i}\underline{\hat j}}~,~~~
\{Q^{(+)}_+,Q^{(+)}_+\}=
2i\CC\Gamma_-\CP_+P_+~,
\nonumber\\&&
\{Q^{(-)}_+,Q^{(-)}_+\}=
2i\CC\Gamma_+\CP_-P_+
-i\frac{\lambda}{\sqrt{2}}\CC\widehat\Gamma^{\bar{\hat i}\bar{\hat j}}i\sigma_2\ell_-\CP_+
 J_{\bar{\hat i}\bar{\hat j}}
-i\frac{\lambda}{\sqrt{2}}\CC\widehat
\Gamma^{\underline{\hat i}\underline{\hat j}}i\sigma_2\ell_-\CP_+
 J_{\underline{\hat i}\underline{\hat j}} ~,
 \nonumber\\&&
\{Q^{(-)}_\pm,Q^{(-)}_\mp\}=
-2i\frac{\lambda}{\sqrt{2}}\CC\widehat\Gamma^{\bar{\hat i}\underline{\hat j}}i\sigma_2
\ell_-
\CP_\mp
 J_{\bar{\hat i}\underline{\hat j}}~, 
  \nonumber\\&&
\{Q^{(+)}_+,Q^{(-)}_+\}=
2i\CC\Gamma^{\bar{\hat i}}\ell_-\CP_+P_{\bar{\hat i}}
+i\lambda\CC\widehat\Gamma^{\bar{\hat i}}i\sigma_2\ell_-\CP_+P^*_{\bar{\hat i}}~, 
  \nonumber\\&&
\{Q^{(+)}_\pm,Q^{(-)}_\mp\}=
2i\CC\Gamma^{\underline{\hat i}}\ell_-\CP_\mp P_{\underline{\hat i}}
+i\lambda\CC\widehat\Gamma^{\underline{\hat i}}i\sigma_2\ell_-\CP_\mp  P^*_{\underline{\hat i}}~.
\label{NH fermionic pp 10}
\end{eqnarray}
In summary we have obtained the NH superalgebra of a pp-wave brane
as (\ref{NH bosonic lo pp 10}), 
(\ref{NH bosonic Lorentz pp 10}) and (\ref{NH fermionic pp 10}).
This superalgebra can be derived from the NH superalgebra of an AdS brane
(\ref{IIB NH bosonic}) and (\ref{IIB NH fermionic})
by an IW contraction.

We note that the NH superalgebra of a pp-wave brane
contains a super-subalgebra generated by
$P_\pm,P_{\bar\ih},P^*_{\bar\ih},J_{\bar\ih\bar\jh},J_{\underline\ih\underline\jh}$
and $Q^{(\pm)}_+$
\begin{eqnarray}
&&
[P_-,P_{\bar{\hat i}}]=-\frac{\lambda^2}{\sqrt{2}}P^*_{\bar{\hat i}}~,~~~
[P_-,P^*_{\bar{\hat i}}]=\frac{1}{\sqrt{2}}P_{\bar{\hat i}}~,~~~
[P_{\bar{\hat i}},P^*_{\bar{\hat i}}]=
-\frac{1}{\sqrt{2}}\eta_{\bar{\hat i}\bar{\hat j}}P_+~,~~~
\nonumber\\&&
[P_{\bar{\hat i}},J_{\bar{\hat j}\bar{\hat k}}]=
\eta_{\bar{\hat i}\bar{\hat j}}P_{\bar{\hat k}}
-\eta_{\bar{\hat i}\bar{\hat k}}P_{\bar{\hat j}}~,~~~
[J_{\bar{\hat i}\bar{\hat j}},P^*_{\bar{\hat k}}]=
\eta_{\bar{\hat j}\bar{\hat k}}P_{\bar{\hat i}}^*
-\eta_{\bar{\hat i}\bar{\hat k}}P_{\bar{\hat j}}^*~,~~~
\nonumber\\&&
[J_{\bar{\hat i}\bar{\hat j}},J_{\bar{\hat k}\bar{\hat l}}]=
\eta_{\bar{\hat j}\bar{\hat k}}J_{\bar{\hat i}\bar{\hat l}}+\text{3-terms}~,~~~
[J_{\underline{\hat i}\underline{\hat j}},J_{\underline{\hat k}\underline{\hat l}}]=
\eta_{\underline{\hat j}\underline{\hat k}}J_{\underline{\hat i}\underline{\hat l}}+\text{3-terms}~,
\nonumber\\&&
[Q^{(-)}_+,P_{\bar{\hat i}}]=
-\frac{1}{2\sqrt{2}}Q^{(+)}_+\Gamma_{\bar{\hat i}}\Gamma_+fi\sigma_2~,~~~
[Q^{(+)}_+,P_-]=
-\frac{1}{\sqrt{2}}Q^{(+)}_+ fi\sigma_2~,~~~
\nonumber\\&&
[Q^{(-)}_+,P^*_{\bar{\hat i}}]=
\frac{1}{2\sqrt{2}}Q^{(+)}_+\Gamma_+\Gamma_{\bar{\hat i}}~,~~~
[Q^{(\bullet)}_+,J_{\bar{\hat i}\bar{\hat j}}]=
-\frac{1}{2}Q^{(\bullet)}_+\Gamma_{\bar{\hat i}\bar{\hat j}}~,~~~
[Q^{(\bullet)}_+,J_{\underline{\hat i}\underline{\hat j}}]=
-\frac{1}{2}Q^{(\bullet)}_+\Gamma_{\underline{\hat i}\underline{\hat j}}~,~~~
\nonumber\\&&
\{Q^{(+)}_+,Q^{(+)}_+\}=
2i\CC\Gamma_-\CP_+P_+~,
\nonumber\\&&
\{Q^{(-)}_+,Q^{(-)}_+\}=
2i\CC\Gamma_+\CP_-P_+
-i\frac{\lambda}{\sqrt{2}}\CC\widehat\Gamma^{\bar{\hat i}\bar{\hat j}}i\sigma_2\ell_-\CP_+
 J_{\bar{\hat i}\bar{\hat j}}
-i\frac{\lambda}{\sqrt{2}}\CC\widehat
\Gamma^{\underline{\hat i}\underline{\hat j}}i\sigma_2\ell_-\CP_+
 J_{\underline{\hat i}\underline{\hat j}} ~,
 \nonumber\\&&
\{Q^{(+)}_+,Q^{(-)}_+\}=
2i\CC\Gamma^{\bar{\hat i}}\ell_-\CP_+P_{\bar{\hat i}}
+i\lambda\CC\widehat\Gamma^{\bar{\hat i}}i\sigma_2\ell_-\CP_+P^*_{\bar{\hat i}}~.
\end{eqnarray}
This is regarded as a supersymmetrization
of the pp-wave algebra which is the isometry 
on the brane worldvolume and the Lorentz symmetry
in the transverse space.
The existence of this super-subalgebra is ensured by the conditions
(\ref{condition 1 pp 10}) and (\ref{condition 2 pp 10}).

\subsection{Euclidean branes}

We consider the case that $(+,-)$ are contained in the Dirichlet direction.
Let us denote Neumann and Dirichlet directions as 
$\bar A=\bar{\hat i}$ and $\underline A=(+,-,\underline{\hat i})$\,,
respectively. 
We derive the NH superalgebra of a Euclidean pp-wave brane
as an IW contraction of the pp-wave superalgebra.

First we consider the bosonic subalgebra.
We rescale generators
in the pp-wave algebra as
\begin{eqnarray}
P_{\underline A}\to\frac{1}{\Omega}P_{\underline A}~,~~~
J_{\bar{\hat i}\underline{\hat j}}\to\frac{1}{\Omega}J_{\bar{\hat i}\underline{\hat j}}~
,~~~
P^*_{\bar{\hat i}}\to\frac{1}{\Omega}P_{\bar{\hat i}}^*~,~~~
\end{eqnarray}
and then take the limit $\Omega\to 0$.
Under the contraction,
we obtain the NH algebra of a Euclidean pp-wave brane
\begin{eqnarray}
&&
[P_-,P_{\bar{\hat i}}]=-\frac{\lambda^2}{\sqrt{2}}P^*_{\bar{\hat i}}~,~~~
[P_-,P^*_{\underline{\hat i}}]=\frac{1}{\sqrt{2}}P_{\underline{\hat i}}~,~~~
[P_{\bar{\hat i}},J_{\bar{\hat j}\underline{\hat k}}]
=\eta_{\bar{\hat i}\bar{\hat j}}P_{\underline{\hat k}}~,~~~
[J_{\bar{\hat i}\underline{\hat j}},P^*_{\underline{\hat k}}]=
\eta_{\underline{\hat j}\underline{\hat k}}P^*_{\bar{\hat i}}
~,~~~
\nonumber\\&&
[P_{\bar{\hat i}},P^*_{\bar{\hat j}}]=
-\frac{1}{\sqrt{2}}\eta_{\bar{\hat i}\bar{\hat j}}P_+~,~~~
[P_{\underline{\hat i}},P^*_{\underline{\hat j}}]=
-\frac{1}{\sqrt{2}}\eta_{\underline{\hat i}\underline{\hat j}}P_+~,~~~
\label{NH bosonic euc pp 10}
\end{eqnarray}
and (\ref{NH bosonic Lorentz pp 10}).

To contract the fermionic part of the pp-wave superalgebra,
we introduce a matrix
\begin{eqnarray}
M=\ell\Gamma^{\bar A_1\cdots\bar A_p}\rho~,~~~
\ell^2(-1)^{[\frac{p+1}{2}]}\rho^2=1
\end{eqnarray}
and decompose $Q^{(\pm)}$ as
\begin{eqnarray}
Q^{(\bullet)}_\pm=\pm Q^{(\bullet)}_\pm M~
\end{eqnarray}
where $p=$odd for the chirality of $Q^{(\pm)}$.
We demand that (\ref{condition 1 pp 10}) and (\ref{condition 2 pp 10})
are satisfied.
The first condition (\ref{condition 1 pp 10})
is satisfied when
\begin{eqnarray}
\pm(-1)^{p+[\frac{p+1}{2}]}=1~,~~~\rho^T=\pm\rho
\end{eqnarray}
so that $\rho^T=\rho$ for $p=1\mod 4$
and  $\rho^T=-\rho$ for $p=3\mod 4$.
It follows that $M'=-M$ and  $\ell=\sqrt{-1}$.
Next, the second condition (\ref{condition 2 pp 10})
is found to be satisfied when
\begin{eqnarray}
\pm(-1)^{p+\mathrm{n}}=1~,~~~
\rho=\left\{
  \begin{array}{l}
    1,i\sigma_2   \\
    \sigma_1, \sigma_3   \\
  \end{array}
\right.
~.
\end{eqnarray}
This restricts the brane configuration
as follows:
(odd,odd)-branes with $\rho=1$
and (even,even)-branes with $\rho=\sigma_1,\sigma_3$ for $p=1\mod 4$,
and (odd,odd)-branes with $\rho=i\sigma_2$ for $p=3\mod 4$.
We summarize the result in Table \ref{euclidean pp-wave branes 10}.
\begin{table}[htbp]
 \begin{center}
  \begin{tabular}{|c|c|c|c|c|c|}
    \hline
     $\rho$  &1-brane    &3-brane    &5-brane    &7-brane    &9-brane    \\
    \hline
     $\sigma_1,\sigma_3$  &(0,2),(2,0)    &    &(2,4),(4,2)    &    &-    \\

    \hline
     $i\sigma_2$  &    &(1,3),(3,1)    &    &-    &    \\
    \hline
     1  &(1,1)    &    &(1,3),(3,1)    &    &(5,5)    \\
             \hline
  \end{tabular}
 \end{center}
  \caption{Euclidean pp-wave branes}
  \label{euclidean pp-wave branes 10}
\end{table}
(even,even)-branes of $p=1\mod 4$
and (odd,odd)-branes of $p=3\mod 4$
are 1/2 BPS D-branes of an open pp-wave superstring
\cite{Bain:2002tq,Sakaguchi:2003py}.

Scaling $Q^{(\bullet)}_\pm$ as
(\ref{Q scale ppNH 10})
and taking the limit $\Omega\to 0$,
we obtain the fermionic part of the NH superalgebra
of a Euclidean pp-wave
brane
\begin{eqnarray}
&&
[Q^{(-)}_\pm,P_{\bar\ih}]=
-\frac{1}{2\sqrt{2}}Q^{(+)}_\pm\Gamma_{\bar\ih}\Gamma_+fi\sigma_2~,~~~
[Q^{(-)}_+,P_{\underline\ih}]=
-\frac{1}{2\sqrt{2}}Q^{(+)}_\mp\Gamma_{\underline\ih}\Gamma_+fi\sigma_2~,~~~
\nonumber\\&&
[Q^{(+)}_+,P_-]=
-\frac{1}{\sqrt{2}}Q^{(+)}_-fi\sigma_2~,~~~
[Q^{(-)}_+,P^*_{\bar\ih}]=
\frac{1}{2\sqrt{2}}Q^{(+)}_-\Gamma_+\Gamma_{\bar\ih}~,~~~
\nonumber\\&&
[Q^{(-)}_\pm,P^*_{\underline\ih}]=
\frac{1}{2\sqrt{2}}Q^{(+)}_\pm\Gamma_+\Gamma_{\underline\ih}~,~~~
[Q^{(\bullet)}_\pm,J_{\bar\ih\bar\jh}]=
-\frac{1}{2}Q^{(\bullet)}_\pm\Gamma_{\bar\ih\bar\jh}~,~~~
[Q^{(\bullet)}_\pm,J_{\underline\ih\underline\jh}]=
-\frac{1}{2}Q^{(\bullet)}_\pm\Gamma_{\underline\ih\underline\jh}~,~~~
\nonumber\\&&
[Q^{(\bullet)}_+,J_{\bar\ih\underline\jh}]=
-\frac{1}{2}Q^{(\bullet)}_-\Gamma_{\bar\ih\underline\jh}~,~~~
\nonumber\\&&
\{Q^{(+)}_\pm,Q^{(+)}_\mp\}=
2i\CC\Gamma_-P_+~,
\nonumber\\&&
\{Q^{(-)}_\pm,Q^{(-)}_\mp\}=
2i\CC\Gamma_+P_-
-2i\frac{\lambda}{\sqrt{2}}\CC\widehat\Gamma^{\bar\ih\underline\jh}i\sigma_2
J_{\bar\ih\underline\jh}~,
\nonumber\\&&
\{Q^{(-)}_+,Q^{(-)}_+\}=
-i\frac{\lambda}{\sqrt{2}}\CC\widehat\Gamma^{\bar\ih\bar\jh}i\sigma_2
J_{\bar\ih\bar\jh}
-i\frac{\lambda}{\sqrt{2}}\CC\widehat\Gamma^{\underline\ih\underline\jh}i\sigma_2
J_{\underline\ih\underline\jh}~,
\nonumber\\&&
\{Q^{(\pm)}_+,Q^{(\mp)}_+\}=
2i\CC\Gamma^{\bar\ih}\ell_\mp P_{\bar\ih}
+i\lambda\CC\widehat\Gamma^{\underline\ih}\ell_\mp i\sigma_2
P^*_{\underline\ih}~,
\nonumber\\&&
\{Q^{(\pm)}_+,Q^{(\mp)}_-\}=
2i\CC\Gamma^{\underline\ih}\ell_\mp P_{\underline\ih}
+i\lambda\CC\widehat\Gamma^{\bar\ih}\ell_\mp i\sigma_2
P^*_{\bar\ih}~.
\label{NH fermionic euc pp 10}
\end{eqnarray}
Summarizing we have obtained the NH superalgebra
of Euclidean pp-wave brane as (\ref{NH bosonic euc pp 10}),
(\ref{NH bosonic Lorentz pp 10}) and (\ref{NH fermionic euc pp 10}).
Obviously, this superalgebra can be derived from the NH superalgebra of an AdS brane
(\ref{IIB NH bosonic}) and (\ref{IIB NH fermionic})
by an IW contraction.

We note that the NH superalgebra contains a super-subalgebra
generated by 
$P_{\bar\ih}$, $P^*_{\underline\ih}$, 
$J_{\bar\ih\bar\jh}$,
$J_{\underline\ih\underline\jh}$
and $Q_+^{(\pm)}$
\begin{eqnarray}
&&
[P_{\bar{\hat i}},J_{\bar{\hat j}\bar{\hat k}}]
=\eta_{\bar{\hat i}\bar{\hat j}}P_{\bar{\hat k}}
-\eta_{\bar{\hat i}\bar{\hat k}}P_{\bar{\hat j}}~,~~~
[J_{\underline{\hat i}\underline{\hat j}},P^*_{\underline{\hat k}}]=
-\eta_{\underline{\hat i}\underline{\hat k}}P^*_{\underline{\hat j}}
+\eta_{\underline{\hat j}\underline{\hat k}}P^*_{\underline{\hat i}}~,~~~
\nonumber\\&&
[J_{\bar{\hat i}\bar{\hat j}},J_{\bar{\hat k}\bar{\hat l}}]=
\eta_{\bar{\hat j}\bar{\hat k}}J_{\bar{\hat i}\bar{\hat l}}+
\text{3-terms}~,~~~
[J_{\underline{\hat i}\underline{\hat j}},J_{\underline{\hat k}\underline{\hat l}}]=
\eta_{\underline{\hat j}\underline{\hat k}}J_{\underline{\hat i}\underline{\hat l}}+
\text{3-terms}~
\nonumber\\
&&
[Q^{(-)}_+,P_{\bar\ih}]=
-\frac{1}{2\sqrt{2}}Q^{(+)}_+\Gamma_{\bar\ih}\Gamma_+fi\sigma_2~,~~~
[Q^{(-)}_+,P^*_{\underline\ih}]=
\frac{1}{2\sqrt{2}}Q^{(+)}_+\Gamma_+\Gamma_{\underline\ih}~,~~~
\nonumber\\&&
[Q^{(\bullet)}_+,J_{\bar\ih\bar\jh}]=
-\frac{1}{2}Q^{(\bullet)}_+\Gamma_{\bar\ih\bar\jh}~,~~~
[Q^{(\bullet)}_+,J_{\underline\ih\underline\jh}]=
-\frac{1}{2}Q^{(\bullet)}_+\Gamma_{\underline\ih\underline\jh}~,~~~
\nonumber\\&&
\{Q^{(-)}_+,Q^{(-)}_+\}=
-i\frac{\lambda}{\sqrt{2}}\CC\widehat\Gamma^{\bar\ih\bar\jh}i\sigma_2
J_{\bar\ih\bar\jh}
-i\frac{\lambda}{\sqrt{2}}\CC\widehat\Gamma^{\underline\ih\underline\jh}i\sigma_2
J_{\underline\ih\underline\jh}~,
\nonumber\\&&
\{Q^{(\pm)}_+,Q^{(\mp)}_+\}=
2i\CC\Gamma^{\bar\ih}\ell_\mp P_{\bar\ih}
+i\lambda\CC\widehat\Gamma^{\underline\ih}\ell_\mp i\sigma_2
P^*_{\underline\ih}~
\end{eqnarray}
which  is regarded as a supersymmetrization
of the Poincar\'e algebra generated by $\{P_{\bar\ih},J_{\bar\ih\bar\jh}\}$
which is the isometry on the brane worldvolume
and the Lorentz symmetry
in the transverse space
generated by $\{P^*_{\underline\ih},J_{\underline\ih\underline\jh}\}$.
The conditions
(\ref{condition 1 pp 10}) and (\ref{condition 2 pp 10})
ensure the existence of this super-subalgebra.

\section{Branes in AdS$_5\times$S$^5$}
A D-brane action \cite{D:curved}
(see \cite{D:flat} for flat D-branes) is composed of
the Dirac-Born-Infeld (DBI) action
and the WZ action
\begin{eqnarray}
S=S_{\DBI}+S_{\WZ}~.
\label{S 10}
\end{eqnarray}
The DBI action  is given, 
suppressing the dilaton and axion factors here, 
as
\begin{eqnarray}
S_{\DBI}=T\int_\Sigma \, \CL_{\DBI}~,~~
\CL_{\DBI}=\sqrt{s\det(g+\CF)}\,d^{p+1}\xi
\end{eqnarray}
where $\CF=F-\CB$ and $F=dA$, and $s=-1$ for a Lorentzian
brane while $s=1$ for a Euclidean brane. 
$T$ is the tension of the brane.
$\CB$ is the pullback of the NS-NS two-form and $A$ is the gauge field
on the worldvolume.
For an F-string, the DBI action is replaced by the Nambu-Goto (NG) action
\begin{eqnarray}
S_{\NG}=T\int_\Sigma \CL_{\NG}~,~~~
\CL_{\NG}=\sqrt{s\det g}\,d^{2}\xi ~.
\end{eqnarray}
The WZ action\footnote{
See \cite{Hatsuda:2004vi}
for the Roiban-Siegel formulation \cite{Roiban:2000yy}
of AdS D-branes.
} is characterized by
supersymmetric closed $(p+2)$-form $h_{p+2}$
\begin{eqnarray}
S_{\WZ}=T\int_\Sigma \CL_{\WZ}~,~~~
h_{p+2}=d\CL_{\WZ}
=\sum_{n=0}\frac{1}{n!}h^{(p+2-2n)}\CF^n~.
\label{h_{p+2}}
\end{eqnarray}
The closedness of $h_{p+2}$
\begin{eqnarray}
0=dh_{p+2}=\sum_{n=0}\frac{1}{n!}\left(
dh^{(p+2-2n)}-h^{(p-2n)}d\CF
\right)
\CF^n
\end{eqnarray}
implies
\begin{eqnarray}
dh^{(p+2-2n)}-h^{(p-2n)}d\CF=0~.
\label{closedness}
\end{eqnarray}

\subsection{CE-cohomology classification} 

In  \cite{DeAzcarraga:1989vh},
it is shown that
the Wess-Zumino (WZ)  terms of $p$-branes in flat spacetime
can be classified  as non-trivial elements of the
Chevalley-Eilenberg (CE) cohomology \cite{CE}.
Let $C^p(\g,\bR)$ be
the vector space of
$p$-cochains of a Lie algebra $\g$.
A $p$-cochain is a linear antisymmetric  map:
$\g\times\cdots\times\g\mapsto \bR$
and 
a coboundary operator $\delta$
with $\delta^2=0$ acts as
$C^p(\g,\bR)\mapsto C^{p+1}(\g,\bR)$.
The CE cohomology group $H^p(\g,\bR)$
is defined by $Z^p/B^p$
where $Z^p$ and $B^p$ are the vector spaces
of $p$-cocycles $c\in Z^p$ satisfying $\delta c=0$
and $p$-coboundaries $c\in B^p$ satisfying $c=\delta c'$
with $c'\in C^{p-1}(\g,\bR)$,
respectively.
In the present context, this is viewed as
the de Rham cohomology group $E^p(G,\bR)$
for left-invariant (LI) $p$-forms on
the supergroup 
$G=$``super-Poincar\'e''/``Lorentz'',
for which a non-trivial element of the cohomology
is a closed LI $p$-form modulo exact LI $p$-forms on $G$.
This is generalized to
D-branes in \cite{Chryssomalakos:1999xd,Sakaguchi:1999fm}
by introducing an additional two form
which corresponds to the modified field strength of
background $\CB$ field.

Here we examine WZ terms
of AdS branes
by using the CE cohomology
on
$\g$ of the supergroup
$G=$PSU(2,2$|$4)/(SO(4,1)$\times$SO(5)), 
i.e.
``super-AdS$_5\times$S$^5$''/``Lorentz''.
We show that except for the $p=1$ case
$h_{p+2}$ can be obtained as 
a Lorentz invariant non-trivial element of
the CE-cohomology on the free differential algebra
which is the 
 MC equations (\ref{MC AdSxS in 10-dim})
corresponding to the super-AdS$_5\times$S$^5$ algebra
(\ref{AdS5xS5 algebra}) 
equipped with
\begin{eqnarray}
d\CF&=&-i\BL^A\bar L\Gamma_A\sigma L
\label{dF in 10-dim}
\end{eqnarray}
where
$\sigma$ is $\sigma_3$ for D-branes
while $-\sigma_1$ for F1- and NS5-branes.

In order not to introduce an additional dimensionful parameter
we assign a dimension to Cartan one-forms as follows
\begin{eqnarray}
  \begin{array}{cccccccc}
&   \BL^A &L^\alpha &\BL^{AB}&\lambda &\CF & h_{p+2}    &h^{(k)}    \\
\text{dim} &   1   & 1/2   &0&-1    &2    &p+1    &k-1    \\
  \end{array}
\end{eqnarray}
where $\dim h_{p+2}=p+1$ because $\dim h_{p+2}=\dim \CL_{\WZ}^{p}
=\dim \CL_{BI}^{p}=p+1$ for structureless fundamental branes.

Suppose that $h^{(k)}$ is of the form $(\BL^A)^n(L^\alpha)^m\lambda^l$,
then $n,~m$ and $l$ must satisfy
\begin{eqnarray}
n+\frac{1}{2}m-l=k-1,~~~
n+m=k,
\label{condition CE h}
\end{eqnarray}
because $h^{(k)}$ is a Lorentz invariant $k$-form of dimension $k-1$.
We require that
$\epsilon_{a_1\cdots a_5}$ and $\epsilon_{a_1'\cdots a_5'}$
are accompanied with $\lambda$;
$\lambda\epsilon_{a_1\cdots a_5}$ and $\lambda\epsilon_{a_1'\cdots a_5'}$,
because $\epsilon_{a_1\cdots a_5}$ and $\epsilon_{a_1'\cdots a_5'}$
disappear in the flat limit $\lambda\to 0$.
Requiring $l\ge 0$ because otherwise $h^{(k)}$ diverges in the 
flat limit.
This  implies $l=-\frac{1}{2}m+1\le 1$ and so we consider $l=0,1$.
Since 
(\ref{condition CE h}) is satisfied for
$(m,n)=(2,k-2)$, $(0,k)$ for $l=0,1$, respectively,
we find that $h^{(k)},~k=1,3,5,...$, has the following form
\begin{eqnarray}
h^{(1)}&=&0,\\
h^{(3)}&=&c^{(3)}_0\BL^a\bar L\Gamma_a\varrho^{(3)}_0L
+c^{(3)}_1\BL^{a'}\bar L\Gamma_{a'}\varrho^{(3)}_1L
,\\
h^{(5)}&=&c^{(5)}_0\BL^{a_1}\BL^{a_2}\BL^{a_3}\bar L\Gamma_{a_1a_2a_3}
 \varrho^{(5)}_0L
+\cdots+
c^{(5)}_3\BL^{a'_1}\BL^{a'_2}\BL^{a'_3}\bar L\Gamma_{a'_1a'_2a'_3}
 \varrho^{(5)}_3L 
\nonumber\\&&
 +b_0\lambda\epsilon_{a_1\cdots a_5}\BL^{a_1}\cdots \BL^{a_5}
 +b_5\lambda\epsilon_{a'_1\cdots a'_5}\BL^{a'_1}\cdots \BL^{a'_5},\\
h^{(7)}&=&c^{(7)}_0\BL^{a_1}\cdots \BL^{a_5}\bar L\Gamma_{a_1\cdots a_5}
 \varrho^{(7)}_0L
+\cdots+
c^{(7)}_5\BL^{a'_1}\cdots \BL^{a'_5}\bar L\Gamma_{a'_1\cdots a'_5}
 \varrho^{(7)}_5L 
 ,\\
h^{(9)}&=&c^{(9)}_0\BL^{a_1}\cdots \BL^{a_7}\bar L\Gamma_{a_1\cdots a_7}
 \varrho^{(9)}_0L
+\cdots +
c^{(9)}_7\BL^{a'_1}\cdots \BL^{a'_7}\bar L\Gamma_{a'_1\cdots a'_7}
 \varrho^{(9)}_7L,
\end{eqnarray}
where
$c_i^{(k)}$ and $b_i$ are constants determined below.
$\varrho^{(k)}_i$ are $2\times 2$ matrices
satisfying
${\varrho^{(k)}}^T=\varrho^{(k)}$ for $k=3,7$
while ${\varrho^{(k)}}^T=-\varrho^{(k)}$ for $k=5,9$,
because $\CC\Gamma^{A_1\cdots A_N}$
is symmetric for $N=1,2\mod 4$
and anti-symmetric otherwise.

It is straightforward to solve (\ref{closedness})
 to determine
coefficients and $\varrho^{(k)}_i$. 
We find 
\begin{eqnarray}
h^{(1)}&=&0,\\
h^{(3)}&=&c
\BL^A\bar L\Gamma_A\varrho L
,\label{h3 10-dim}\\
h^{(5)}&=&\frac{
c}{3!}
\Bigl[
\BL^{A_1}\BL^{A_2}\BL^{A_3}\bar L\Gamma_{A_1A_2A_3}i\sigma_2L
+\frac{i}{5}\lambda\bigl(
 \epsilon_{a_1\cdots a_5}\BL^{a_1}\cdots \BL^{a_5}
 -\epsilon_{a'_1\cdots a'_5}\BL^{a'_1}\cdots \BL^{a'_5}\bigr)\Bigr],
 ~~~\label{h5 10-dim}\\
h^{(7)}&=&\frac{
c}{5!}
\BL^{A_1}\cdots \BL^{A_5}\bar L\Gamma_{A_1\cdots A_5}\varrho L
,\label{h7 10-dim}\\
h^{(9)}&=&\frac{
c}{7!}
\BL^{A_1}\cdots \BL^{A_7}\bar L
 \Gamma_{A_1\cdots A_7}
 i\sigma_2L
 ~.
\label{h9 10-dim}
\end{eqnarray}
In Appendix \ref{appendix:kappa},
$c=c_0^{(3)}$ is determined by the $\kappa$-invariance
\cite{D:curved}
of the total action $S$
as $c=i$ and $1$ for Lorentzian and Euclidean branes respectively:
$c=\sqrt{s}$.
$\varrho$ is $\sigma_1(\sigma_3)$ for $\sigma=\sigma_3(-\sigma_1)$ respectively.
The closedness (\ref{closedness}) is ensured by
the Fierz identities
\begin{eqnarray}
&&
(\CC\Gamma_A)_{(\alpha\beta}(\CC\Gamma^A\varrho)_{\gamma\delta)}=0~,
\nonumber\\&&
(\CC\Gamma_C)_{(\alpha\beta}(\CC\Gamma^{ABC}i\sigma_2)_{\gamma\delta)}
+2(\CC\Gamma^{[A}\varrho)_{(\alpha\beta}(\CC\Gamma^{B]}\sigma)_{\gamma\delta)}=0
~,
\nonumber\\&&
(\CC\Gamma_{B})_{(\alpha\beta}(\CC\Gamma^{A_1\cdots A_4B}\varrho)_{\gamma\delta)}
+4(\CC\Gamma^{[A_1A_2A_3}i\sigma_2)_{(\alpha\beta}
 (\CC\Gamma^{A_4]}\sigma)_{\gamma\delta)}=0
 ~,
\nonumber\\&&
(\CC\Gamma_B)_{(\alpha\beta}(\CC\Gamma^{A_1\cdots A_6B}i\sigma_2)_{\gamma\delta)}
+6(\CC\Gamma^{[A_1\cdots A_5}\varrho)_{(\alpha\beta}(\CC\Gamma^{A_6]}\sigma)_{\gamma\delta)}=0
~.
\end{eqnarray}
In summary, closed $(p+2)$-forms $h_{p+2}$ are
composed in terms of $h^{(k)}$ found above as in (\ref{h_{p+2}}). 
The actions $S$ for F1- and D3-branes
coincide with those obtained in \cite{F1:AdS} and \cite{D3:AdS},
respectively.

We show that $h_{p+2}$ is a non-trivial element of the cohomology
except for $h_3$.
If $h_{p+2}$ is exact, there exists $b_{p+1}$ such as $h_{p+2}=db_{p+1}$.
Since 
\begin{eqnarray}
h_3=db_2~,~~~b_2=-c\lambda^{-1}\bar L\CI\varrho i\sigma_2 L
\end{eqnarray}
$h_3$ is a trivial element of the cohomology \cite{Hatsuda:2002hz,Hatsuda:2002iu}.
Next we show that
$h_{p+2}$ with $p=3,5,7$ is not exact.
Let us examine 
a term of the form 
$\frac{1}{(\frac{p-1}{2})!}h^{(3)}\CF^{\frac{p-1}{2}}$
contained in $h_{p+2}$.
We note that $\CF$ can be written as\footnote{
This implies that
$h_{2}=h^{(2)}+h^{(0)}\CF$
with $h^{(0)}=i\hat c\lambda$ and 
$h^{(2)}=-\hat c\bar L\CI\sigma i\sigma_2L$
can be a nontrivial element of the cohomology.
It is interesting to examine the 0-brane action with the WZ term $h_2$.
}
\begin{eqnarray}
\CF=ic\lambda^{-1}\bar L\CI\sigma i\sigma_2 L~
\end{eqnarray}
up to an exact form, 
and that there does not exist a one-form supercurrent $f$ such that $\CF=df$. 
So $b_{p+1}$ must contain a term of the form 
$\bar L\CI\varrho i\sigma_2 L\CF^{\frac{p-1}{2}}$.
Differentiating it, we have
$\bar L\CI\varrho i\sigma_2 L~L^A\bar L\Gamma_A\sigma L\CF^{\frac{p-3}{2}}$
in addition to $\frac{1}{(\frac{p-1}{2})!}h^{(3)}\CF^{\frac{p-1}{2}}$.
For $h_{p+2}$ to be exact,
this term must be canceled by the differential of a term
which is a $(p+1)$-form with $p-1$ $L^\alpha$'s.
From the MC equation (\ref{MC AdSxS in 10-dim}),
we see that there does not exist such a term.
Thus $h_{p+2}$ with $p=3,5,7$ obtained above are non-trivial elements of the cohomology.

\subsection{$(p+1)$-dimensional form of the WZ term}
In this subsection, we give the $(p+1)$-dimensional 
form of the WZ term $h_{p+2}$.
We follow \cite{F1:AdS,D3:AdS}
in which the $(p+1)$-dimensional 
form of the WZ term of F1- and D3-branes
are given.

The LI Cartan one-forms satisfy
the following differential equations
\begin{eqnarray}
\partial_t\hat \BL^{A}&=&
2i\bar\theta\Gamma^{A}\hat L~,
\label{diff 1}\\
\partial_t\hat L^\alpha&=&
d\theta
+\frac{\lambda}{2}\hat \BL^A\widehat\Gamma_Ai\sigma_2\theta
+\frac{1}{4}\hat\BL^{AB}\Gamma_{AB}\theta~,
\label{diff 2}\\
\partial_t\hat\BL^{AB}&=&
-2i\lambda\bar\theta\widehat\Gamma^{AB}i\sigma_2\hat L
\label{diff 3}
\end{eqnarray}
where
a ``hat'' on a supercurrent
implies that $\theta$ is rescaled as $\theta\to t\theta$.
First we note that
\begin{eqnarray}
\partial_td\hat\CF&=&
-\partial_td\hat\CB
=-2id(\hat\BL^A\hat{ \bar L}\Gamma_A\sigma\theta)~.
\end{eqnarray}
This is solved by
\begin{eqnarray}
\CB=2i\int^1_0\!\!dt\,
\hat\BL^A\hat{ \bar L}\Gamma_A\sigma\theta
+B^{(2)}~.
\label{B}
\end{eqnarray}
where $B^{(2)}$ is a bosonic 2-form satisfying $dB^{(2)}=0$.
Thus we obtain
\begin{eqnarray}
\CF&=&F-2i\int^1_0\!\!dt\,
\hat\BL^A\hat{ \bar L}\Gamma_A\sigma\theta
-B^{(2)}~,
\label{CF}\\
\partial_t\hat\CF&=&
-2i(\hat\BL^A\hat{ \bar L}\Gamma_A\sigma\theta)~.
\label{partiat CF}
\end{eqnarray}
For D-brane actions, we choose $B^{(2)}=0$.

By using (\ref{diff 1})-(\ref{diff 3}) and (\ref{partiat CF}),
one sees that
the closed  $(p+1)$-form $h_{p+2}$
satisfies
\begin{eqnarray}
\partial_t\hat h_{p+2}&=&
d b_{p+1}~,
\end{eqnarray}
where
\begin{eqnarray}
b_{p+1}&=&[\CC\wedge \e^{\hat\CF}]_{p+1}~,~~~
\CC=\bigoplus_{\ell=\text{even}}\CC^{(\ell)}~,\nonumber\\
\CC^{(2n)}&=&
\frac{2\sqrt{s}}{(2n-1)!}\hat\BL^{A_1}\cdots\hat\BL^{A_{2n-1}}
\hat{\bar L}\Gamma_{A_1\cdots A_{2n-1}}(\sigma)^{n}i\sigma_2\theta
~.
\end{eqnarray}
It follows that
\begin{eqnarray}
\int_{B}h_{p+2}
=\int_{\Sigma}\CL_{\WZ}
=\int_{\Sigma} \left[
\int_0^1 \!\!d t\,  b_{p+1}
+C^{(p+1)}
\right]~
\label{WZ p+1}
\end{eqnarray}
where $\partial B=\Sigma$,
and
$C^{(p+1)}$ is a bosonic $(p+1)$-form satisfying
\begin{eqnarray}
\left.h_{p+2}\right|_{\text{bosonic}}=dC^{(p+1)}~.
\end{eqnarray}
Letting $p=1$ and $\sigma=-\sigma_1$,
we reproduce 
the WZ term of an F-string.

\section{Non-relativistic Branes in AdS$_5\times$S$^5$}

In \cite{Gomis:2005pg},
the non-relativistic F-string in AdS$_5\times$S$^5$
is examined. 
There the leading contributions
of the NG and the WZ parts
in the non-relativistic limit
cancel each other,
and the next-to-leading terms 
contribute to the non-relativistic F-string action.
Thus, in order to extract 
non-relativistic brane actions,
we need to know the next-to-leading order terms
in the  limit $\Omega\to 0$.
Let us consider the scaling
\begin{eqnarray}
&&X^{\underline A}\to \Omega X^{\underline A}~,~~~
\theta_-\to \Omega\theta_-~,
\label{coordinate scaling}
\\&&
T=\Omega^{-2}T_{\NR}~,~~~
F=\Omega F_1~.
\end{eqnarray}
(\ref{coordinate scaling}) 
is consistent with the scaling (\ref{Omega bosonic}) and (\ref{Omega fermionic}).
It is straightforward to see
that by substituting (\ref{coordinate scaling})
into the concrete expression of the supercurrents
given in Appendix \ref{parametrization 10},
$\BL^{A}$ and $L$ are expanded as
\begin{eqnarray}
&&
\BL^{\bar A}=\sum_{n=0}\Omega^{2n}\BL^{\bar A}_{2n}~,~~~
\BL^{\underline A}=\sum_{n=0}\Omega^{2n+1}\BL^{\underline A}_{2n+1}~,
\nonumber\\&&
L_+=\sum_{n=0}\Omega^{2n}L_{+2n}~,~~~
L_-=\sum_{n=0}\Omega^{2n+1}L_{-2n+1}~.
\label{expansion PQ 10}
\end{eqnarray}
Expand $\BL^{ A B}$ as
\begin{eqnarray}
\BL^{\bar A\bar B}=\sum_{n=0}\Omega^{2n}\BL^{\bar A\bar B}_{2n}~,~~
\BL^{\underline A\underline B}=\sum_{n=0}\Omega^{2n}\BL^{\underline A\underline B}_{2n}~,~~
\BL^{\bar A\underline B}=\sum_{n=0}\Omega^{2n+1}\BL^{\bar A\underline B}_{2n+1}~,
\label{expansion J 10}
\end{eqnarray}
and substitute (\ref{expansion PQ 10}) and (\ref{expansion J 10})
into the MC equation (\ref{MC AdSxS in 10-dim})
for the super-AdS$_5\times$S$^5$ algebra,
then the LI Cartan one-forms
$\{
\BL_0^{\bar A},
\BL_1^{\underline A},
L_{+0},
L_{-1},
\BL^{\bar A\bar B}_0,
\BL^{\underline A\underline B}_0,
\BL^{\bar A\underline B}_1
\}$
form
the MC equations (\ref{MC NH 10})-(\ref{MC NH 10 last})
for the NH superalgebra.
\footnote{
As will be seen below,
the non-relativistic actions
are composed of
$\{\BL^{\bar A}_0,\BL^{\bar A}_2,\BL^{\underline A}_1,L_{+0},L_{+2},L_{-1}\}$.
So these actions are not invariant
under the NH superalgebra,
but under
an expanded superalgebra \cite{Hatsuda:2002hz,Hatsuda:2001pp}
(see also \cite{deAzcarraga:2002xi,deAzcarraga:2004zj})
which is  a generalization of the IW contraction\cite{IW},
generated by generators dual to 
$\{
\BL_m^{\bar A},
\BL_m^{\underline A},
L_{\pm m},
\BL^{\bar A\bar B}_m,
\BL^{\underline A\underline B}_m,
\BL^{\bar A\underline B}_m
\,|\, 0\le m\le 2\}$.
}

We consider the non-relativistic limit of the AdS branes
obtained in the previous section.
In the following subsections, 
we will show that
when we introduce
\begin{eqnarray}
M=\sqrt{-s}\Gamma^{\bar A_0\cdots \bar A_p}\otimes \rho
\end{eqnarray}
with $\rho=\sigma_1(i\sigma_2)$ for D$p$-branes with $p=1(3)\mod 4$,
respectively,
and with
$\rho=\sigma_3$ for F1 and NS5,
AdS $p$-brane actions  admit
expansion
\begin{eqnarray}
S=T_{\NR}\int_{\Sigma}\Big[
\Omega^{-2}(\CL_{\{{\NG \atop \DBI}}^{\div}+\CL_{\WZ}^{\div})
+\CL_{\{{\NG \atop \DBI}}^{\fin}+\CL_{\WZ}^{\fin}
+O(\Omega^2)
\Big]~.
\end{eqnarray}
For the consistent non-relativistic limit $\Omega\to 0$,
the divergent term $\int (\CL_{\{{\NG \atop \DBI}}^{\div}+\CL_{\WZ}^{\div})$
should cancel out.
First, we show that
\begin{eqnarray}
d\CL_{\{{\NG \atop \DBI}}^{\div}+h_{p+2}^{\div}=0~.
\end{eqnarray}
This implies that
the divergent terms  with $\theta$ cancel out,
since $h_{p+2}^{\div}$ is composed of only terms with $\theta$.
Next, we consider
the bosonic terms of $\CL_{\{{\NG \atop \DBI}}^{\div}+\CL_{\WZ}^{\div}$
\begin{eqnarray}
\frac{1}{(p+1)!}\epsilon_{\bar A_0\cdots \bar A_p}e^{\bar A_0}_0\cdots e^{\bar A_p}_0
+
C_0^{(p+1)}
\end{eqnarray}
where $C^{(p+1)}_0$ is the leading contribution of
$C^{(p+1)}$ in (\ref{WZ p+1}).
This is deleted by choosing 
$C_0^{(p+1)}=-\frac{1}{(p+1)!}\epsilon_{\bar A_0\cdots \bar A_p}e^{\bar A_0}_0\cdots e^{\bar A_p}_0
$.
It is easy to see that
$dC^{(p+1)}_0=0$
by using the expressions given in Appendix \ref{parametrization 10}.
Thus the bosonic divergent terms
also cancel out.
As a result,
we derive the non-relativistic brane action
\begin{eqnarray}
S_{\NR}=T_{\NR}\int_\Sigma\CL_{\NR}~,~~~
\CL_{\NR}=\CL_{\{{\NG \atop \DBI}}^{\fin}+\CL_{\WZ}^{\fin}
\label{S_NR}
\end{eqnarray}
which is drastically simplified by gauge fixing the $\kappa$-symmetry by $\theta_+=0$.
We examine each AdS branes in turn below.

\subsection{F-string}
First, we consider an F-string.
The 3-form
$h_3$ is given in (\ref{h3 10-dim}) with $\varrho=\sigma_3$.
The gluing matrix $M$ is 
\begin{eqnarray}
M=\sqrt{-s}\Gamma^{\bar A_0\bar A_1}\otimes \rho~,~~~\rho=\sigma_1,\sigma_3,1.
\label{M string}
\end{eqnarray}
Since
\begin{eqnarray}
M'\Gamma_{\bar A}\varrho=\Gamma_{\bar A}M\varrho=\pm\Gamma_{\bar A}\varrho M~,~~~
\rho=\left\{
  \begin{array}{l}
\sigma_3,1       \\
\sigma_1  \\
  \end{array}
\right.~,
\end{eqnarray}
$h_3$  is expanded as
\begin{eqnarray}
Th_3&=&T_{\NR}\Omega^{-2}h_3^{\div}+T_{\NR}h_3^{\fin}+O(\Omega^2)~,
\label{h_3 expansion}\\
h_3^{\div}&=&
\sqrt{s}
\BL^{\bar A}_0\bar L_{+0}\Gamma_{\bar A}\varrho L_{+0}~,
\label{h_3 div}\\
h_3^{\fin}&=&
\sqrt{s}\Big[
\BL^{\bar A}_0\bar L_{-1}\Gamma_{\bar A}\varrho L_{-1}
+
\BL^{\bar A}_2\bar L_{+0}\Gamma_{\bar A}\varrho L_{+0}
\nonumber\\&&
+2
\BL^{\bar A}_0\bar L_{+0}\Gamma_{\bar A}\varrho L_{+2}
+2
\BL^{\underline A}_1\bar L_{+0}\Gamma_{\underline A}\varrho L_{-1}
\Big]
~,
\label{h_3 fin}
\end{eqnarray}
for $\rho=\sigma_3,1 $, while $h_3$ is of order $\Omega$ for $\rho=\sigma_1$.
On the other hand, the NG part is expanded as
\begin{eqnarray}
T\CL_{\NG}&=&
T_{\NR}\Omega^{-2}\CL_{\NG}^{\div}+T_{\NR}\CL_{\NG}^{\fin}
+O(\Omega^2)~,
\label{L_NG expansion}\\
\CL_{\NG}^{\div}&=&\sqrt{s\det g_0}d^2\xi
=\det((\BL^{\bar A}_{0})_{i})d^2\xi
=\frac{1}{2}\epsilon_{\bar A\bar B}(\BL_{0}^{\bar A})
(\BL_{0}^{\bar B})~,
\label{L_NG div}\\
\CL_{\NG}^{\fin}&=&\frac{1}{2}\sqrt{s\det g_0}\,
g_0^{ij} 
(g_2)_{ij}
\,d^2\xi~,
\label{L_NG fin}
\end{eqnarray}
with $\epsilon_{\bar A_0\bar A_1}=1$ and 
\begin{eqnarray}
(g_0)_{ij}&=&(\BL_0^{\bar A})_i(\BL_0^{\bar B})_j\eta_{\bar A\bar B}~,
\label{g_0}\\
(g_2)_{ij}&=&
2(\BL_0^{\bar A})_{(i}(\BL_2^{\bar B})_{j)}\eta_{\bar A\bar B}
+(\BL_1^{\underline A})_i(\BL_1^{\underline B})_j\eta_{\underline A\underline B}
~.
\label{g_2}
\end{eqnarray}
The leading contribution
satisfies \cite{Gomis:2005pg}
\begin{eqnarray}
d\CL_{\NG}^{\div}=
\epsilon_{\bar A\bar B}i\bar L_{+0}\Gamma^{\bar A}L_{+0}\BL_{0}^{\bar B}
=-\sqrt{s}\BL_0^{\bar A}\bar L_{+0}\Gamma_{\bar A}\rho L_{+0}
\end{eqnarray}
where
we have used 
(\ref{MC NH 10}) and
$L_+=ML_+$.
This cancels out $h_3^{\div}$ 
in (\ref{h_3 div})
only when $\varrho
=\rho$
\begin{eqnarray}
d\CL_{\NG}^{\div}
+h_3^{\div}=0~.
\end{eqnarray}
This implies that
$\theta$-dependent terms in $\CL_{\NG}^{\div}+\CL_{\WZ}^{\div}$
cancel each other.
The bosonic term  of $\CL_{\NG}^{\div}$ in (\ref{L_NG div}),
$\frac{1}{2}\epsilon_{\bar A\bar B}e^{\bar A}_0e^{\bar B}_0$,
is deleted by choosing
$C_0^{(2)}$ in (\ref{WZ p+1})
as
\begin{eqnarray}
C_0^{(2)}=
-\frac{1}{2}\epsilon_{\bar A\bar B}e^{\bar A}_0e^{\bar B}_0~
\end{eqnarray}
which satisfies $dC_0^{(2)}=0$.
Thus, the gluing matrix (\ref{M string}) 
with $\rho=\sigma_3$ leads to the consistent
non-relativistic limit of the F-string.
The non-relativistic  F-string action is
(\ref{S_NR})
with (\ref{L_NG fin}) and
\begin{eqnarray}
\CL_{\WZ}^{\fin}&=&
\int_0^1\!\!dt\,
\label{WZ F1 fin p+1}
2\sqrt{s}\Big[
\hat \BL^{\bar A}_0
(\hat{\bar L}_{-1}\Gamma_{\bar A}\varrho\theta_-
 +\hat{\bar L}_{+2}\Gamma_{\bar A}\varrho\theta_+)
\nonumber\\&&
+\hat\BL^{\underline A}_1
(\hat{\bar L}_{-1}\Gamma_{\underline A}\varrho\theta_+
 +\hat{\bar L}_{+0}\Gamma_{\underline A}\varrho\theta_-)
+\hat\BL^{\bar A}_2
\hat{\bar L}_{+0}\Gamma_{\bar A}\varrho\theta_+
\Big]~.
\end{eqnarray}

We fix the $\kappa$-gauge symmetry of the action
by $\theta_+=0$
(see Appendix \ref{appendix:kappa}).
Then we have
\begin{eqnarray}
&&
\BL^{\bar A}_0=e^{\bar A}_0~,~~~
\BL^{\bar A}_2=e^{\bar A}_2+i\bar\theta_-\Gamma^{\bar A}\mathrm{D}\theta_-~,~~~
\BL^{\underline A}_1=e^{\underline A}_1~,
\nonumber\\&&
L_{-1}=\mathrm{D}\theta_-~,~~~~
\mathrm{D}\theta_-\equiv
d\theta_-
+\frac{\lambda}{2}e^{\bar A}_0\widehat\Gamma_{\bar A}i\sigma_2\theta_-
+\frac{1}{4}\omega^{\bar A\bar B}_0\Gamma_{\bar A\bar B}\theta_-~,
\nonumber\\&&
(g_0)_{ij}=(e^{\bar A}_0)_i(e^{\bar B}_0)_j\eta_{\bar A\bar B}~.
\end{eqnarray}
In the static gauge, $x^{\bar A}=\xi^i$,
$(e^{\bar A}_0)_i$ is the vielbein on the AdS brane worldvolume.
Thanks to the $\kappa$-gauge fixing,
we can perform the $t$-integration in (\ref{WZ F1 fin p+1}) easily.
$\CL_{\NG}^{\fin}$ is reduced to
\begin{eqnarray}
\CL_{\NG}^{\fin}&=&
d^2\xi\sqrt{s\det g_0}
\Big[
g_0^{ij}(e^{\bar A}_0)_i(e^{\bar B}_2)_j\eta_{\bar A\bar B}
+\frac{1}{2}g_0^{ij}(e^{\underline A}_1)_i(e^{\underline B}_1)_j\eta_{\bar A\bar B}
+ig_0^{ij}
\bar\theta_-\gamma_i
\mathrm{D}_j\theta_-
\Big]
\end{eqnarray}
where $\gamma_i=(e^{\bar A}_0)_i\Gamma_{\bar A}$.
By parameterizing the group manifold as in Appendix \ref{parametrization 10},
it is rewritten as
\begin{eqnarray}
\CL_{\NG}^{\fin}&=&
d^2\xi\sqrt{s\det g_0}
\Big[
\frac{1}{2}g_0^{ij}\partial_iy^{\underline A}\partial_jy^{\underline B}
\eta_{\underline A\underline B}
+\frac{\lambda^2}{2}(my^2-n{y'}^2)
+i
\bar\theta_-\gamma^i
\mathrm{D}_i\theta_-
\Big]
\label{NR action NG F1}
\end{eqnarray}
for an $(m,n)$-brane
with $(m,n)$=$(2,0)$, $(0,2)$.
On the other hand, $\CL_{\WZ}^{\fin}$
is reduced to
\begin{eqnarray}
\CL_{\WZ}^{\fin}&=&
\sqrt{s}e^{\bar A}_0\mathrm{D}\bar\theta_-\Gamma_{\bar A}\varrho\theta_-
=
d^2\xi\sqrt{s\det g_0}\Big[
-i\mathrm{D}_i\bar\theta_-\gamma^i\theta_-
\Big]
\label{NR action WZ F1}
\end{eqnarray}
where we have used $\theta_-=-M\theta_-$ in the second equality.
Combining these results,
we obtain the non-relativistic action
\begin{eqnarray}
S_{\NR}^{F1}&=&T_{\NR}\int\!\!d^2\xi\sqrt{s\det g_0}\Big[
\frac{1}{2}g_0^{ij}\partial_iy^{\underline A}\partial_jy_{\underline A}
+\frac{\lambda^2}{2}({m}y^2-{n}{y'}^2)
+2i
\bar\theta_-\gamma^i
\mathrm{D}_i\theta_-
\Big]~.
\end{eqnarray}
This is a free field action of scalars and fermions
propagating on (2,0)- or (0,2)-brane worldvolume.
For the case of a  Lorentzian (2,0)-brane,
this reproduces
the non-relativistic 
AdS$_2$ brane action obtained in \cite{Gomis:2005pg}.

\subsection{D-string}
Secondly, we consider a D-string, 
for which $\varrho=\sigma_1$ and $\sigma=\sigma_3$. 
The gluing matrix $M$ is given in (\ref{M string}).
Since
\begin{eqnarray}
M'\Gamma_{\bar A}\varrho=\Gamma_{\bar A}M\varrho=\pm\Gamma_{\bar A}\varrho M~,~~~
\rho=\left\{
  \begin{array}{l}
\sigma_1,1       \\
\sigma_3  \\
  \end{array}
\right.~,
\end{eqnarray}
$h_3$  is expanded as
(\ref{h_3 expansion}) with $\varrho=\sigma_1$
for $\rho=\sigma_1,1 $, while $h_3$ is of order $\Omega$ for $\rho=\sigma_3$.
We note that
for $\rho=\sigma_1$, $\CF$ is of order $\Omega$
\begin{eqnarray}
\CF&=&\Omega \CF_1+O(\Omega^3)~,
\label{F Omega}\\
\CF_1&=&
F_1
-2i\int_0^1\!\!d t\Big[
\hat\BL^{\bar A}_0( \hat{\bar L}_{+0}\Gamma_{\bar A}\sigma\theta_-
+ \hat{\bar L}_{-1}\Gamma_{\bar A}\sigma\theta_+)
+\hat\BL^{\underline A}_1\hat {\bar L}_{+0}\Gamma_{\underline A}\sigma\theta_+
\Big]
\label{F_1}
\end{eqnarray}
since
\begin{eqnarray}
M'\Gamma_{\bar A}\sigma=-\Gamma_{\bar A}\sigma M~.
\end{eqnarray}
So, the DBI part is expanded as
\begin{eqnarray}
T\CL_{\DBI}&=&T_{\NR}\Omega^{-2}\CL_{\DBI}^{\div}+T_{\NR}\CL_{\DBI}^{\fin}
+O(\Omega^4)~,
\label{L_DBI expansion}\\
\CL_{\DBI}^{\div}&=&\sqrt{s\det g_0}\,d^2\xi~,
\label{L_DBI div}\\
\CL_{\DBI}^{\fin}&=&
\frac{1}{2}\sqrt{s\det g_0}\left(
g_0^{ij}(g_2)_{ij}
-\frac{1}{2} g_0^{ik}(\CF_1)_{kj}g_0^{jl}(\CF_1)_{li}
\right)\,d^2\xi
\label{L_DBI fin}
\end{eqnarray}
where $g_0$, $g_2$ and $\CF_1$
are given in 
(\ref{g_0}), (\ref{g_2}) and (\ref{F_1}),
respectively.
For $\rho=1,\sigma_3$, $\CF$ is of order $\Omega^0$.
As was done for the F-string case, 
the $h_3^{\div}$ in (\ref{h_3 div}) with $\varrho=\sigma_1$ 
and the leading contribution of 
the fermionic part of the DBI action cancel each other.
By choosing $C_0^{(2)}=-\frac{1}{2}\epsilon_{\bar A\bar B}e^{\bar A}_0e^{\bar B}_0$, 
the bosonic terms of
the divergent part cancel out.
Thus, the gluing matrix with $\rho=\sigma_1$ leads to the consistent
non-relativistic limit of the D-string.

The non-relativistic D-string action is given 
by (\ref{S_NR}) with
(\ref{L_DBI fin}) and
\begin{eqnarray}
\CL_{\WZ}^{\fin}&=&\int_0^1\!\!d t\,
2\sqrt{s}\Big[
\hat\BL^{\bar A}_0
(\hat{\bar L}_{-1}\Gamma_{\bar A}\varrho\theta_-
 +\hat{\bar L}_{+2}\Gamma_{\bar A}\varrho\theta_+)
\nonumber\\&&
+\hat\BL^{\underline A}_1
(\hat{\bar L}_{-1}\Gamma_{\underline A}\varrho\theta_+
 +\hat{\bar L}_{+0}\Gamma_{\underline A}\varrho\theta_-)
+\hat\BL^{\bar A}_2
\hat{\bar L}_{+0}\Gamma_{\bar A}\varrho\theta_+
\Big]~.
\end{eqnarray}

Let us gauge fix the $\kappa$-gauge symmetry by choosing $\theta_+=0$.
This makes it easy to perform the $t$-integration in $\CL_{\WZ}^{\fin}$.
The $t$-integration in $\CF_1$ in (\ref{F_1})
disappears and we have $\CF_1=F_1$.
In the similar way in the F-string case, we obtain
the non-relativistic D-string action
\begin{eqnarray}
S_{\NR}^{D1}&=&T_{\NR}\int\!\!d^2\xi\sqrt{s\det g_0}\Big[
\frac{1}{2}g_0^{ij}\partial_iy^{\underline A}\partial_jy_{\underline A}
+\frac{\lambda^2}{2}({m}y^2-{n}{y'}^2)
\nonumber\\&&
+2i
\bar\theta_-\gamma^i
\mathrm{D}_i\theta_-
+\frac{1}{4}(F_1)_{ij}(F_1)^{ij}
\Big]~.
\end{eqnarray}
This is a free field action of scalars, fermions
and a gauge field
propagating on (2,0)- or (0,2)-brane worldvolume.

\subsection{D3-brane}

Thirdly, we consider a D3-brane for which $\varrho=\sigma_1$ and $\sigma=\sigma_3$.
The gluing matrix is
\begin{eqnarray}
M=\sqrt{-s}\Gamma^{\bar A_0\cdots\bar A_3}\otimes i\sigma_2~.
\end{eqnarray}
Since
\begin{eqnarray}
M'\Gamma_{\bar B_1\cdots \bar B_3}i\sigma_2
=\Gamma_{\bar B_1\cdots \bar B_3}i\sigma_2 M~,~~
M'\Gamma_{\bar A}\varrho
=-\Gamma_{\bar A}\varrho M~,~~
M'\Gamma_{\bar A}\sigma
=-\Gamma_{\bar A}\sigma M~,
\end{eqnarray}
$\CF$ and $h_3$ are of order $\Omega$ as in (\ref{F Omega})
and 
the WZ part is expanded as 
\begin{eqnarray}
Th_{5}&=&T_{\NR}\Omega^{-2}h_{5}^{\div}+T_{\NR}h_{5}^{\fin}
+O(\Omega^4)~,\\
h_{5}^{\div}&=&
\frac{\sqrt{s}}{3!}
\BL^{\bar A_1}_0\BL^{\bar A_2}_0\BL^{\bar A_3}_0
\bar L_{+0}\Gamma_{\bar A_1\cdots \bar A_3}i\sigma_2 L_{+0}~,
\label{h_5 div}\\
h_{5}^{\fin}&=&
h^{(5)}_2+h^{(3)}_1\CF_1~
\label{h_5 fin}
\end{eqnarray}
with
\begin{eqnarray}
h^{(5)}_2&=&\frac{\sqrt{s}}{3!}\Big[
\BL^{\bar A_1}_0\BL^{\bar A_2}_0 \BL^{\bar A_3}_0
\bar L_{-1}\Gamma_{\bar A_1\cdots\bar A_3}\sigma_1 L_{-1}
+3\BL^{\bar A_1}_0\BL^{\bar A_{2}}_0 \BL^{\bar A_3}_2
\bar L_{+0}\Gamma_{\bar A_1\cdots\bar A_3}\sigma_1 L_{+0}
\nonumber\\&&
+2\BL^{\bar A_1}_0\BL^{\bar A_2}_0\BL^{\bar A_3}_0
\bar L_{+0}\Gamma_{\bar A_1\cdots\bar A_3}\sigma_1 L_{+2}
+6\BL^{\bar A_1}_0\BL^{\bar A_{2}}_0 \BL^{\underline A_3}_1
\bar L_{+0}\Gamma_{\bar A_1\cdots\bar A_{2}\underline A_3}\sigma_1 L_{-1}
\nonumber\\&&
+6\BL^{\bar A_1}_0\BL^{\underline A_{2}}_1 \BL^{\underline A_3}_1
\bar L_{+0}\Gamma_{\bar A_1\underline A_2\underline A_3}\sigma_1 
L_{+0}
\nonumber\\&&
+
4i\lambda
(\delta^{(3,1)}
\epsilon_{\bar a_1\bar a_2\bar a_3\underline a_4\underline a_5}
\BL_0^{\bar a_1}\BL^{\bar a_2}_0\BL^{\bar a_3}_0\BL^{\underline a_4}_1
\BL^{\underline a_5}_1
-\delta^{(1,3)}
\epsilon_{\bar a_1'\bar a_2'\bar a_3'\underline a_4'\underline a_5'}
\BL^{\bar a_1'}_0\BL^{\bar a_2'}_0\BL^{\bar a_3'}_0
\BL^{\underline a_4'}_1\BL^{\underline a_5'}_1)
\Big]~,~~~
\label{h(5)_2}\\
h^{(3)}_1&=&
\sqrt{s}\Big[
2\BL^{\bar A}_0\bar L_{+0}\Gamma_{\bar A}\varrho L_{-1}
+\BL^{\underline A}_1\bar L_{+0}\Gamma_{\underline A}\varrho L_{+0}
\Big]~,
\label{h(3)_1}
\end{eqnarray}
where
$\delta^{(m,n)}=1$ for a $(m,n)$-brane
and  $\delta^{(m,n)}=0$ for others.
This implies that the bosonic $4$-form $C^{(4)}$ is expanded as
\begin{eqnarray}
TdC^{(4)}&=&T_{\NR}\Omega^{-2}dC^{(4)}_0+T_{\NR}dC^{(4)}_2 +O(\Omega^4)~,\\
dC^{(4)}_0&=&0~,\\
dC^{(4)}_2&=&\frac{\sqrt{s}}{3!}4i\lambda
\big(\delta^{(3,1)}
\epsilon_{\bar a_1\bar a_2\bar a_3\underline a_4\underline a_5}
e_0^{\bar a_1}e^{\bar a_2}_0e^{\bar a_3}_0e^{\underline a_4}_1
e^{\underline a_5}_1
-\delta^{(1,3)}
\epsilon_{\bar a_1'\bar a_2'\bar a_3'\underline a_4'\underline a_5'}
e^{\bar a_1'}_0e^{\bar a_2'}_0e^{\bar a_3'}_0
e^{\underline a_4'}_1e^{\underline a_5'}_1
\big)~.~~~
\label{D3 bosonic C_2}
\end{eqnarray}
On the other hand, 
as $\CF$ is of order $\Omega$,
the DBI part is expanded as in (\ref{L_DBI expansion}).
As was done in the $p=1$ case, we find
\begin{eqnarray}
d(\sqrt{s\det g_0}\,d^{4}\xi)=
d(\det((\BL^{\bar A}_{0})_{i})d^{4}\xi)=
-\frac{\sqrt{s}}{3!}
\BL^{\bar A_1}_0\cdots\BL^{\bar A_3}_0\bar L_{+0}\Gamma_{\bar A_1\cdots \bar A_3}
i\sigma_2 L_{+0}
~.
\end{eqnarray}
Thus 
the fermionic part contained in
$\CL_\WZ^{\div}$  and $\CL_{\DBI}^{\div}$
cancel each other.
In addition, the bosonic terms deleted by choosing
\begin{eqnarray}
C^{(4)}_0=
-\frac{1}{4!}\epsilon_{\bar A_0\cdots\bar A_3}e^{\bar A_0}_0\cdots e^{\bar A_3}_0~
\end{eqnarray}
which satisfies $dC^{(4)}_0 =0$.
Thus the matrix $M$ leads to the consistent non-relativistic 
limit of the AdS D3-brane.

The non-relativistic D3-brane action is given as 
(\ref{S_NR}) with (\ref{L_DBI fin}) and
\begin{eqnarray}
\CL_{\WZ}^{\rm D3}&=&\int_0^1\!\!d t\,
\Big[
 \CC^{(4)}_2
+
\CC^{(2)}_1
\hat\CF_1
\Big]
+C^{(4)}_2
\end{eqnarray}
with
\begin{eqnarray}
\CC^{(4)}_2&=&
2c\Big[\frac{1}{3!}
\hat\BL^{\bar A_1}_0\hat\BL^{\bar A_2}_0\hat\BL^{\bar A_3}_0
(\hat{\bar L}_{-1}\Gamma_{\bar A_1\bar A_2\bar A_3}i\sigma_2\theta_-
 +\hat{\bar L}_{+2}\Gamma_{\bar A_1\bar A_2\bar A_3}i\sigma_2\theta_+)
\nonumber\\&&
+\frac{1}{2}\hat\BL^{\bar A_1}_0\hat\BL^{\bar A_2}_0\hat\BL^{\underline A_3}_1
(\hat{\bar L}_{-1}\Gamma_{\bar A_1\bar A_2\underline A_3}i\sigma_2\theta_+
 +\hat{\bar L}_{+0}\Gamma_{\bar A_1\bar A_2\underline A_3}i\sigma_2\theta_-)
\nonumber\\&&
+\hat\BL^{\bar A_1}_0\hat\BL^{\bar A_2}_0\hat\BL^{\bar A_3}_2
\hat{\bar L}_{+0}\Gamma_{\bar A_1\bar A_2\bar A_3}i\sigma_2\theta_+
+\hat\BL^{\bar A_1}_0\hat\BL^{\underline A_2}_1\hat\BL^{\underline A_3}_1
\hat{\bar L}_{+0}\Gamma_{\bar A_1\underline A_2\underline A_3}i\sigma_2\theta_+
\Big]~,
\nonumber\\
\CC^{(2)}_1&=&
2c\big(
\hat\BL^{\bar A}_0(\hat{\bar L}_{+0}\Gamma_{\bar A}\varrho\theta_-
+\hat{\bar L}_{-1}\Gamma_{\bar A}\varrho\theta_+)
+\hat\BL^{\underline A}_1\hat{\bar L}_{+0}\Gamma_{\underline A}\varrho\theta_+
\big)~.
\end{eqnarray}
The bosonic contribution $C_2^{(4)}$ is
\begin{eqnarray}
\int_\Sigma\!\!C_2^{(4)}&=&
4i\sqrt{s}\lambda \int_{\Sigma}\left[
\delta^{(3,1)}
\vol_{\Sigma_3}
\epsilon_{\underline a\underline b}dy^{\underline a}y^{\underline b}
-
\delta^{(1,3)}
\vol_{\Sigma_3'}
\epsilon_{\underline a'\underline b'}dy^{\underline a'}y^{\underline b'}
\right]~,
\nonumber\\
&=&4i\sqrt{s}\lambda\int\!\!d^4\xi\sqrt{s\det g_0}\Big[
\delta^{(3,1)}\epsilon_{\underline a\underline b}
 \partial_{\xi'}y^{\underline a}y^{\underline b}
-
\delta^{(1,3)}\epsilon_{\underline a'\underline b'}
 \partial_{\xi}y^{\underline a'}y^{\underline b'}
\Big]
\end{eqnarray}
where $\Sigma_m\times\Sigma_n'$ is the ($m,n$)-brane worldvolume,
and $\vol_{\Sigma_\ell}=\frac{1}{\ell!}\epsilon_{\bar a_1\cdots \bar a_\ell}e^{\bar a_1}_0
\cdots e^{\bar a_\ell}_0$.
$\xi(\xi')$ represents the worldvolume direction in AdS$_5$(S$^5$ respectively).
By fixing the $\kappa$-symmetry as $\theta_+=0$,
the non-relativistic action is
simplified as
\begin{eqnarray}
S_{\NR}^{D3}&=&T_{\NR}\int\!\!d^4\xi\sqrt{s\det g_0}\Big[
\frac{1}{2}g_0^{ij}\partial_iy^{\underline A}\partial_jy^{\underline B}
\eta_{\underline A\underline B}
+\frac{\lambda^2}{2}({m}y^2-{n}{y'}^2)
\nonumber\\&&
+2i
\bar\theta_-\gamma^i
\mathrm{D}_j\theta_-
+\frac{1}{4}(F_1)_{ij}(F_1)^{ij}
\Big]
+
T_{\NR}\int_\Sigma\!\! C_2^{(4)}~.
\end{eqnarray}

\subsection{D5-brane}

Fourthly, we consider a D5-brane 
for which $\varrho=\sigma_1$ and $\sigma=\sigma_3$.
The gluing matrix is
\begin{eqnarray}
M=\sqrt{-s}\Gamma^{\bar A_0\cdots\bar A_5}\otimes \rho~,~~~
\rho=\sigma_1,\sigma_3,1~.
\label{M D5}
\end{eqnarray}
Since
\begin{eqnarray}
&&M'\Gamma_{\bar B_1\cdots \bar B_5}\varrho
=\pm\Gamma_{\bar B_1\cdots \bar B_5}\varrho M~,~~~
M'\Gamma_{\bar A}\varrho
=\pm\Gamma_{\bar A}\varrho M~,~~~
\rho=\textstyle\left\{
  \begin{array}{l}
\sigma_1,1       \\
\sigma_3       \\
  \end{array}
\right.
\\
&&M'\Gamma_{\bar B_1\cdots \bar B_3}i\sigma_2
=\pm\Gamma_{\bar B_1\cdots \bar B_5}i\sigma_2 M~,~~~
\rho=\textstyle\left\{
  \begin{array}{l}
1      \\
 \sigma_1,\sigma_3      \\
  \end{array}
\right.
\\
&&M'\Gamma_{\bar A}\sigma
=\pm\Gamma_{\bar A}\sigma M~,~~~
\rho=\textstyle\left\{
  \begin{array}{l}
\sigma_3,1       \\
\sigma_1       \\
  \end{array}
\right.,
\end{eqnarray}
$\CF$ is of order $\Omega$
only for $\rho=\sigma_1$.
In this case
the WZ part is expanded as
\begin{eqnarray}
Th_{7}&=&T_{\NR}\Omega^{-2}h_{7}^{\div}+T_{\NR}h_{7}^{\fin}
+O(\Omega^4)~,\\
h_{7}^{\div}&=&
\frac{\sqrt{s}}{5!}
\BL^{\bar A_1}_0\cdots\BL^{\bar A_5}_0
\bar L_{+0}\Gamma_{\bar A_1\cdots \bar A_5}\varrho L_{+0}
~,
\label{h_7 div}\\
h_{7}^{\fin}&=&
h^{(7)}_2
+h^{(5)}_1\CF_1
+\frac{1}{2}h^{(3)}_0\CF_1^2
\label{h_7 fin}
\end{eqnarray}
and the DBI part is expanded as (\ref{L_DBI expansion}).
We find that
$h^{(7)}_2$, $h^{(5)}_1$ and $h^{(3)}_0$
are given as
\begin{eqnarray}
h^{(7)}_2&=&\frac{\sqrt{s}}{5!}\Big[
\BL^{\bar A_1}_0\cdots \BL^{\bar A_5}_0
\bar L_{-1}\Gamma_{\bar A_1\cdots\bar A_5}\varrho L_{-1}
+5\BL^{\bar A_1}_0\cdots \BL^{\bar A_4}_0 \BL^{\bar A_5}_2
\bar L_{+0}\Gamma_{\bar A_1\cdots\bar A_5}\varrho L_{+0}
\nonumber\\&&
+2\BL^{\bar A_1}_0\cdots  \BL^{\bar A_5}_0
\bar L_{+0}\Gamma_{\bar A_1\cdots\bar A_5}\varrho L_{+2}
+10\BL^{\bar A_1}_0\cdots \BL^{\bar A_{4}}_0 \BL^{\underline A_5}_1
\bar L_{+0}\Gamma_{\bar A_1\cdots\bar A_{4}\underline A_5}\varrho L_{-1}
\nonumber\\&&
+20\BL^{\bar A_1}_0\cdots\BL^{\bar A_3}_0 
\BL^{\underline A_{4}}_1 \BL^{\underline A_5}_1
\bar L_{+0}\Gamma_{\bar A_1\cdots\bar A_3\underline A_4\underline A_5}
\varrho L_{+0}
\Big]~,
\label{h(7)_0}\\
h^{(5)}_1&=&
\frac{\sqrt{s}}{3!}\Big[
2\BL^{\bar A_1}_0\BL^{\bar A_2}_0\BL^{\bar A_3}_0
\bar L_{+0}\Gamma_{\bar A_1\bar A_2\bar A_3}i\sigma_2 L_{-1}
+
3\BL^{\bar A_1}_0\BL^{\bar A_2}_0\BL^{\underline A_3}_1
\bar L_{+0}\Gamma_{\bar A_1\bar A_2\underline A_3}i\sigma_2 L_{+0}
\nonumber\\&&
+\delta^{(4,2)}{i}\lambda
\epsilon_{\bar a_1\cdots\bar a_4\underline a_5}
\BL^{\bar a_1}_0\cdots\BL^{\bar a_4}_0\BL^{\underline a_5}_1
-\delta^{(2,4)}{i}\lambda
\epsilon_{\bar a_1'\cdots\bar a_4'\underline a_5'}
\BL^{\bar a_1'}_0\cdots\BL^{\bar a_4'}_0\BL^{\underline a_5'}_1
\Big]
~,
\label{h(5)_1}\\
h^{(3)}_0&=&
\sqrt{s}\BL^{\bar A}_0\bar L_{+0}\Gamma_{\bar A}\varrho L_{+0}~.
\end{eqnarray}
This implies that the bosonic 6-form $C^{(6)}$ is expanded as
\begin{eqnarray}
TdC^{(6)}&=&T_{\NR}\Omega^{-2}dC^{(6)}_0+T_{\NR}dC^{(6)}_2 +O(\Omega^4)~,\\
dC^{(6)}_0&=&0~,\\
dC^{(6)}_2&=&
\frac{i\sqrt{s}}{3!}\lambda\Big[
\delta^{(4,2)}
\epsilon_{\bar a_1\cdots\bar a_4\underline a_5}e^{\bar a_1}_0\cdots e^{\bar a_4}_0
e^{\underline a_5}_1
-
\delta^{(2,4)}
\epsilon_{\bar a_1'\cdots\bar a_4'\underline a_5'}e^{\bar a_1'}_0\cdots e^{\bar a_4'}_0
e^{\underline a_5'}_1
\Big]F_1~.
\label{D5 bosonic C_2}
\end{eqnarray}

Because
\begin{eqnarray}
d(\sqrt{s\det g_0}\,d^{6}\xi)=
-i\frac{\sqrt{s}}{5!}
\BL^{\bar A_1}_0\cdots\BL^{\bar A_5}_0\bar L_{+0}\Gamma_{\bar A_1\cdots \bar A_5}
\varrho L_{+0}
~,
\end{eqnarray}
 $h^{\div}_{7}$
 in (\ref{h_7 div}) and the fermionic term in the DBI part
$\CL_{\DBI}^{\div}$
cancel each other.
As before, one sees that
the bosonic terms are also deleted by choosing
$C_0^{(6)}=-\frac{1}{6!}\epsilon_{\bar A_0\cdots\bar A_5}e^{\bar A_0}_0\cdots e^{\bar A_5}_0$.
Summarizing we have shown that the matrix $M$ with $\varrho=\sigma_1$
leads to the consistent non-relativistic limit of the AdS D5-brane\footnote{
It is now obvious that for NS5-brane with $\varrho=\sigma_3$ and $\sigma=-\sigma_1$
the gluing matrix (\ref{M D5}) with $\rho=\sigma_3$
leads to the consistent non-relativistic NS5-brane.}.

The non-relativistic D5-brane action
is given as (\ref{S_NR})
with
(\ref{L_DBI fin}) and 
\begin{eqnarray}
\CL_{\WZ}^{\rm D5}&=&\int_0^1\!\!d t
\Big[
\CC^{(6)}_2
+
\CC^{(4)}_1
\hat\CF_1
+
\frac{1}{2}
\CC^{(2)}_0
\hat\CF_1^2
\Big]
+
C_2^{(6)}
\end{eqnarray}
with
\begin{eqnarray}
\CC^{(6)}_2&=&
2c\Big[\frac{1}{5!}
\hat\BL^{\bar A_1}_0\cdots\hat\BL^{\bar A_5}_0
(\hat{\bar L}_{-1}\Gamma_{\bar A_1\cdots\bar A_5}\varrho\theta_-
 +\hat{\bar L}_{+2}\Gamma_{\bar A_1\cdots\bar A_5}\varrho\theta_+)
\nonumber\\&&
+\frac{1}{4!}\hat\BL^{\bar A_1}_0\cdots\hat\BL^{\bar A_4}_0\hat\BL^{\underline A_5}_1
(\hat{\bar L}_{-1}\Gamma_{\bar A_1\cdots\bar A_4\underline A_5}\varrho\theta_+
 +\hat{\bar L}_{+0}\Gamma_{\bar A_1\cdots\bar A_4\underline A_5}\varrho\theta_-)
\nonumber\\&&
+\frac{1}{4!}\hat\BL^{\bar A_1}_0\cdots\hat\BL^{\bar A_4}_0\hat\BL^{\bar A_5}_2
\hat{\bar L}_{+0}\Gamma_{\bar A_1\cdots\bar A_5}\varrho\theta_+
\nonumber\\&&
+\frac{1}{3!}\hat\BL^{\bar A_1}_0\hat\BL^{\bar A_2}_0\hat\BL^{\bar A_3}_0
 \hat\BL^{\underline A_4}_1\hat\BL^{\underline A_5}_1
\hat{\bar L}_{+0}\Gamma_{\bar A_1\bar A_2\bar A_3\underline A_4\underline A_5}
\varrho\theta_+
\Big]~,
\nonumber\\
\CC^{(4)}_1&=&
2c\Big[
\frac{1}{3!}\hat\BL^{\bar A_1}_0\hat\BL^{\bar A_2}_0\hat\BL^{\bar A_3}_0
(\hat{\bar L}_{+0}\Gamma_{\bar A_1\bar A_2\bar A_3}i\sigma_2\theta_-
+\hat{\bar L}_{-1}\Gamma_{\bar A_1\bar A_2\bar A_3}i\sigma_2\theta_+)
\nonumber\\&&
+\frac{1}{2}\hat\BL^{\bar A_1}_0\hat\BL^{\bar A_2}_0\hat\BL^{\underline A_3}_1
 \hat{\bar L}_{+0}\Gamma_{\bar A_1\bar A_2\underline A_3}i\sigma_2\theta_+
\Big]~,
\nonumber\\
\CC^{(2)}_0&=&
2c\Big[
\hat\BL^{\bar A}_0\hat{\bar L}_{+0}\Gamma_{\bar A}\varrho\theta_+
\Big]~.
\end{eqnarray}
The bosonic contribution is
\begin{eqnarray}
\int_\Sigma\!\!C_2^{(6)}&=&
4i\sqrt{s}\lambda\int_{\Sigma}\Big[
\delta^{(4,2)}
\vol_{\Sigma_4}
y
F_1
-
\delta^{(2,4)}
\vol_{\Sigma_4'}
y'
F_1
\Big]~
\nonumber\\
&=&
-4i\sqrt{s}\lambda\int\!\!d^6\xi\sqrt{s\det g_0}\Big[
\delta^{(4,2)}\partial_{i'}y(^*A_1)^{i'}
-
\delta^{(2,4)}\partial_{i}y(^*A_1)^{i}
\Big]
\end{eqnarray}
where $y(y')$  is the transverse direction in AdS$_5$(S$^5$), 
and $i(i')$ represents the worldvolume directions in  AdS$_5$(S$^5$).
$*$ means the Hodge dual in $\Sigma_2$ or $\Sigma_2'$.
The $\kappa$-gauge symmetry is fixed by $\theta_+=0$,
and the non-relativistic action is
simplified as
\begin{eqnarray}
S_{\NR}^{D5}&=&T_{\NR}\int\!\!d^6\xi\sqrt{s\det g_0}\Big[
\frac{1}{2}g_0^{ij}\partial_iy^{\underline A}\partial_jy_{\underline A}
+\frac{\lambda^2}{2}({m}y^2-{n}{y'}^2)
\nonumber\\&&
+2i
\bar\theta_-\gamma^i
\mathrm{D}_i\theta_-
+\frac{1}{4}(F_1)_{ij}(F_1)^{ij}
\Big]
+T_{\NR}\int_\Sigma\!\!C_2^{(6)}
~.
\end{eqnarray}

\subsection{D7-brane}
Finally, let us consider a D7-brane for which
$\varrho=\sigma_1$ and $\sigma=\sigma_3$.
By using the gluing matrix
\begin{eqnarray}
M=\sqrt{-s}\Gamma^{\bar A_0\cdots\bar A_6}\otimes i\sigma_2
\end{eqnarray}
one derives
\begin{eqnarray}
&&
M'\Gamma_{\bar B_1\cdots \bar B_7}i\sigma_2
=\Gamma_{\bar B_1\cdots \bar B_7}i\sigma_2 M~,~~~
M'\Gamma_{\bar B_1\cdots \bar B_5}\varrho
=-\Gamma_{\bar B_1\cdots \bar B_5}\varrho M~,
\nonumber\\&&
M'\Gamma_{\bar B_1\cdots \bar B_3}i\sigma_2
=\Gamma_{\bar B_1\cdots \bar B_3}i\sigma_2 M~,~~~
M'\Gamma_{\bar A}\varrho
=-\Gamma_{\bar A}\varrho M~,
\nonumber\\&&
M'\Gamma_{\bar A}\sigma
=-\Gamma_{\bar A}\sigma M~.
\end{eqnarray}
These imply that
$\CF$ is of order $\Omega$
and
the WZ  part is expanded as
\begin{eqnarray}
Th_{9}&=&T_{\NR}\Omega^{-2}h_{9}^{\div}+T_{\NR}h_{9}^{\fin}
+O(\Omega^4)~,\\
h_{9}^{\div}&=&
\frac{\sqrt{s}}{7!}
\BL^{\bar A_1}_0\cdots\BL^{\bar A_7}_0
\bar L_{+0}\Gamma_{\bar A_1\cdots \bar A_7}i\sigma_2 L_{+0}~,
\label{h_9 div}\\
h_{9}^{\fin}&=&
h^{(9)}_2
+h^{(7)}_1\CF_1
+\frac{1}{2}h^{(5)}_0\CF_1^2~,
\label{h_9 fin}
\end{eqnarray}
and the DBI is as in (\ref{L_DBI expansion}).
It is straightforward to see that
$h^{(9)}_2$, $h^{(7)}_1$
and $h^{(5)}_0$
are given as
\begin{eqnarray}
h^{(9)}_2&=&\frac{\sqrt{s}}{7!}\Big[
\BL^{\bar A_1}_0\cdots \BL^{\bar A_7}_0
\bar L_{-1}\Gamma_{\bar A_1\cdots\bar A_7}i\sigma_2 L_{-1}
+7\BL^{\bar A_1}_0\cdots \BL^{\bar A_{6}}_0 \BL^{\bar A_7}_2
\bar L_{+0}\Gamma_{\bar A_1\cdots\bar A_7}i\sigma_2 L_{+0}
\nonumber\\&&
+2\BL^{\bar A_1}_0\cdots  \BL^{\bar A_7}_0
\bar L_{+0}\Gamma_{\bar A_1\cdots\bar A_7}i\sigma_2 L_{+2}
+14\BL^{\bar A_1}_0\cdots \BL^{\bar A_6}_0 \BL^{\underline A_7}_1
\bar L_{+0}\Gamma_{\bar A_1\cdots\bar A_{6}\underline A_7}i\sigma_2 L_{-1}
\nonumber\\&&
+42\BL^{\bar A_1}_0\cdots\BL^{\bar A_5}_0  
\BL^{\underline A_6}_1 \BL^{\underline A_7}_1
\bar L_{+0}\Gamma_{\bar A_1\cdots\bar A_{5}
\underline A_{6}\underline A_7}i\sigma_2 L_{+0}
\Big]~,
\label{h(9)_2}\\
h^{(7)}_1&=&
\frac{\sqrt{s}}{5!}\Big[
2\BL^{\bar A_1}_0\cdots\BL^{\bar A_5}_0
\bar L_{+0}\Gamma_{\bar A_1\cdots\bar A_5}\varrho L_{-1}
+
5\BL^{\bar A_1}_0\cdots\BL^{\bar A_4}_0\BL^{\underline A_5}_1
\bar L_{+0}\Gamma_{\bar A_1\cdots\bar A_4\underline A_5}\varrho L_{+0}
\Big]~,
\label{h(7)_1}\\
h_0^{(5)}&=&\frac{\sqrt{s}}{3!}\Big[
\BL^{\bar A_1}_0\cdots\BL^{\bar A_{3}}_0
\bar L_{+0}\Gamma_{\bar A_1\cdots \bar A_{3}}i\sigma_2 L_{+0}
\nonumber\\&&
+\frac{i\lambda}{5}(
\delta^{(5,3)}
\epsilon_{\bar a_1\bar a_2\bar a_3\bar  a_4\bar  a_5}
\BL^{\bar a_1}_0\BL^{\bar a_2}_0\BL^{\bar a_3}_0\BL^{\bar  a_4}_0
\BL^{\bar  a_5}_0
-\delta^{(3,5)}
\epsilon_{\bar a_1'\bar a_2'\bar a_3'\bar  a_4'\bar  a_5'}
\BL^{\bar a_1'}_0\BL^{\bar a_2'}_0\BL^{\bar a_3'}_0
\BL^{\bar  a_4'}_0\BL^{\bar  a_5'}_0
)
\Big]~.
~~~
\label{h(5)_0 D7}
\end{eqnarray}
This implies that the bosonic 8-form $C^{(8)}$ is expanded as
\begin{eqnarray}
TdC^{(8)}&=&T_{\NR}\Omega^{-2}dC^{(8)}_0+T_{\NR}dC^{(8)}_2 +O(\Omega^4)~,\\
dC^{(8)}_0&=&0~,\\
dC^{(8)}_2&=&
\frac{2i\sqrt{s}}{5!}\lambda\Big[
\delta^{(5,3)}
\epsilon_{\bar a_1\cdots\bar a_5}e^{\bar a_1}_0\cdots e^{\bar a_5}_0
-
\delta^{(3,5)}
\epsilon_{\bar a_1'\cdots\bar a_5'}e^{\bar a_1'}_0\cdots e^{\bar a_5'}_0
\Big](F_1)^2~.
\label{D7 bosonic C_2}
\end{eqnarray}

As before, we find
\begin{eqnarray}
d(\sqrt{s\det g_0}\,d^{8}\xi)=
d(\det((\BL^{\bar A}_{0})_{i})d^{8}\xi)=
-i\frac{\sqrt{s}}{7!}
\BL^{\bar A_1}_0\cdots\BL^{\bar A_7}_0\bar L_{+0}\Gamma_{\bar A_1\cdots \bar A_7}
i\sigma_2 L_{+0}
~.
\end{eqnarray}
This implies that $h^{\div}_{9}$ and the fermionic terms in 
$\CL_{\DBI}^{\div}$
cancel each other.
The bosonic terms are also deleted by choosing 
$C_0^{(8)}=-
\frac{1}{8!}
\epsilon_{\bar A_0\cdots\bar A_7}e^{\bar A_0}_0\cdots e^{\bar A_7}_0$.
Thus we find that the matrix $M$ leads to the consistent non-relativistic limit of
the AdS D7-brane.

The non-relativistic D7-brane action
is given as (\ref{S_NR})
with (\ref{L_DBI fin})
and
\begin{eqnarray}
\CL_{\WZ}^{\rm D7}&=&\int_0^1\!\!d t
\Big[
\CC^{(8)}_2
+
\CC^{(6)}_1
\hat\CF_1
+
\frac{1}{2}
\CC^{(4)}_0
\hat\CF_1^2
\Big]
+C_2^{(8)}
\end{eqnarray}
with
\begin{eqnarray}
\CC^{(8)}_2&=&
2c\Big[\frac{1}{7!}
\hat\BL^{\bar A_1}_0\cdots\hat\BL^{\bar A_7}_0
(\hat{\bar L}_{-1}\Gamma_{\bar A_1\cdots\bar A_7}i\sigma_2\theta_-
 +\hat{\bar L}_{+2}\Gamma_{\bar A_1\cdots\bar A_7}i\sigma_2\theta_+)
\nonumber\\&&
+\frac{1}{6!}\hat\BL^{\bar A_1}_0\cdots\hat\BL^{\bar A_6}_0\hat\BL^{\underline A_7}_1
(\hat{\bar L}_{-1}\Gamma_{\bar A_1\cdots\bar A_6\underline A_7}i\sigma_2\theta_+
 +\hat{\bar L}_{+0}\Gamma_{\bar A_1\cdots\bar A_6\underline A_7}i\sigma_2\theta_-)
\nonumber\\&&
+\frac{1}{6!}\hat\BL^{\bar A_1}_0\cdots\hat\BL^{\bar A_6}_0\hat\BL^{\bar A_7}_2
\hat{\bar L}_{+0}\Gamma_{\bar A_1\cdots\bar A_7}i\sigma_2\theta_+
\nonumber\\&&
+\frac{1}{5!}\hat\BL^{\bar A_1}_0\cdots\hat\BL^{\bar A_5}_0\hat\BL^{\underline A_6}_1
 \hat\BL^{\underline A_7}_1
\hat{\bar L}_{+0}\Gamma_{\bar A_1\cdots\bar A_5\underline A_6\underline A_7}i\sigma_2\theta_+
\Big]~,
\nonumber\\
\CC^{(6)}_1&=&
2c\Big[
\frac{1}{5!}\hat\BL^{\bar A_1}_0\cdots\hat\BL^{\bar A_5}_0
(\hat{\bar L}_{+0}\Gamma_{\bar A_1\cdots\bar A_5}\varrho\theta_-
+\hat{\bar L}_{-1}\Gamma_{\bar A_1\cdots\bar A_5}\varrho\theta_+)
\nonumber\\&&
+\frac{1}{4!}\hat\BL^{\bar A_1}_0\cdots\hat\BL^{\bar A_4}_0\hat\BL^{\underline A_5}_1
 \hat{\bar L}_{+0}\Gamma_{\bar A_1\cdots\bar A_4\underline A_5}\varrho\theta_+
\Big]~,
\nonumber\\
\CC^{(4)}_0&=&
2c\Big[
\frac{1}{3!}
\hat\BL^{\bar A_1}_0\cdots \hat\BL^{\bar A_3}_0
 \hat{\bar L}_{+0}\Gamma_{\bar A_1\cdots\bar A_3}
i\sigma_2\theta_+
\Big]~.
\end{eqnarray}
The bosonic contribution is
\begin{eqnarray}
\int_\Sigma\!\!C_2^{(8)}&=&
-2i\sqrt{s}\lambda\int_\Sigma\Big[
\delta^{(5,3)}
\vol_{\Sigma_5}
A_1F_1
-
\delta^{(3,5)}
\vol_{\Sigma_5'}
A_1F_1
\Big]
\nonumber\\
&=&-2i\sqrt{s}\lambda\int\!\!d^8\xi\sqrt{s\det g_0}\Big[
\delta^{(5,3)}(A_1)_{i'}(^*F_1)^{i'}
-
\delta^{(3,5)}(A_1)_{i}(^*F_1)^{i}
\Big]
\end{eqnarray}
where $i(i')$ represents worldvolume directions in AdS$_5$(S$^5$),
and $*$ means the Hodge dual in $\Sigma_3$ or $\Sigma_3'$.
The $\kappa$-gauge symmetry is fixed by $\theta_+=0$,
and then the non-relativistic action is
simplified as
\begin{eqnarray}
S_{\NR}^{D7}&=&T_{\NR}\int\!\!d^8\xi\sqrt{s\det g_0}\Big[
\frac{1}{2}g_0^{ij}\partial_iy^{\underline A}\partial_jy_{\underline A}
+\frac{\lambda^2}{2}({m}y^2-{n}{y'}^2)
\nonumber\\&&
+2i
\bar\theta_-\gamma^i
\mathrm{D}_i\theta_-
+\frac{1}{4}(F_1)_{ij}(F_1)^{ij}
\Big]
+T_{\NR}\int_\Sigma\!\!C_2^{(8)}
~.
\end{eqnarray}

\bigskip

In summary,
we have derived non-relativistic AdS D$p$-brane actions
in AdS$_5\times$S$^5$.
In the flat limit $\lambda \to 0$, these actions
for Lorentzian branes
are reduced to the non-relativistic D$p$-brane actions
in flat spacetime derived in \cite{Gomis:2005bj}.


\section{NH Superalgebra of
Branes in AdS$_{4/7}\times$S$^{7/4}$}

The super-isometry algebra of the AdS$_{q+2}\times$S$^{9-q}$ ($q=2,5$)
solution of the eleven-dimensional supergravity is
generated by translation $P_A$,
Lorentz rotation $J_{AB}=(J_{ab},J_{a'b'})$ 
and 32-component Majorana supercharge $Q$
as
\begin{eqnarray}
&&
[P_a,P_b]=4\epsilon^2\lambda^2J_{ab}~,~~~
[P_{a'},P_{b'}]=-\epsilon^2\lambda^2J_{a'b'}~,
\nonumber\\&&
[J_{ab},P_{c}]=\eta_{bc}P_{a}-\eta_{ac}P_{b}~,~~~
[J_{a'b},P_{c'}]=\eta_{b'c'}P_{a'}-\eta_{a'c'}P_{b'}~,~~~
\nonumber\\&&
[J_{ab},J_{cd}]=\eta_{bc}J_{ad}+\text{3-terms}~,~~~
[J_{a'b'},J_{c'd'}]=\eta_{b'c'}J_{a'd'}+\text{3-terms}~,~~~
\nonumber\\&&
[P_{A},Q]=-\frac{\lambda}{2} Q\widehat\Gamma_{A}~,~~~
[J_{AB},Q]=\frac{1}{2} Q\Gamma_{AB}~,~~~
\nonumber\\&&
\{Q,Q\}=-2\CC\Gamma^AP_A
+\lambda\CC\widehat\Gamma^{AB}J_{AB}
\label{AdS algebra in 11-dim}
\end{eqnarray}
where $\widehat\Gamma^A=(2\CI\Gamma^a,\CI\Gamma^{a'})$
and 
$\widehat\Gamma^{AB}=(2\CI\Gamma^{ab},-\CI\Gamma^{a'b'})$,
and $\epsilon^2=1$ for $q=2$ and $\epsilon^2=-1$ for $q=5$.
For $q=2$, this superalgebra is the super-AdS$_4\times$S$^7$ algebra,
osp(8$|$4),
with the vector index of AdS$_4$, $a=0,1,2,3$
and that of S$^7$, $a'=4,5,...,9,\natural$.
On the other hand, for $q=5$,
this superalgebra is the super-AdS$_7\times$S$^4$ algebra,
osp(8$^*|$4)
with the vector index of S$^4$, $a=\natural,1,2,3$
and that of AdS$_7$, $a'=4,5,...,9,0$.
We use the almost positive metric $\eta_{\mu\nu}$.
We define $\lambda$ and $\CI$ as
\begin{eqnarray}
&&
\lambda=\frac{1}{R},~~~
R^2=2k R_{S}^2=\frac{1}{k}R_{AdS}^2,~~~
k=\left\{
  \begin{array}{ll}
    1/2,&q=2   \\
    2,&q=5   \\
  \end{array}
\right.
\nonumber\\&&
\CI=\Gamma^{\sharp123},~~~
\Gamma^\sharp=\left\{
  \begin{array}{ll}
\Gamma^0 &q=2    \\
-\Gamma^\natural &q=5    \\
  \end{array}
\right.
\end{eqnarray}
where $R_S$ and $R_{AdS}$ are the radii of S$^{9-q}$ and 
that of AdS$_{q+2}$, respectively.
The gamma matrix $\Gamma^A\in$ Spin(1,10) 
and the charge conjugation matrix $\CC$
satisfy (\ref{Clifford}).

Letting $g$ be a group element of the supergroup
of the superalgebra
(\ref{AdS algebra in 11-dim}),
the LI Cartan one-form is defined as
\begin{eqnarray}
\Omega=g^{-1}dg=\BL^AP_A
+\frac{1}{2}\BL^{AB}J_{AB}
+Q_\alpha L^\alpha~.
\label{generator Omega 11-dim}
\end{eqnarray}
The commutation relations 
${[}T_{\hat A},T_{\hat B}\}=f_{\hat A\hat B}{}^{\hat C}T_{\hat C}$,
$T_{\hat A}=\{P_A,J_{AB},Q_I\}$,
are equivalent to the
Maurer-Cartan (MC) equation
\begin{eqnarray}
d\Omega=-\Omega\wedge\Omega~.
\end{eqnarray}
The MC equations corresponding to the superalgebra (\ref{AdS algebra in 11-dim})
are derived as
\begin{eqnarray}
d\BL^A&=&
-\eta_{BC}\BL^{AB}\BL^{C}
-\bar L\Gamma^{A}L~,
\nonumber\\
d\BL^{ab}&=&
-4\epsilon^2\lambda^2\BL^a\BL^b
-\eta_{cd}\BL^{ca}\BL^{bd}
+2\lambda\bar L\CI\Gamma^{ab}L~,
\nonumber\\
d\BL^{a'b'}&=&
+\epsilon^2\lambda^2\BL^{a'}\BL^{b'}
-\eta_{c'd'}\BL^{c'a'}\BL^{b'd'}
-\lambda\bar L\CI\Gamma^{a'b'}L~,
\nonumber\\
dL^\alpha&=&
\frac{\lambda}{2}\BL^{A}\widehat\Gamma_{A}L
-\frac{1}{4}\BL^{AB}\Gamma_{AB}L~.
\label{MC 11}
\end{eqnarray}

We introduce a matrix $M$
\begin{eqnarray}
M=\ell\Gamma^{\bar A_0\cdots \bar A_p}~,~~~
\ell^2(-1)^{[\frac{p+1}{2}]}s=1~
\label{M in 11-dim}
\end{eqnarray}
where $\{\bar A_0,\cdots, \bar A_p\}$
are directions along which the brane worldvolume
extends,
so $A=(\bar A,\underline A)$.
Let an AdS brane extend along $m$ directions in AdS$_4$ or S$^4$
and $n$ directions in S$^7$ or AdS$_7$,
then
the AdS brane worldvolume admits
AdS$_m$(H$^m)\times$S$^n$ or S$^m\times$AdS$_n$(H$^n$)
isometry algebra for a Lorentzian (a Euclidean) brane,
respectively.
After contraction, the isometry of the transverse space
is reduced to the Poincar\'e algebra iso$(4-m)\times$iso$(7-n)$
(iso$(3-m,1)\times$iso$(7-n)$)
or
iso$(4-m)\times$iso$(7-n)$
(iso$(4-m)\times$iso$(6-n,1)$)
for a Lorentzian (a Euclidean) brane.
We require that the contracted superalgebra
contains a super subalgebra, 
the supersymmetrization
of the direct product of the isometry algebra
on the AdS brane worldvolume
and the Lorentz symmetry in the transverse space,
so$(m-1,2)\times$so$(n+1)\times$so$(4-m)\times$so$(7-n)$
(
so$(m,1)\times$so$(n+1)\times$so$(3-m,1)\times$so$(7-n)$
)
for a Lorentzian (a Euclidean) brane in AdS$_4\times$S$^7$,
and
so$(m+1)\times$so$(n-1,2)\times$so$(4-m)\times$so$(7-n)$
(
so$(m+1)\times$so$(n,1)\times$so$(4-m)\times$so$(6-n,1)$
)
for a Lorentzian (a Euclidean) brane in S$^4\times$AdS$_7$,
respectively.
This is satisfied if
\begin{eqnarray}
M'\Gamma^{\bar A}=\Gamma ^{\bar A}M~,~~~
M'\widehat\Gamma^{\bar A\bar B}=\widehat\Gamma^{\bar A\bar B}M~,~~
\label{condition 11}
\end{eqnarray}
where
\begin{eqnarray}
M'=C^{-1}M^TC=(-1)^{p+1+[\frac{p+1}{2}]}M~.
\end{eqnarray}
The first condition is satisfied if $p=1,2\mod 4$.
Since
\begin{eqnarray}
M'\widehat\Gamma^{\bar A\bar B}=
(-1)^{p+1+\mathrm{d}+[\frac{p+1}{2}]}\widehat\Gamma^{\bar A\bar B}M
\end{eqnarray}
where $\mathrm{d}$ is the number of the Dirichlet directions contained in 
$\{\sharp,1,2,3\}$,
these are satisfied by
(odd,odd)-branes ($p=1\mod 4$)
and (even,even)-branes ($p=3\mod 4$).
We depict branes in Table \ref{M-branes}.
\begin{table}[tb]
 \begin{center}
  \begin{tabular}{|c|c|c|c|c|c|}
    \hline
    1-brane  &2-brane &5-brane  &6-brane  & 9-brane &10-brane  \\
    \hline
    (1,1)  &(0,3),~(2,1) &(1,5),~(3,3)   &(0,7),~(2,5),~(4,3) &(3,7)  &(4,7)  \\
    \hline
  \end{tabular}
 \end{center}
 \caption{Branes in AdS$_{4/7}\times$S$^{7/4}$}
\label{M-branes}
\end{table}
The 10-brane is just AdS$_{4/7}\times$S$^{7/4}$ itself as
$M=1$. The $p$-branes with $p=1\mod 4$ are 1/2 BPS Dirichlet branes of
an open supermembrane in AdS$_{4/7}\times$S$^{7/4}$ 
\cite{Sakaguchi:2003hk,Sakaguchi:2004bu}. The brane
probe analysis for M-branes \cite{Kim} is also consistent
with this result.

We derive the NH superalgebra for these branes
as IW contractions of the super-AdS$_{4/7}\times$S$^{7/4}$ algebra.
First, we rescale 
generators as
\begin{eqnarray}
&&
P_{\underline A}\to\frac{1}{\Omega }P_{\underline A}~,~~~
J_{\bar A\underline B}\to\frac{1}{\Omega } J_{\bar A\underline B}~,~~~
Q_-\to\frac{1}{\Omega } Q_-~
\end{eqnarray}
where we have decomposed $Q$ as
\begin{eqnarray}
Q=Q_++Q_-~,~~~ Q_\pm\CP_\pm=Q_\pm~,~~~
\CP_\pm=\frac{1}{2}(1+M)~.
\end{eqnarray}
Substituting these into (\ref{AdS algebra in 11-dim})
and then taking the limit $\Omega\to 0$,
we derive the  NH superalgebra for an AdS brane
\begin{eqnarray}
&&
[P_{\bar a},P_{\bar b}]=4\epsilon^2\lambda^2J_{\bar a\bar b}~,~~~
[P_{\bar a'},P_{\bar b'}]=-\epsilon^2\lambda^2J_{\bar a'\bar b'}~,~~~
\nonumber\\&&
[P_{\bar a},P_{\underline b}]=4\epsilon^2\lambda^2J_{\bar a\underline b}~,~~~
[P_{\bar a'},P_{\underline b'}]=-\epsilon^2\lambda^2J_{\bar a'\underline b'}~,~~~
\nonumber\\&&
[J_{\bar A\bar B},P_{\bar C}]
 =\eta_{\bar B\bar C}P_{\bar A}-\eta_{\bar A\bar C}P_{\bar B}~,~~~
[J_{\bar A\underline B},P_{\bar C}]
 =-\eta_{\bar A\bar C}P_{\underline B}~,
\nonumber\\&&
[J_{\underline A\underline B},P_{\underline C}]
 =\eta_{\underline B\bar C}P_{\underline A}
 -\eta_{\underline A\underline C}P_{\underline B}~,~~~ 
 \nonumber\\&&
[J_{\bar A\bar B},J_{\bar C\bar D}]=
\eta_{\bar B\bar C}J_{\bar A\bar D}+\text{3-terms}~,~~~
[J_{\underline A\underline B},J_{\underline C\underline D}]=
\eta_{\underline B\underline C}J_{\underline A\underline D}+\text{3-terms}~,~~~
\nonumber\\&&
[J_{\bar A\bar B},J_{\bar C\underline D}]=
 \eta_{\bar B\bar C}J_{\bar A\underline D}
 -\eta_{\bar A\bar C}J_{\bar B\underline D}~,~~~
[J_{\underline A\underline B},J_{\bar C\underline D}]=
 \eta_{\underline B\underline D}J_{\bar C\underline A}
 -\eta_{\underline A\underline D}J_{\bar C\underline B}~,
 \nonumber\\&&
[P_{\bar A},Q_\pm]=-\frac{\lambda}{2} Q_\pm\widehat\Gamma_{\bar A}~,~~~
[P_{\underline A},Q_+]=-\frac{\lambda}{2} Q_-\widehat\Gamma_{\underline A}~,
 \nonumber\\&&
[J_{\bar A\bar B},Q_\pm]=\frac{1}{2} Q_\pm\Gamma_{\bar A\bar B}~,~~~
[J_{\underline A\underline B},Q_\pm]=\frac{1}{2} Q_\pm\Gamma_{\underline A\underline B}~,~~~
[J_{\bar A\underline B},Q_+]=\frac{1}{2} Q_-\Gamma_{\bar A\underline B}~,~~~
\nonumber\\&&
\{Q_+,Q_+\}=
-2\CC\Gamma^{\bar A}\CP_+P_{\bar A}
+\lambda\CC\widehat\Gamma^{\bar A\bar B}\CP_+J_{\bar A\bar B}
+\lambda\CC\widehat\Gamma^{\underline A\underline B}\CP_+J_{\underline A\underline B}
~,
\nonumber\\&&
\{Q_+,Q_-\}=
-2\CC\Gamma^{\underline A}\CP_-P_{\underline A}
+2\lambda\CC\widehat\Gamma^{\bar A\underline B}\CP_-J_{\bar A\underline B}~.
\label{NH 11}
\end{eqnarray}

We note that this superalgebra contains
two bosonic algebras
and a superalgebra as subalgebras.
One is
the isometry of the ($m,n$)-brane worldvolume,
generated by $\{P_{\bar A}, J_{\bar A\bar B}\}$,
the AdS$_m$(H$^m$)$\times$S$^n$ algebra
for a Lorentzian (a Euclidean) brane in AdS$_4\times$S$^7$
and 
the S$^m\times$AdS$_n$(H$_n$) algebra
for a Lorentzian (a Euclidean) brane in S$^4\times$AdS$_7$.
Another is the isometry in the transverse space
generated by $\{P_{\underline A}, J_{\underline A\underline B}\}$,
the Poincar\'e algebra 
iso$(4-m)\times$iso$(7-n)$
(
iso$(3-m,1)\times$iso$(7-n)$
)
in AdS$_4\times$S$^7$
and 
iso$(4-m)\times$iso$(7-n)$
(
iso$(4-m)\times$iso$(6-n,1)$
)
in S$^4\times$AdS$_7$
for a Lorentzian (a Euclidean) brane.
The other is the superalgebra
generated by
$\{P_{\bar A}, J_{\bar A\bar B},J_{\underline A\underline B},Q_+\}$
\begin{eqnarray}
&&
[P_{\bar a},P_{\bar b}]=4\epsilon^2\lambda^2J_{\bar a\bar b}~,~~~
[P_{\bar a'},P_{\bar b'}]=-\epsilon^2\lambda^2J_{\bar a'\bar b'}~,~~~
[J_{\bar A\bar B},P_{\bar C}]
 =\eta_{\bar B\bar C}P_{\bar A}-\eta_{\bar A\bar C}P_{\bar B}~,~~~
\nonumber\\&&
[J_{\bar A\bar B},J_{\bar C\bar D}]=
\eta_{\bar B\bar C}J_{\bar A\bar D}+\text{3-terms}~,~~~
[J_{\underline A\underline B},J_{\underline C\underline D}]=
\eta_{\underline B\underline C}J_{\underline A\underline D}+\text{3-terms}~,~~~
 \nonumber\\&&
[P_{\bar A},Q_+]=-\frac{\lambda}{2} Q_+\widehat\Gamma_{\bar A}~,~~~
[J_{\bar A\bar B},Q_+]=\frac{1}{2} Q_+\Gamma_{\bar A\bar B}~,~~~
[J_{\underline A\underline B},Q_+]=\frac{1}{2} Q_+\Gamma_{\underline A\underline B}~,~~~
\nonumber\\&&
\{Q_+,Q_+\}=
-2\CC\Gamma^{\bar A}\CP_+P_{\bar A}
+\lambda\CC\widehat\Gamma^{\bar A\bar B}\CP_+J_{\bar A\bar B}
+\lambda\CC\widehat\Gamma^{\underline A\underline B}\CP_+J_{\underline A\underline B}
~,
\end{eqnarray}
which is the supersymmetrization of the algebra,
so$(m-1,2)\times$so$(n+1)\times$so$(4-m)\times$so$(7-n)$ (
so$(m,1)\times$so$(n+1)\times$so$(3-m,1)\times$so$(7-n)$ ) for a
Lorentzian (a Euclidean) brane in AdS$_4\times$S$^7$, and
so$(m+1)\times$so$(n-1,2)\times$so$(4-m)\times$so$(7-n)$ (
so$(m+1)\times$so$(n,1)\times$so$(4-m)\times$so$(6-n,1)$ ) for a
Lorentzian (a Euclidean) brane in S$^4\times$AdS$_7$.  For a (4,7)-brane
the superalgebra is obviously osp(8$|4$) or osp(8$^*|4$).  Since the
dimension of the bosonic subalgebra is 18 for (0,3)- and (3,3)-branes,
20 for (1,1)-, (2,1)-, (1,5)- and (2,5)-branes, 22 for a (4,3)-brane,
and 34 for (0,7)- and (3,7)-branes, one may guess the superalgebra as
those including variants of osp(4$|$2)$\times$osp(4$|$2),
osp(6$|$2)$\times$so(2$|$2), sp(4$|$2)$\times$osp(4$|$2) and
osp(8$|$2)$\times$su(2), respectively.  The existence of this
superalgebra is ensured by (\ref{condition 11}).

The NH superalgebra (\ref{NH 11}) is equivalent to the MC equation
\begin{eqnarray}
d\BL^{\bar A}&=&
-\eta_{\bar B\bar C}\BL^{\bar A\bar B}\BL^{\bar C}
-\bar L_+\Gamma^{\bar A}L_+
~,
\label{MC NH 11}\\
d\BL^{\underline A}&=&
-\eta_{\bar B\bar C}\BL^{\underline A\bar B}\BL^{\bar C}
-\eta_{\underline{B}\underline C}\BL^{\underline A\underline B}
 \BL^{\underline C}
-\bar L_+\Gamma^{\underline  A}L_-
-\bar L_-\Gamma^{\underline  A}L_+
~,\\
d\BL^{\bar a\bar b}&=&
-4\epsilon^2\lambda^2\BL^{\bar a}\BL^{\bar b}
-\eta_{\bar c\bar d}\BL^{\bar c \bar a}\BL^{\bar b \bar d}
+2\lambda \bar L_+\CI\Gamma^{\bar a\bar b} L_+
~,\\
d\BL^{\bar a'\bar b'}&=&
+\lambda^2\BL^{\bar a'}\BL^{\bar b'}
-\eta_{\bar c'\bar d'}\BL^{\bar c' \bar a'}\BL^{\bar b' \bar d'}
-\lambda \bar L_+\CI \Gamma^{\bar a'\bar b'} L_+
~,\\
d\BL^{\underline A\underline B}&=&
-\eta_{\underline C\underline D}\BL^{\underline C \underline D}
 \BL^{\underline B \underline D}
+\lambda \bar L_+\widehat \Gamma^{\underline A\underline B} L_+
~,\\
d\BL^{\bar a\underline b}&=&
-4\epsilon^2\lambda^2\BL^{\bar a}\BL^{\underline b}
-\eta_{\bar c\bar d}\BL^{\bar c \bar a}\BL^{\underline b \bar d}
-\eta_{\underline c\underline d}\BL^{\underline c \bar a}\BL^{\underline b \underline d}
\nonumber\\&&
+2\lambda \bar L_+\CI\Gamma^{\bar a\underline b} L_-
+2\lambda \bar L_-\CI\Gamma^{\bar a\underline b} L_+
~,\\
d\BL^{\bar a'\underline b'}&=&
+\epsilon^2\lambda^2\BL^{\bar a'}\BL^{\underline b'}
-\eta_{\bar c'\bar d'}\BL^{\bar c' \bar a'}\BL^{\underline b' \bar d'}
-\eta_{\underline c'\underline d'}\BL^{\underline c' \bar a'}\BL^{\underline b' \underline d'}
\nonumber\\&&
- \lambda \bar L_+\CI \Gamma^{\bar a'\underline b'} L_-
-  \lambda \bar L_-\CI \Gamma^{\bar a'\underline b'} L_+
~,\\
dL_+&=&
\frac{\lambda}{2}\BL^{\bar A}\widehat\Gamma_{\bar A}L_+
-\frac{1}{4}\BL^{\bar A\bar B}\Gamma_{\bar A\bar B}L_+
-\frac{1}{4}\BL^{\underline A\underline B}
 \Gamma_{\underline A\underline B}L_+
 ~,\\
dL_-&=&
\frac{\lambda}{2}\BL^{\bar A}\widehat\Gamma_{\bar A}L_-
\frac{\lambda}{2}\BL^{\underline A}
 \widehat\Gamma_{\underline A}L_+
\nonumber\\&&
-\frac{1}{4}\BL^{\bar A\bar B}\Gamma_{\bar A\bar B}L_-
-\frac{1}{4}\BL^{\underline A\underline B}
 \Gamma_{\underline A\underline B}L_-
- \frac{1}{2}\BL^{\bar A\underline B}
 \Gamma_{\bar A\underline B}L_+
 ~.
 \label{MC NH 11 last}
\end{eqnarray}
The MC equation above can be obtained by
rescaling Cartan one-forms in (\ref{MC 11}) as
\begin{eqnarray}
\BL^{\underline A}\to\Omega\BL^{\underline A}~,~~
\BL^{\bar A\underline B}\to\Omega\BL^{\bar A\underline B}~,~~
L_-\to\Omega L_-~,
\end{eqnarray}
and taking the limit $\Omega\to 0$.

\section{NH superalgebra of Branes in M pp-wave}

We define
\begin{eqnarray}
&&
P_\pm=\frac{1}{\sqrt{2}}(P_\natural\pm P_0)~,~~~
P_\ih^*=\left(
P^*_i=\left\{
  \begin{array}{l}
 J_{i0}      \\
 J_{i\natural}
  \end{array}
\right.,
P^*_{i'}=\left\{
  \begin{array}{l}
 J_{i'\natural}      \\
 J_{i'0}
  \end{array}
\right.
\right)~~~
\text{for}~~\left\{
  \begin{array}{l}
    \text{AdS$_4\times$S$^7$}   \\
    \text{AdS$_7\times$S$^4$}     \\
  \end{array}
\right.
\nonumber\\&&
Q^{(\pm)}=Q^{(\pm)}\ell_\pm~,~~~
\ell_\pm=\frac{1}{2}\Gamma_\pm\Gamma_\mp~,~~~
\Gamma_\pm=\frac{1}{\sqrt{2}}(\Gamma_\natural\pm\Gamma_0)~,\\
&&
\CI=\Gamma^{\sharp 123}~,~~~
\Gamma^\sharp=\left\{
  \begin{array}{l}
  \Gamma^0     \\
  -\Gamma^\natural     \\
  \end{array}
\right.~~~\text{for}~~\left\{
  \begin{array}{l}
    \text{AdS$_4\times$S$^7$}   \\
    \text{AdS$_7\times$S$^4$}     \\
  \end{array}
\right.~
\end{eqnarray}
where where $i=1,2,3$ and $i'=4,5,6,7,8,9$.

Scaling generators in the super-AdS$_{4/7}\times$S$^{7/4}$ algebra
as
\begin{eqnarray}
P_+\to\frac{1}{\Lambda^2}P_+~,~~~
P_\ih\to\frac{1}{\Lambda}P_\ih~,~~~
P_\ih^*\to\frac{1}{\Lambda}P_\ih^*~,~~~
Q^{(+)}\to\frac{1}{\Lambda}Q^{(+)}
\end{eqnarray}
and taking the limit $\Lambda\to 0$ limit \cite{Hatsuda:2002kx},
we obtain the M pp-wave superalgebra
\begin{eqnarray}
&&
[P_-,P_i]=\frac{4\lambda^2}{\sqrt{2}}P^*_i~,~~~
[P_-,P_{i'}]=\frac{\lambda^2}{\sqrt{2}}P^*_{i'}~,~~~
[P_-,P^*_{\ih}]=-\frac{1}{\sqrt{2}}P_{\ih}~,~~~
\nonumber\\&&
[P_\ih,P^*_\jh]=
\frac{1}{\sqrt{2}}\eta_{\ih\jh}P_+~,~~~
[P_\ih,J_{\jh\kh}]=\eta_{\ih\jh}P_{\kh}-\eta_{\ih\kh}P_{\jh}~,~~~
[P^*_\ih,J_{\jh\kh}]=\eta_{\ih\jh}P^*_{\kh}-\eta_{\ih\kh}P^*_{\jh}~,~~~
\nonumber\\&&
[J_{\ih\jh},J_{\kh\lh}]=
\eta_{\jh\kh}J_{\ih\lh}+\text{3-terms}~,
\nonumber\\&&
[P_-,Q^{(+)}]=
-\frac{3\lambda}{2\sqrt{2}}Q^{(+)}f~,~~~
[P_-,Q^{(-)}]=
-\frac{\lambda}{2\sqrt{2}}Q^{(-)}f~,~~~
\nonumber\\&&
[P_{i},Q^{(-)}]=
-\frac{\lambda}{\sqrt{2}}Q^{(+)}f\Gamma_{i}\Gamma_+~,~~~
[P_{i'},Q^{(-)}]=
-\frac{\lambda}{2\sqrt{2}}Q^{(+)}f\Gamma_{i'}\Gamma_+~,~~~
\nonumber\\&&[J_{\ih\jh},Q^{(\pm)}]=
\frac{1}{2}Q^{(\pm)}\Gamma_{\ih\jh}~,~~~
[P^*_{\ih},Q^{(-)}]=
\frac{1}{2\sqrt{2}}Q^{(+)}\Gamma_{\ih}\Gamma_+~,~~~
\nonumber\\&&
\{Q^{(+)},Q^{(+)}\}=
-2\CC\Gamma_-P_+~,
\nonumber\\&&
\{Q^{(-)},Q^{(-)}\}=
-2\CC\Gamma_+P_-
-\frac{\lambda}{\sqrt{2}}\CC\widehat\Gamma^{\ih\jh}J_{\ih\jh}
\nonumber\\&&
\{Q^{(\pm)},Q^{(\mp)}\}=
-2\CC\Gamma^\ih\ell_\mp P_\ih
-4\lambda\CC f\Gamma^{i}\ell_\mp P^*_{i}
\mp 2\lambda\CC f\Gamma^{i'}\ell_\mp P^*_{i'}~,
\end{eqnarray}
where $\widehat\Gamma^{\ih\jh}=(-2\Gamma_+f\Gamma^{ij},\Gamma_+f\Gamma^{i'j'})$
and $f=\Gamma^{123}$.
The bosonic subalgebra is the semi-direct product of
the Heisenberg algebra generated by $\{P_\ih,P^*_\ih\}$
with an outer automorphism $P_-$
and the Lorentz symmetry generated by $J_{\ih\jh}$.

\subsection{Lorentzian branes}

We consider a Lorentzian pp-wave brane
for which $(+,-)$ directions are contained in the Neumann directions.
We denote Neumann and Dirichlet directions,  
$\bar A=(+,-,\bar\ih)$ and $\underline A=\underline\ih$\,, respectively.  

We derive NH superalgebras of Lorentzian pp-wave branes
as IW contractions of the super-AdS$_{4/7}\times$S$^{7/4}$ algebra.

First we consider the bosonic subalgebra.
The contraction is taken by
rescaling generators
as
\begin{eqnarray}
&&
P_{\underline A}\to\frac{1}{\Omega}P_{\underline A}~,~~~
J_{\bar A\underline B}\to\frac{1}{\Omega}J_{\bar A\underline B}~,~~~
P^*_{\underline\ih}\to\frac{1}{\Omega}P^*_{\underline\ih}~,
\end{eqnarray}
and taking the limit $\Omega\to 0$.
One obtains the NH algebra of an M pp-wave brane
\begin{eqnarray}
&&
[P_-,P_{\bar i}]=
\frac{4\lambda^2}{\sqrt{2}}P^*_{\bar i}~,~~~
[P_-,P_{\underline i}]=
\frac{4\lambda^2}{\sqrt{2}}P^*_{\underline i}~,~~~
[P_-,P_{\bar i'}]=
\frac{\lambda^2}{\sqrt{2}}P^*_{\bar i'}~,~~~
[P_-,P_{\underline i'}]=
\frac{\lambda^2}{\sqrt{2}}P^*_{\underline i'}~,~~~
\nonumber\\&&
[P_-,P^*_{\bar \ih}]=-\frac{1}{\sqrt{2}}P_{\bar \ih}~,~~~
[P_-,P^*_{\underline \ih}]=-\frac{1}{\sqrt{2}}P_{\underline \ih}~,~~~
[P_{\bar\ih},P^*_{\bar\jh}]=\frac{1}{\sqrt{2}}\eta_{\ih\jh}P_+~,~~~
\nonumber\\&&
[P_{\bar\ih},J_{\bar\jh\underline\kh}]=
\eta_{\bar\ih\bar\jh}P_{\underline\kh}~,~~~
[P^*_{\bar\ih},J_{\bar\jh\underline\kh}]=
\eta_{\bar\ih\bar\jh}P^*_{\underline\kh}~,~~~
\label{NH bos lor M pp}
\end{eqnarray}
and
\begin{eqnarray}
&&
[P_{\bar\ih},J_{\bar\jh\bar\kh}]=\eta_{\bar\ih\bar\jh}P_{\bar\kh}
-\eta_{\bar\ih\bar\kh}P_{\bar\jh}~,~~~
[P_{\underline\ih},J_{\underline\jh\underline\kh}]=
\eta_{\underline\ih\underline\jh}P_{\underline\kh}
-\eta_{\underline\ih\underline\kh}P_{\underline\jh}~,~~~
\nonumber\\&&
[P^*_{\bar\ih},J_{\bar\jh\bar\kh}]=
\eta_{\bar\ih\bar\jh}P^*_{\bar\kh}
-\eta_{\bar\ih\bar\kh}P^*_{\bar\jh}~,~~~
[P^*_{\underline\ih},J_{\underline\jh\underline\kh}]=
\eta_{\underline\ih\underline\jh}P^*_{\underline\kh}
-\eta_{\underline\ih\underline\kh}P^*_{\underline\jh}~,~~~
\nonumber\\&&
[J_{\bar\ih\bar\jh},J_{\bar\kh\bar\lh}]=
\eta_{\bar\jh\bar\kh}J_{\bar\ih\bar\lh}+\text{3-terms}~,~~~
[J_{\underline\ih\underline\jh},J_{\underline\kh\underline\lh}]=
\eta_{\underline\jh\underline\kh}J_{\underline\ih\underline\lh}+\text{3-terms}~,~~~
\nonumber\\&&
[J_{\bar\ih\bar\jh},J_{\bar\kh\underline\lh}]=
\eta_{\bar\jh\bar\kh}J_{\bar\ih\underline\lh}
-\eta_{\bar\ih\bar\kh}J_{\bar\jh\underline\lh}
~,~~~
[J_{\underline\ih\underline\jh},J_{\bar\kh\underline\lh}]=
\eta_{\underline\ih\underline\lh}J_{\underline\jh\bar\kh}
-\eta_{\underline\jh\underline\lh}J_{\underline\ih\bar\kh}
~.
\label{NH bos lorentz lor M pp}
\end{eqnarray}

Next we consider the fermionic part.
We decompose $Q^{(\bullet)}$ as
\begin{eqnarray}
Q^{(\bullet)}_\pm=\pm Q^{(\bullet)}_\pm M \qquad \mbox{with} \quad 
M=\ell\Gamma^{+-\bar A_1\cdots\bar A_{p-1}}~,~~~
M^2=\ell^2(-1)^{[\frac{p-1}{2}]}=1
\end{eqnarray}
which satisfies
\begin{eqnarray}
M'=\CC^{-1}M^T\CC=(-1)^{p+1+[\frac{p+1}{2}]}M~.
\end{eqnarray}
We demand that
\begin{eqnarray}
M'\Gamma^{\bar A}=\Gamma^{\bar A}M~,
\label{condition 1 ppNH 11}\\
M'\widehat\Gamma^{\bar \ih\bar\jh}=\widehat\Gamma^{\bar \ih\bar\jh}M~.
\label{condition 2 ppNH 11}
\end{eqnarray}
Since
\begin{eqnarray}
M'\Gamma^{\bar A}=(-1)^{1+[\frac{p+1}{2}]}\Gamma^{\bar A}M~,
\end{eqnarray}
the first condition (\ref{condition 1 ppNH 11})
is satisfied when $p=1,2\mod 4$.
The second condition  (\ref{condition 2 ppNH 11})
restricts the directions along which a pp-wave brane extends.
Since
\begin{eqnarray}
M'\widehat\Gamma^{\bar\ih\bar\jh}=
(-1)^{1+[\frac{p+1}{2}]+\mathrm{n}}\widehat\Gamma^{\bar\ih\bar\jh}M
\end{eqnarray}
where $\mathrm{n}$ is the number of the Neumann directions contained in $\{1,2,3\}$,
we find that $\mathrm{n}=$even for $p=1,2\mod 4$.
In Table \ref{lorentzian M pp-wave branes} we summarize the result.
The $p$-branes with $p=1\mod 4$ are Dirichlet branes of an open supermembrane
in M pp-wave \cite{Sugiyama:2002rs,Sakaguchi:2003ah}.
\begin{table}[htbp]
 \begin{center}
  \begin{tabular}{|c|c|c|c|c|c|}
    \hline
     1-brane  &2-brane    &5-brane    &6-brane    &9-brane    &10-brane    \\
    \hline
     ($+,-$)  &($+,-;0,1$)    &($+,-;0,4$)     &($+,-;0,5$)     &($+,-;2,6$)     & -   \\
       &    &($+,-;2,2$)    &($+,-;2,3$)    &    &    \\
    \hline
  \end{tabular}
 \end{center}
\caption{Lorentzian M pp-wave branes}
\label{lorentzian M pp-wave branes}
\end{table}

Scaling $Q^{(\bullet)}_\pm$ as
\begin{eqnarray}
Q^{(\bullet)}_-\to\frac{1}{\Omega}Q^{(\bullet)}_-
\end{eqnarray}
and taking the limit $\Omega\to 0$,
we obtain the fermionic part of the NH superalgebra
\begin{eqnarray}
&&
[P_-,Q^{(+)}_\pm]=
-\frac{3\lambda}{2\sqrt{2}}Q^{(+)}_\pm f~,~~~
[P_-,Q^{(-)}_\pm]=
-\frac{\lambda}{2\sqrt{2}}Q^{(-)}_\pm f~,~~~
\nonumber\\&&
[P_{\bar i},Q^{(-)}_\pm]=
-\frac{\lambda}{\sqrt{2}}Q^{(+)}_\pm f\Gamma_{\bar i}\Gamma_+~,~~~
[P_{\underline i},Q^{(-)}_+]=
-\frac{\lambda}{\sqrt{2}}Q^{(+)}_- f\Gamma_{\underline i}\Gamma_+~,~~~
\nonumber\\&&
[P_{\bar i'},Q^{(-)}_\pm]=
-\frac{\lambda}{2\sqrt{2}}Q^{(+)}_\pm f\Gamma_{\bar i'}\Gamma_+~,~~~
[P_{\underline i'},Q^{(-)}_+]=
-\frac{\lambda}{2\sqrt{2}}Q^{(+)}_- f\Gamma_{\underline i'}\Gamma_+~,~~~
\nonumber\\&&
[J_{\bar\ih\bar\jh},Q^{(\bullet)}_\pm]=
\frac{1}{2}Q^{(\bullet)}_\pm\Gamma_{\bar\ih\bar\jh}~,~~~
[J_{\underline\ih\underline\jh},Q^{(\bullet)}_\pm]=
\frac{1}{2}Q^{(\bullet)}_\pm\Gamma_{\underline\ih\underline\jh}~,~~~
[J_{\bar\ih\underline\jh},Q^{(\pm)}_+]=
\frac{1}{2}Q^{(\pm)}_-\Gamma_{\bar\ih\underline\jh}~,~~~
\nonumber\\&&
[P^*_{\bar\ih},Q^{(-)}_\pm]=
\frac{1}{2\sqrt{2}}Q^{(+)}_\pm\Gamma_{\bar\ih}\Gamma_+~,~~~
[P^*_{\underline\ih},Q^{(-)}_+]=
\frac{1}{2\sqrt{2}}Q^{(+)}_-\Gamma_{\underline\ih}\Gamma_+~,~~~
\nonumber\\&&
\{Q^{(+)}_+,Q^{(+)}_+ \}=
-2\CC\Gamma_-\CP_+P_+~,
\nonumber\\&&
\{Q^{(-)}_+,Q^{(-)}_+ \}=
-2\CC\Gamma_+\CP_+P_-
+\frac{\lambda}{\sqrt{2}}\CC\widehat\Gamma^{\bar\ih\bar\jh}\CP_+J_{\bar\ih\bar\jh}
+\frac{\lambda}{\sqrt{2}}\CC\widehat\Gamma^{\underline\ih\underline\jh}\CP_+
J_{\underline\ih\underline\jh}~,
\nonumber\\&&
\{Q^{(-)}_\pm,Q^{(-)}_\mp \}=
\sqrt{2}\lambda
\CC\widehat\Gamma^{\bar\ih\underline\jh}J_{\bar\ih\underline\jh}~,
\nonumber\\&&
\{Q^{(\pm)}_+,Q^{(\mp)}_+ \}=
-2\CC\Gamma^{\bar\ih}\ell_\mp\CP_+P_{\bar\ih}
-4\lambda\CC f\Gamma^{\bar i}\ell_\mp\CP_+ P^*_{\bar i}
\mp 2\lambda\CC f\Gamma^{\bar i'}\ell_\mp\CP_+ P^*_{\bar i'}~,
\nonumber\\&&
\{Q^{(\pm)}_+,Q^{(\mp)}_- \}=
-2\CC\Gamma^{\underline\ih}\ell_\mp\CP_-P_{\underline\ih}
-4\lambda\CC f\Gamma^{\underline i}\ell_\mp\CP_- P^*_{\underline i}
\mp 2\lambda\CC f\Gamma^{\underline i'}\ell_\mp\CP_- P^*_{\underline i'}~.
\label{NH fermi lor M pp}
\end{eqnarray}
Summarizing we have derived the NH superalgebra
of an M pp-wave brane as
(\ref{NH bos lor M pp}), (\ref{NH bos lorentz lor M pp})
and (\ref{NH fermi lor M pp}).

We note that the NH superalgebra of a Lorentzian M pp-wave brane
contains a super-subalgebra generated by
$P_\pm$, $P_{\bar \ih}$, $P^*_{\bar \ih}$, $J_{\bar\ih\bar\jh}$, 
$J_{\underline\ih\underline\jh}$
and $Q^{(\pm)}_+$
\begin{eqnarray}
&&
[P_-,P_{\bar i}]=
\frac{4\lambda^2}{\sqrt{2}}P^*_{\bar i}~,~~~
[P_-,P_{\bar i'}]=
\frac{\lambda^2}{\sqrt{2}}P^*_{\bar i'}~,~~~
[P_-,P^*_{\bar \ih}]=-\frac{1}{\sqrt{2}}P_{\bar \ih}~,~~~
\nonumber\\&&
[P_{\bar\ih},P^*_{\bar\jh}]=\frac{1}{\sqrt{2}}\eta_{\ih\jh}P_+~,~~~
[P_{\bar\ih},J_{\bar\jh\bar\kh}]=\eta_{\bar\ih\bar\jh}P_{\bar\kh}
-\eta_{\bar\ih\bar\kh}P_{\bar\jh}~,~~~
[P^*_{\bar\ih},J_{\bar\jh\bar\kh}]=
\eta_{\bar\ih\bar\jh}P^*_{\bar\kh}
-\eta_{\bar\ih\bar\kh}P^*_{\bar\jh}~,~~~
\nonumber\\&&
[J_{\bar\ih\bar\jh},J_{\bar\kh\bar\lh}]=
\eta_{\bar\jh\bar\kh}J_{\bar\ih\bar\lh}+\text{3-terms}~,~~~
[J_{\underline\ih\underline\jh},J_{\underline\kh\underline\lh}]=
\eta_{\underline\jh\underline\kh}J_{\underline\ih\underline\lh}+\text{3-terms}~,~~~
\nonumber\\&&
[P_-,Q^{(+)}_+]=
-\frac{3\lambda}{2\sqrt{2}}Q^{(+)}_; f~,~~~
[P_-,Q^{(-)}_+]=
-\frac{\lambda}{2\sqrt{2}}Q^{(-)}_+ f~,~~~
\nonumber\\&&
[P_{\bar i},Q^{(-)}_+]=
-\frac{\lambda}{\sqrt{2}}Q^{(+)}_+ f\Gamma_{\bar i}\Gamma_+~,~~~
[P_{\bar i'},Q^{(-)}_+]=
-\frac{\lambda}{2\sqrt{2}}Q^{(+)}_+ f\Gamma_{\bar i'}\Gamma_+~,~~~
\nonumber\\&&
[J_{\bar\ih\bar\jh},Q^{(\bullet)}_+]=
\frac{1}{2}Q^{(\bullet)}_+\Gamma_{\bar\ih\bar\jh}~,~~~
[J_{\underline\ih\underline\jh},Q^{(\bullet)}_+]=
\frac{1}{2}Q^{(\bullet)}_+\Gamma_{\underline\ih\underline\jh}~,~~~
[P^*_{\bar\ih},Q^{(-)}_+]=
\frac{1}{2\sqrt{2}}Q^{(+)}_+\Gamma_{\bar\ih}\Gamma_+~,~~~
\nonumber\\&&
\{Q^{(+)}_+,Q^{(+)}_+ \}=
-2\CC\Gamma_-\CP_+P_+~,
\nonumber\\&&
\{Q^{(-)}_+,Q^{(-)}_+ \}=
-2\CC\Gamma_+\CP_+P_-
+\frac{\lambda}{\sqrt{2}}\CC\widehat\Gamma^{\bar\ih\bar\jh}\CP_+J_{\bar\ih\bar\jh}
+\frac{\lambda}{\sqrt{2}}\CC\widehat\Gamma^{\underline\ih\underline\jh}\CP_+
J_{\underline\ih\underline\jh}~,
\nonumber\\&&
\{Q^{(\pm)}_+,Q^{(\mp)}_+ \}=
-2\CC\Gamma^{\bar\ih}\ell_\mp\CP_+P_{\bar\ih}
-4\lambda\CC f\Gamma^{\bar i}\ell_\mp\CP_+ P^*_{\bar i}
\mp 2\lambda\CC f\Gamma^{\bar i'}\ell_\mp\CP_+ P^*_{\bar i'}~.
\end{eqnarray}
This is the supersymmetrization of the pp-wave algebra
which is the isometry on the brane worldvolume
and the Lorentz symmetry in the transverse space.
The conditions (\ref{condition 1 ppNH 11}) and (\ref{condition 2 ppNH 11})
ensure the existence of this superalgebra.

\subsection{Euclidean branes}

We consider a Euclidean pp-wave brane for which $(+,-)$ directions are
contained in the Dirichlet directions.  We denote Neumann and Dirichlet
directions as 
$\bar A=\bar\ih$ and 
$\underline A=(+,-,\underline\ih)$\,, respectively. 

We derive NH superalgebras of Euclidean pp-wave branes
as IW contractions of the super-AdS$_{4/7}\times$S$^{7/4}$ algebra.
First we consider the bosonic subalgebra.
The contraction is taken by
rescaling generators
as
\begin{eqnarray}
&&
P_{\underline A}\to\frac{1}{\Omega}P_{\underline A}~,~~~
J_{\bar A\underline B}\to\frac{1}{\Omega}J_{\bar A\underline B}~,~~~
P^*_{\bar\ih}\to\frac{1}{\Omega}P^*_{\bar\ih}~,
\end{eqnarray}
and taking the limit $\Omega\to 0$.
One obtains the NH algebra of a Euclidean M pp-wave brane
\begin{eqnarray}
&&
[P_-,P_{\bar i}]=
\frac{4\lambda^2}{\sqrt{2}}P^*_{\bar i}~,~~~
[P_-,P_{\bar i'}]=
\frac{\lambda^2}{\sqrt{2}}P^*_{\bar i'}~,~~~
[P_-,P^*_{\underline \ih}]=
-\frac{1}{\sqrt{2}}P_{\underline \ih}~,~~~
[P_{\bar\ih},J_{\bar\jh\underline\kh}]=
\eta_{\bar\ih\bar\jh}P_{\underline\kh}~,~~~
\nonumber\\&&
[P_{\bar\ih},P^*_{\bar\jh}]=
\frac{1}{\sqrt{2}}\eta_{\bar\ih\bar\jh}P_+~,~~~
[P_{\underline\ih},P^*_{\underline\jh}]=
\frac{1}{\sqrt{2}}\eta_{\underline\ih\underline\jh}P_+~,~~~
[P^*_{\underline\ih},J_{\bar\jh\underline\kh}]=
-\eta_{\underline\ih\underline\kh}P^*_{\bar\jh}~,~~~
\label{NH bos euc M pp}
\end{eqnarray}
and (\ref{NH bos lorentz lor M pp}).

Next we consider the fermionic part of the NH superalgebra.
We decompose $Q^{(\bullet)}$ as
\begin{eqnarray}
Q^{(\bullet)}_\pm=\pm Q^{(\bullet)}_\pm M~,~~~
M=\ell\Gamma^{\bar A_0\cdots\bar A_p}~,~~~
M^2=\ell^2(-1)^{[\frac{p+1}{2}]}=1~.
\end{eqnarray}
We demand that the conditions (\ref{condition 1 ppNH 11}) and (\ref{condition 2 ppNH 11})
are satisfied.
The first condition (\ref{condition 1 ppNH 11}) 
implies that $p=1,2\mod 4$
as
\begin{eqnarray}
M'\Gamma^{\bar A}=(-1)^{1+[\frac{p+1}{2}]}\Gamma^{\bar A}M~.
\end{eqnarray}
On the other hand, since
\begin{eqnarray}
M'\widehat\Gamma^{\bar\ih\bar\jh}=
(-1)^{\mathrm{n}+[\frac{p+1}{2}]}\widehat\Gamma^{\bar\ih\bar\jh} M
\end{eqnarray}
where $\mathrm{n}$ is the number of the Neumann directions contained in $\{1,2,3\}$,
the second condition is satisfied when $\mathrm{n}=$odd for $p=1,2\mod 4$.
We summarize the result in Table \ref{euclidean M pp-wave branes}.
The $p$-branes with $p=1\mod 4$ are Dirichlet branes
of an open supermembrane in M pp-wave \cite{Sugiyama:2002rs,Sakaguchi:2003ah}.
\begin{table}[htbp]

 \begin{center}
  \begin{tabular}{|c|c|c|c|c|c|}
    \hline
     1-brane  &2-brane    &5-brane    &6-brane    &9-brane    &10-brane    \\
    \hline
     (1,1)  &(1,2), (3,0)    &(1,5), (3,3)    &(1,6), (3,4)    &-    & -   \\
    \hline
  \end{tabular}
 \end{center}
  \caption{Euclidean M pp-wave branes}
  \label{euclidean M pp-wave branes}
\end{table}

Scaling $Q^{(\bullet)}_\pm$ as (\ref{Q scale ppNH 10}) and taking the limit $\Omega\to 0$,
we obtain the fermionic part of the NH superalgebra
of a Euclidean M pp-wave brane
\begin{eqnarray}
&&
[P_-,Q^{(+)}_+]=
-\frac{3\lambda}{2\sqrt{2}}Q^{(+)}_-f~,~~~
[P_-,Q^{(-)}_+]=
-\frac{\lambda}{2\sqrt{2}}Q^{(-)}_-f~,~~~
\nonumber\\&&
[P_{\bar i},Q^{(-)}_\pm]=
-\frac{\lambda}{\sqrt{2}}Q^{(+)}_\pm f\Gamma_{\bar i}\Gamma_+~,~~~
[P_{\underline i},Q^{(-)}_+]=
-\frac{\lambda}{\sqrt{2}}Q^{(+)}_- f\Gamma_{\underline i}\Gamma_+~,~~~
\nonumber\\&&
[P_{\bar i'},Q^{(-)}_\pm]=
-\frac{\lambda}{2\sqrt{2}}Q^{(+)}_\pm f\Gamma_{\bar i'}\Gamma_+~,~~~
[P_{\underline i'},Q^{(-)}_+]=
-\frac{\lambda}{2\sqrt{2}}Q^{(+)}_- f\Gamma_{\underline i'}\Gamma_+~,~~~
\nonumber\\&&
[J_{\bar\ih\bar\jh},Q^{\bullet}_\pm]=
\frac{1}{2}Q^{(\bullet)}_\pm\Gamma_{\bar\ih\bar\jh}~,~~~
[J_{\underline\ih\underline\jh},Q^{\bullet}_\pm]=
\frac{1}{2}Q^{(\bullet)}_\pm\Gamma_{\underline\ih\underline\jh}~,~~~
[J_{\bar\ih\underline\jh},Q^{\bullet}_+]=
\frac{1}{2}Q^{(\bullet)}_-\Gamma_{\bar\ih\underline\jh}~,~~~
\nonumber\\&&
[P^*_{\bar\ih},Q^{(-)}_+]=
\frac{1}{2\sqrt{2}}Q^{(+)}_-\Gamma_{\bar\ih}\Gamma_+~,~~~
[P^*_{\underline\ih},Q^{(-)}_\pm]=
\frac{1}{2\sqrt{2}}Q^{(+)}_\pm\Gamma_{\underline\ih}\Gamma_+~,~~~
\nonumber\\&&
\{Q^{(+)}_\pm,Q^{(+)}_\mp\}=
-2\CC\Gamma_-\CP_\mp P_+~,
\nonumber\\&&
\{Q^{(-)}_+,Q^{(-)}_+\}=
-\frac{\lambda}{\sqrt{2}}\CC\widehat\Gamma^{\bar\ih\bar\jh}J_{\bar\ih\bar\jh}
-\frac{\lambda}{\sqrt{2}}\CC\widehat\Gamma^{\underline\ih\underline\jh}
 J_{\underline\ih\underline\jh}
~,
\nonumber\\&&
\{Q^{(-)}_\pm,Q^{(-)}_\mp\}=
-2\CC\Gamma_+\CP_\mp P_-
-\sqrt{2}\lambda\CC\widehat\Gamma^{\bar\ih\underline\jh}J_{\bar\ih\underline\jh}~,
\nonumber\\&&
\{Q^{(\pm)}_+,Q^{(\mp)}_+\}=
-2\CC\Gamma^{\bar\ih}\ell_\mp\CP_+ P_{\bar\ih}
-4\lambda\CC f\Gamma^{\underline i}\ell_\mp\CP_+ P^*_{\underline i}
\mp 2\lambda\CC f\Gamma^{\underline i'}\ell_\mp\CP_+ P^*_{\underline i'}
~,
\nonumber\\&&
\{Q^{(\pm)}_+,Q^{(\mp)}_-\}=
-2\CC\Gamma^{\underline\ih}\ell_\mp\CP_- P_{\underline \ih}
-4\lambda\CC f\Gamma^{\bar i}\ell_\mp\CP_- P^*_{\bar i}
\mp 2\lambda\CC f\Gamma^{\bar i'}\ell_\mp\CP_- P^*_{\bar i'}
~.
\label{NH fermi euc M pp}
\end{eqnarray}
Summarizing we have derived the NH superalgebra of a Euclidean M pp-wave brane
as (\ref{NH bos euc M pp}), (\ref{NH bos lorentz lor M pp}) and 
(\ref{NH fermi euc M pp}).

We note that there exists a super-subalgebra of the NH superalgebra
generated by $P_{\bar \ih}$, $P^*_{\underline \ih}$, $J_{\bar\ih\bar\jh}$,
$J_{\underline\ih\underline\jh}$ and  $Q^{(\pm)}_+$
\begin{eqnarray}
&&[P_{\bar\ih},J_{\bar\jh\bar\kh}]=
\eta_{\bar\ih\bar\jh}P_{\bar\kh}
-\eta_{\bar\ih\bar\kh}P_{\bar\jh}~,~~~
[P^*_{\underline\ih},J_{\underline\jh\underline\kh}]=
\eta_{\underline\ih\underline\jh}P^*_{\underline\kh}
-\eta_{\underline\ih\underline\kh}P^*_{\underline\jh}~,~~~
\nonumber\\&&
[J_{\bar\ih\bar\jh},J_{\bar\kh\bar\lh}]=
\eta_{\bar\jh\bar\kh}J_{\bar\ih\bar\lh}+\text{3-terms}~,~~~
[J_{\underline\ih\underline\jh},J_{\underline\kh\underline\lh}]=
\eta_{\underline\jh\underline\kh}J_{\underline\ih\underline\lh}+\text{3-terms}~,~~~
\nonumber\\&&
[P_{\bar i},Q^{(-)}_+]=
-\frac{\lambda}{\sqrt{2}}Q^{(+)}_+ f\Gamma_{\bar i}\Gamma_+~,~~~
[P_{\bar i'},Q^{(-)}_+]=
-\frac{\lambda}{2\sqrt{2}}Q^{(+)}_+ f\Gamma_{\bar i'}\Gamma_+~,~~~
\nonumber\\&&
[J_{\bar\ih\bar\jh},Q^{\bullet}_+]=
\frac{1}{2}Q^{(\bullet)}_+\Gamma_{\bar\ih\bar\jh}~,~~~
[J_{\underline\ih\underline\jh},Q^{\bullet}_+]=
\frac{1}{2}Q^{(\bullet)}_+\Gamma_{\underline\ih\underline\jh}~,~~~
[P^*_{\underline\ih},Q^{(-)}_+]=
\frac{1}{2\sqrt{2}}Q^{(+)}_+\Gamma_{\underline\ih}\Gamma_+~,~~~
\nonumber\\&&
\{Q^{(-)}_+,Q^{(-)}_+\}=
-\frac{\lambda}{\sqrt{2}}\CC\widehat\Gamma^{\bar\ih\bar\jh}J_{\bar\ih\bar\jh}
-\frac{\lambda}{\sqrt{2}}\CC\widehat\Gamma^{\underline\ih\underline\jh}
 J_{\underline\ih\underline\jh}
~,
\nonumber\\&&
\{Q^{(\pm)}_+,Q^{(\mp)}_+\}=
-2\CC\Gamma^{\bar\ih}\ell_\mp\CP_+ P_{\bar\ih}
-4\lambda\CC f\Gamma^{\underline i}\ell_\mp\CP_+ P^*_{\underline i}
\mp 2\lambda\CC f\Gamma^{\underline i'}\ell_\mp\CP_+ P^*_{\underline i'}
~.
\end{eqnarray}
This is the supersymmetrization of the Poincar\'e algebra
generated by $\{P_{\bar\ih},J_{\bar\ih\bar\jh}\}$
which is the isometry on the brane worldvolume
and the Lorentz symmetry in the transverse space
generated by $\{P^*_{\underline\ih},J_{\underline\ih\underline\jh}\}$.
The conditions (\ref{condition 1 ppNH 11}) and (\ref{condition 2 ppNH 11})
ensure the existence of this super-subalgebra.

\section{Branes in AdS$_{4/7}\times$S$^{7/4}$}

The action for an M2-brane 
\cite{M2:curved} is composed of the NG action
and the WZ action
\begin{eqnarray}
&&S=T\int_\Sigma [\CL_{\NG}+\CL_{\WZ}]\,, \quad 
\CL_{\NG}= d^{p+1}\xi \sqrt{s\det g_{ij}}\,, 
\label{action 11}
\end{eqnarray}
where $s=-1$ for a Lorentzian brane and $s=1$ for a Euclidean brane.
$T$ is the tension of the brane.
For an M5-brane case, the self-duality of the two-form gauge field $B$
on the brane
is imposed on the field equations,
or the NG action is replaced by the PST action \cite{PST}
\begin{eqnarray}
\CL_{\PST}&=&\sqrt{s\det(g_{ij}-i\alpha^2\CH^*_{ij})}
+\alpha^2\frac{\sqrt{s\det g}}{4}\CH^{*ij}\CH_{ij}\,,\\ 
\CH_{ij}&=&\CH_{ijk}v^k\,,
\quad 
\CH^{*ij}=\CH^{*ijk}v_k\,,\quad
v_i=\frac{\partial_ia}{\sqrt{g^{jk}\partial_ja\partial_ka}}\,, \nonumber \\
\CH &=& H+\CC_3\,, \quad
\CH^{*ijk}=\frac{1}{3!\sqrt{s\det g}}\epsilon^{ijklmn}\CH_{lmn}\,, \quad
H=dB\, \nonumber
\end{eqnarray}
where $\CC_3$ is a pullback of the three-form gauge field,
and
$\alpha^2=i\sqrt{s}$.
Here the PST scalar field $a$ is contained in the M5-brane case as a
modification of the usual DBI action. 
The WZ term is known to be characterized by
manifestly supersymmetric $(p+2)$-form $h_{p+2}=d\CL_{\WZ}$\,, 
which is composed of the pullback of the supercurrents,
$\BL^A$ and $L^\alpha$\,, 
on the supergroup manifold and the modified field strength $\CH$.
The $(p+2)$-form $h_{p+2}$ is closed but not exact on the superspace,
because $\CL_{\WZ}$ is not superinvariant but quasi-superinvariant.
Expanding $h_{p+2}$ with respect to $\CH$
\begin{eqnarray}
h_{p+2}(\BL^A,L^\alpha,\CH)=
h^{(p+2)}(\BL^A,L^\alpha)
-\frac{c}{2}\CH h^{(p-1)}(\BL^A,L^\alpha)\,,
\end{eqnarray}
where $c$ is a constant determined below,
the closedness condition $dh_{p+2}=0$ is expressed as
\begin{eqnarray}
dh^{(p-1)} &=& 0\,, \label{closed1} \\
dh^{(p+2)}-\frac{c}{2}d\CH h^{(p-1)} &=& 0\,. \label{closed2}
\end{eqnarray}

\subsection{CE-cohomology classification}

We show that M$p$-brane actions in AdS$_{q+2}\times$S$^{9-q}$ ($q=2,5$)
can be classified as non-trivial elements of the CE-cohomology on the
differential algebra, MC equations (\ref{MC 11}) for the
super-AdS$_{q+2}\times$S$^{9-q}$ algebra.

In order to avoid 
an additional dimensionful parameter, we
assign dimensions as
\begin{eqnarray}
  \begin{array}{ccccc}
   & \BL^A   &L^\alpha     &\lambda    &\CH
\\
 \mbox{dim}   & 1  &1/2    &-1    &3
\\
  \end{array}
  ~~.
\end{eqnarray}
For structureless branes, the dimension of $S_{\WZ}$
must be equal to the dimension of $S_{\NG}$, from which
we find $\dim h_{p+2}=p+1$ because
$\dim h_{p+2}=\dim \CL_{\WZ}=\dim \CL_{\NG}=p+1$,
and thus $\dim h^{(k)}=k-1$.
$h^{(k)}$ is composed of $\BL^A$, $L^\alpha$ and $\lambda$,
and thus we can write 
$h^{(k)}$
as
$\lambda^l(\BL^A)^m(L^\alpha)^n$.
The integers $l,m$ and $n$
are restricted by the properties of $h^{(k)}$,
$\dim h^{(k)}=k-1$ and $\deg h^{(k)}=k$, as
\begin{eqnarray}
-l+m+\frac{1}{2}n=k-1,~~~
m+n=k.
\label{l,m,n}
\end{eqnarray}
For consistent flat limit, we demand $l\ge 0$
because $\lambda$ is related to the inverse of the radii
of $AdS_{q+2}$ and S$^{9-q}$.
In addition, we require that
$\epsilon_{a_1\cdots a_4}$ and $\epsilon_{a_1'\cdots a_7'}$
are accompanied with $\lambda$;
$\lambda\epsilon_{a_1\cdots a_4}$ and $\lambda\epsilon_{a_1'\cdots a_7'}$,
because $\epsilon_{a_1\cdots a_4}$ and $\epsilon_{a_1'\cdots a_7'}$
disappear in the flat limit.
Noting that (\ref{l,m,n}) implies $l=1-\frac{1}{2}n\le 1$,
we consider the cases with $l=0$ and $1$.
It is easy to see that $(m,n)=(k-2,2)$ for $l=0$
while  $(m,n)=(k,0)$ for $l=1$.
In the former case,
terms of the form $\BL^{A_1}\cdots \BL^{A_{k-2}}\bar L\Gamma_{A_1\cdots A_{k-2}}L$
are candidates for $h^{(k)}$.
These terms are non-trivial only for $k=3,4$ mod $4$,
because $C\Gamma_{A_1\cdots A_{k-2}}$ is symmetric if $k-2=1,2$ mod $4$.
In the latter case,
$\lambda\epsilon_{a_1\cdots a_4}\BL^{a_1}\cdots \BL^{a_4}$
and $\lambda\epsilon_{a_1'\cdots a_7'}\BL^{a_1'}\cdots \BL^{a_7'}$
are candidates for $h^{(4)}$ and $h^{(7)}$, respectively.
We summarize the non-trivial candidates for $h^{(k)}$
\begin{eqnarray}
h^{(3)}  &:&~\BL^A\bar L\Gamma_A L    \\
h^{(4)}  &:&~\BL^A \BL^B\bar L\Gamma_{AB}L ,~~
\lambda\epsilon_{a_1\cdots a_4}\BL^{a_1}\cdots \BL^{a_4}
   \\
h^{(7)}  &:&~
\BL^{A_1}\cdots \BL^{A_5}\bar L\Gamma_{A_1\cdots A_5}L ,~~
\lambda \epsilon_{a_1'\cdots a_7'}\BL^{a_1'}\cdots \BL^{a_7'}
   \\
h^{(8)}  &:&~\BL^{A_1}\cdots \BL^{A_6} \bar L\Gamma_{A_1\cdots A_6}L 
\end{eqnarray}
where $\BL^A\bar L\Gamma_A L $ stands for two candidates $\BL^a\bar L\Gamma_a L$
and
$\BL^{a'}\bar L\Gamma_{a'} L$, and so on.
For example,
$h^{(4)}$ is
of the form
\begin{eqnarray}
h^{(4)}=
c_1\BL^{a} \BL^{b}\bar L\Gamma_{ab}L
+c_2\BL^{a} \BL^{a'}\bar L\Gamma_{aa'}L
+c_3\BL^{a'} \BL^{b'}\bar L\Gamma_{a'b'}L
+c_4\lambda\epsilon_{a_1\cdots a_4}\BL^{a_1}\cdots \BL^{a_4}~.
\end{eqnarray}
Next we are going to find  $h^{(k)}$ satisfying (\ref{closed1})
and (\ref{closed2}).
The first step for this is to find a closed form $dh^{(k)}=0$ in (\ref{closed1}).
$h^{(k)}$ can be a closed form only when $k=4$. 
This is due to the Fierz identity
\begin{eqnarray}
(\CC\Gamma_{AB})_{(\alpha\beta}(\CC\Gamma^B)_{\gamma\delta)}=0~.
\label{Fierz h4 AdS}
\end{eqnarray}
The coefficients are fixed by the closedness condition
$dh^{(4)}=0$
as
\begin{eqnarray}
h^{(4)}=
 c\Big[\frac{1}{2}\BL^A \BL^B\bar L\Gamma_{AB}L
 -\frac{6\lambda}{4!}\epsilon_{a_1\cdots a_4}\BL^{a_1}\cdots \BL^{a_4}
 \Big]
\label{h(4)}
\end{eqnarray}
where
$\epsilon_{0123}=-\epsilon_{\natural 123}=+1$.
As seen in Appendix \ref{appendix:kappa},
the overall coefficient $c$
is fixed by the requirement of the $\kappa$-invariance
\cite{M2:curved,M2:AdS}
of the total action $S$ as
$c=-1$ for Lorentzian brane
and $c=i$ for Euclidean brane:
$c=i\sqrt{s}$.
Using $h^{(4)}$ above, the closed four-form $h_4$ is constructed as
\begin{eqnarray}
h_4=h^{(4)}.
\label{h4}
\end{eqnarray}
Because $h_4$ is not exact on the superspace as will be shown below,
we find that the M2-brane action in AdS$_{4/7}\times$S$^{7/4}$
is a non-trivial element of CE cohomology
of the differential algebra (\ref{MC 11}),
MC equations for super-AdS$_{4/7}\times$S$^{7/4}$
algebra.
The obtained action is consistent with
one given in  \cite{M2:AdS}.

Next we introduce $d\CH$ to the differential algebra 
(\ref{MC 11}).
Since $\CH=dB+\CC_3$ and $h_4=c'd\CC_3$ with a constant $c'$,
$d\CH$ is given by
\begin{eqnarray}
c'd\CH=h_4=h^{(4)}
 ~
\label{dF}
\end{eqnarray}
where $h^{(4)}$ is given in (\ref{h(4)}).
If we can construct $h^{(7)}$ satisfying (\ref{closed2}),
then $h_7$
turns out to be a closed seven-form.
We find that using (\ref{h(4)}) and (\ref{dF})
the condition (\ref{closed2})
fixes coefficients
of a linear combination of candidates as
\begin{eqnarray}
h^{(7)}=
c^2\Big[\frac{1}{5!}\BL^{A_1}\cdots \BL^{A_5}
 \bar L\Gamma_{A_1\cdots A_5}L
-\frac{6\lambda}{7!}\epsilon_{a_1'\cdots a_7'}\BL^{a_1'}\cdots \BL^{a_7'}
\Big]
\label{h(7)}
\end{eqnarray}
where $\epsilon_{4...9\natural}=-\epsilon_{4...90}=+1$
and $c'\equiv-c$.
To see this we have used the Fierz identity,
\begin{eqnarray}
(\CC\Gamma^{A_1\cdots A_5})_{(\alpha\beta}(\CC\Gamma_{A_5})_{\gamma\delta)}
-3(\CC\Gamma^{[A_1A_2})_{(\alpha\beta}
 (\CC\Gamma^{A_3A_4]})_{\gamma\delta)}=0.
\end{eqnarray}
The closed seven-form is constructed using (\ref{h(4)}) and (\ref{h(7)}) as
\begin{eqnarray}
h_7=h^{(7)}-\frac{c}{2}h^{(4)}\CH.
\label{h7}
\end{eqnarray}
Because $h_7$ is not exact on the superspace as will be shown below,
we find that M5-brane action in AdS$_{4/7}\times$S$^{7/4}$
is
characterized as a non-trivial element of CE cohomology
on the differential algebra (\ref{MC 11})
and (\ref{dF}).
The constant $c^2$ is determined by the requirement that
the total action is $\kappa$-invariant
\cite{PST,M5:AdS} 
as
$c^2=-1$ and $i$ for
Lorentzian and Euclidean branes respectively,
i.e. $c^2=\alpha^2=i\sqrt{s}$.
See Appendix \ref{appendix:kappa}.
The obtained action is consistent with
one given in \cite{M5:AdS}.

We show that the four- and seven-forms obtained above
are not exact.
Suppose that $h_4$ is exact, then there must exist $b^{(3)}$ such that
$h^{(4)}=db^{(3)}$.
$b^{(3)}$ can be written as $\lambda^l(\BL^A)^m(L^\alpha)^n$
where integers $l, m$ and $n$ are restricted by the properties of $b^{(3)}$,
$\dim b^{(3)}=3$ and $\deg b^{(3)}=3$.
This implies that $l\le 0$.
We find that there is no candidate for $l= 0$.
For $l=-1$,
we find two candidates,
$\lambda^{-1}\BL^{a}\bar L\Gamma_{a}L$ and
$\lambda^{-1}\BL^{a'}\bar L\Gamma_{a'}L$,
but any linear combination of them does not satisfy $h^{(4)}=db^{(3)}$.
It is obvious that terms with $l\le -2$ do not satisfy $h^{(4)}=db^{(3)}$.
Thus, $h_4$ is not exact.
Next, suppose that $h_7$ is exact, then
there exists $b_6$ such that $h_7=db_6$.
This implies, expanding $b_6(\BL^\mu,L^\alpha,\CH)$ as
$b^{(6)}(\BL^A,L^\alpha)+\frac{1}{2}\CH b^{(3)}(\BL^A,L^\alpha)$,
that
\begin{eqnarray}
h^{(7)}=db^{(6)}-\frac{c}{2}d\CH b^{(3)},~~~
h^{(4)}=-db^{(3)}.
\label{b}
\end{eqnarray}
Because we have shown that $h^{(4)}$ is not exact,
there dose not exist $b^{(3)}$ satisfying (\ref{b}).
Thus, we have shown that
$h_7$ is not exact.

Summarizing we find that actions of M2- and M5-branes 
in AdS$_{4/7}\times$S$^{7/4}$
are characterized as non-trivial elements of the CE cohomology.

\subsection{$(p+1)$-dimensional form of the WZ term}
We derive $(p+1)$-dimensional form of the WZ-term
following \cite{M5:AdS}.

The supervielbein and super spin connection
are given in Appendix \ref{parametrization 11}.
These satisfy the following differential equations
\begin{eqnarray}
\partial_t\hat\BL^{A}&=&
-2\bar\theta\Gamma^A\hat L~,\\
\partial_t\hat\BL^{AB}&=&
2\lambda\bar\theta\widehat\Gamma^{AB} \hat L~,\\
\partial_t\hat L&=&
d\theta-\frac{\lambda}{2}\hat\BL^A\widehat \Gamma_A\theta
+\frac{1}{4}\hat\BL^{AB}\Gamma_{AB}\theta~,
\end{eqnarray}
where the symbols with ``hat'' implies that the fermionic variable $\theta$
is rescaled as $\theta \rightarrow t\theta$\,.  

By using these equations, one finds that
\begin{eqnarray}
\partial_t\hat h_4&=&
db_3~,~~~
b_3=-c\hat \BL^A\hat\BL^B\hat{\bar L}\Gamma_{AB}\theta~.
\end{eqnarray}
This implies that
\begin{eqnarray}
h_4&=&d\CC_3~,~~~
\CC_3=\int^1_0\!\!dt
\, b_3
+C^{(3)}
\end{eqnarray}
where $C^{(3)}$ is a bosonic 3-form satisfying $dC^{(3)} =h_4|_{\text{bosonic}}$.
It follows from
\begin{eqnarray}
-c\CH
=dB+\int^1_0\!\!dt
\, b_3
+C^{(3)}~
\end{eqnarray}
that
\begin{eqnarray}
-c\partial_t\hat \CH&=&
b_3~.
\end{eqnarray}
In the similar way, one derives
\begin{eqnarray}
\partial_t \hat h_7&=&
d(b_6-\frac{c}{2}b_3\hat \CH)~,~~~
b_6=
c^2\frac{2}{5!}\hat\BL^{A_1}\cdots\hat\BL^{A_5}
\hat{\bar L}\Gamma_{A_1\cdots A_5}\theta~
\end{eqnarray}
so that
\begin{eqnarray}
h_7&=&d\CC_6~,~~~
\CC_6=\int^1_0\!\! dt\,
(b_6-\frac{c}{2}b_3\hat\CH)
+C^{(6)}
\end{eqnarray}
where $C^{(6)}$ is a bosonic 6-form satisfying
$d C^{(6)}=h_7|_{\text{bosonic}}$.

Summarizing the ($p+1$)-dimensional form of the WZ term
is given as
\begin{eqnarray}
\CL_{\WZ}^{M2}&=&
-i\sqrt{s}
\int^1_0\!\!dt
\, \hat \BL^A\hat\BL^B\hat{\bar L}\Gamma_{AB}\theta
+C^{(3)}
~,\\
\CL_{\WZ}^{M5}&=&i\sqrt{s}
\int^1_0\!\! dt\,
\left(
\frac{2}{5!}\hat\BL^{A_1}\cdots\hat\BL^{A_5}
\hat{\bar L}\Gamma_{A_1\cdots A_5}\theta
-\frac{1}{2}\hat \BL^A\hat\BL^B\hat{\bar L}\Gamma_{AB}\theta\,\hat\CH
\right)
+C^{(6)}
~.
\end{eqnarray}

\section{Non-relativistic Branes in AdS$_{4/7}\times$S$^{7/4}$}
We consider the non-relativistic limit of the branes in AdS$_{4/7}\times$S$^{7/4}$
obtained in the previous section.

We scale coordinates and the tension as
\begin{eqnarray}
&&X^{\underline A}\to\Omega X^{\underline A}~,~~~
\theta_-\to\Omega \theta_-~,
\label{scale 11}\\
&&
T= T_{\NR}\Omega^{-2}~,~~~
H=\Omega H_1~.
\end{eqnarray}
(\ref{scale 11}) is consistent with (\ref{generator Omega 11-dim}).
As can be seen from the concrete expression of the supercurrents
 given in Appendix \ref{parametrization 11},
the supercurrents are expanded as (\ref{expansion PQ 10}).
Expanding $\BL^{AB}$ as in (\ref{expansion J 10})
and substituting (\ref{expansion PQ 10}) and (\ref{expansion J 10})
into the MC equation (\ref{MC 11}),
one finds that 
Cartan one forms 
$\{\BL^{\bar A}_m, \BL^{\underline A}_m,\BL^{\bar A\bar B}_m, \BL^{\bar A\underline B}_m,
\BL^{\underline A\underline B}_m,
L_{\pm m}
\,|\,m=0, 1\}$ form
the MC equation (\ref{MC NH 11})-(\ref{MC NH 11 last}).

\subsection{M2-brane}
First we consider an M2-brane.
The NG part
$\CL_{\NG}$ is expanded as in (\ref{L_NG expansion})
with (\ref{L_NG div}) and (\ref{L_NG fin}).
By using (\ref{MC NH 11}) and $L_+=ML_+$ with (\ref{M in 11-dim}),
one derives
\begin{eqnarray}
d\CL_{\NG}^{\div}=
d(\det (\BL_0^{\bar A})_id^{3}\xi)
&=&-\frac{i\sqrt{s}}{2!}\BL^{\bar A_1}_0
\BL^{\bar A_2}_0\bar L_{+0}
\Gamma_{\bar A_1
\bar A_2}L_{+0}~.
\end{eqnarray}
Since
\begin{eqnarray}
M'\Gamma^{\bar A\bar B}=\Gamma^{\bar A\bar B}M~,
\end{eqnarray}
the four-form  $h_{4}$ is expanded as
\begin{eqnarray}
Th_{4}&=&
T_{\NR}\Omega^{-2}h_{4}^{\div}+T_{\NR}h_4^{\fin}+O(\Omega^4)~,
\end{eqnarray}
with
\begin{eqnarray}
h_{4}^{\div}&=&\frac{i\sqrt{s}}{2}\BL^{\bar A}_0\BL^{\bar B}_0\bar L_{+0}\Gamma_{\bar A\bar B}L_{+0}
~,\\
h_{4}^{\fin}&=&
\frac{i\sqrt{s}}{2}\Big[
\BL^{\bar A}_0\BL^{\bar B}_0\bar L_{-1}\Gamma_{\bar A\bar B}L_{-1}
+2\BL^{\bar A}_0\BL^{\bar B}_0\bar L_{+0}\Gamma_{\bar A\bar B}L_{+2}
+2\BL^{\bar A}_0\BL^{\bar B}_2\bar L_{+0}\Gamma_{\bar A\bar B}L_{+0}
\nonumber\\&&
+4\BL^{\bar A}_0\BL^{\underline B}_1\bar L_{+0}\Gamma_{\bar A\underline B}L_{-1}
+\BL^{\underline A}_1\BL^{\underline B}_1
 \bar L_{+0}\Gamma_{\underline A\underline B}L_{+0}
\nonumber\\&&
-6\lambda\delta^{(2,1)}\epsilon_{\bar a_1\bar a_2\underline a_3\underline a_4}
e^{\bar a_1}_0e^{\bar a_2}_0e^{\underline a_3}_1e^{\underline a_4}_1
\Big]~.
\label{h4 fin}
\end{eqnarray}
This implies that the bosonic 3-form $C^{3}$
is expanded as
\begin{eqnarray}
TdC^{(3)}&=&T_{\NR}\Omega^{-2}dC^{(3)}_0+T_{\NR}dC^{(3)}_2+O(\Omega^2)~,\\
dC^{(3)}_0&=&0~,\\
dC^{(3)}_2&=&-3i\sqrt{s}\lambda
\delta^{(2,1)}
\epsilon_{\bar a_1\bar a_2\underline a_3\underline a_4}
e^{\bar a_1}_0e^{\bar a_2}_0e^{\underline a_3}_1e^{\underline a_4}_1~.
\end{eqnarray}

Since
\begin{eqnarray}
d(\det (\BL_0^{\bar A})_id^{3}\xi)
+h_{4}^{\div}=0~,
\end{eqnarray}
the fermionic contribution of $\CL_{\NG}^{\div}$ and $\CL_{\WZ}^{\div}$
cancel each other.
In order to delete
the bosonic terms of $\CL_{\NG}^{\div}+\CL_{\WZ}^{\div}$,
we choose
\begin{eqnarray}
C_0^{(3)}=-
\frac{1}{3!}\epsilon_{\bar A_0\bar A_1\bar A_2}e^{\bar A_0}_0e^{\bar A_1}_0e^{\bar A_2}_0
~.
\end{eqnarray}
It follows
from the expressions given in 
Appendix \ref{parametrization 11}
that $dC_0^{(3)}=0$.
In summary, we find that
the gluing matrix $M$
leads to the consistent non-relativistic limit of
the M2-brane in AdS$_{4/7}\times$S$^{7/4}$.

The non-relativistic M2-brane action is given as
\begin{eqnarray}
S_{\NR}&=&T_{\NR}\int_\Sigma
[\CL_{\NG}^{\fin}+\CL_{\WZ}^{\fin}]
\end{eqnarray}
with
(\ref{L_NG fin})
and
\begin{eqnarray}
\CL_{\WZ}^{\fin}&=&
-i\sqrt{s}\int_0^1\!\!dt\,\Big[
\hat\BL_0^{\bar A}\hat\BL_0^{\bar B}(\hat{\bar L}_{-1}\Gamma_{\bar A\bar B}\theta_-
+\hat{\bar L}_{+2}\Gamma_{\bar A\bar B}\theta_+
)+2\hat\BL_{0}^{\bar A}\hat\BL_2^{\bar B}\hat{\bar L}_{+0}\Gamma_{\bar A\bar B}\theta_+
\nonumber\\&&
+2\hat\BL_0^{\bar A}\hat\BL_1^{\underline B}(
\hat{\bar L}_{-1}\Gamma_{\bar A\underline B}\theta_+
+\hat{\bar L}_{+0}\Gamma_{\bar A\underline B}\theta_-
)
+\hat\BL_1^{\underline A}\hat\BL_1^{\underline B}
\hat{\bar L}_{+0}\Gamma_{\underline A\underline B}\theta_+
\Big]
+C_2^{(3)}
\end{eqnarray}
The bosonic contribution is
\begin{eqnarray*}
\int_\Sigma\! C_2^{(3)}&=&
-3i\sqrt{s}\lambda\delta^{(2,1)}
\int_\Sigma
\vol_{\Sigma_2}
\epsilon_{\underline a\underline b}y^{\underline a}d y^{\underline b}
~
\nonumber\\
&=&
-3i\sqrt{s}\lambda\delta^{(2,1)}
\int\!\!d^3\xi\sqrt{s\det g_0}
\epsilon_{\underline a\underline b}y^{\underline a}\partial_{i'}y^{\underline b}
\end{eqnarray*}
where $i'$ represent  worldvolume directions in S$^7$ or AdS$_7$.

The $t$-integration is easily done after fixing the $\kappa$-gauge symmetry
by $\theta_+=0$ (see Appendix \ref{appendix:kappa}), 
which leads to
\begin{eqnarray}
&&
\BL_0^{\bar A}=e^{\bar A}_0~,~~~
\BL_2^{\bar A}=e^{\bar A}_2
-\bar\theta_-\Gamma^{\bar A}\mathrm{D}\theta_-~,~~~
\BL_1^{\underline A}=e^{\underline A}_1~,
\nonumber\\&&
L_{-1}=\mathrm{D}\theta_-
=d\theta_--\frac{\lambda}{2}e^{\bar A}_0\widehat\Gamma_{\bar A}\theta_-
+\frac{1}{4}\omega^{\bar A\bar B}\Gamma_{\bar A\bar B}\theta_-~,
\nonumber\\&&
(g_0)_{ij}=(e^{\bar A}_0)_i(e^{\bar B}_0)_j\eta_{\bar A\bar B}~.
\end{eqnarray}
$(e^{\bar A}_0)_i$ is the vielbein on the worldvolume
in the static gauge, $x^{\bar A}=\xi^i$.
Substituting these into the non-relativistic action
we obtain
\begin{eqnarray}
S_{\NR}&=&T_{\NR}\int
\!\!d^3\xi
\sqrt{s\det g_0}\Big[
g_0^{ij}\partial_iy^{\underline A}\partial_iy^{\underline B}\eta_{\underline A\underline B}
+\frac{\epsilon^2\lambda^2}{2}(4{m}y^2-{n}{y'}^2)
-2\bar\theta_-\gamma^i\mathrm{D}_i\theta_-
\Big]
\nonumber\\&&
+T_{\NR}\int_\Sigma\! C_2^{(3)}~.
\end{eqnarray}
In the flat limit $\lambda\to 0$,
this reproduces the non-relativistic action 
given in \cite{Gomis:2004pw}.

\subsection{M5-brane}
Next we consider an  M5-brane
for which $c^2=i\sqrt{s}$.
In this case the gluing matrix
\begin{eqnarray}
M=\sqrt{-s}\Gamma^{\bar A_0\cdots\bar A_5}~,~~~
M'=-M~
\label{M M5}
\end{eqnarray}
satisfies
\begin{eqnarray}
M'\Gamma_{\bar B_1\bar B_2}=-\Gamma_{\bar B_1\bar B_2}M~,
\end{eqnarray}
so that
 $\CH$ is of order $\Omega$
\begin{eqnarray}
\CH&=&\Omega \CH_1+O(\Omega^3)~,
\nonumber\\
\CH_1&=&
H_1
+\int_0^1\!\!dt\,
\Big[
\hat \BL^{\bar A}_0\hat \BL_0^{\bar B}(\hat {\bar L}_{+0}\Gamma_{\bar A\bar B}\theta_-
+\hat {\bar L}_{-1}\Gamma_{\bar A\bar B}\theta_+)
+2\hat \BL_0^{\bar A}\hat \BL_1^{\underline B}
 \hat {\bar L}_{+0}\Gamma_{\bar A\underline B}\theta_+
\Big]
~.
\label{F1 11}
\end{eqnarray}
The PST part $\CL_{\PST}$ is expanded as
\begin{eqnarray}
T\CL_{\PST}=T_{\NR}\Omega^{-2}\CL_{\PST}^{\div}+T_{\NR}\CL_{\PST}^{\fin}
+O(\Omega^4)
\label{PST div}
\end{eqnarray}
with
\begin{eqnarray}
\CL_{\PST}^{\div}&=&
\sqrt{s\det g_0}d^6\xi
~,\\
\CL_{\PST}^{\fin}&=&
\sqrt{s\det g_0}d^6\Big[
\frac{1}{2}g_0^{ij}(g_2)_{ij}
+\frac{1}{2}(\CH_1^-)_{ij}(\CH_1^*)^{ij}
\Big]~
\label{PST fin}
\end{eqnarray}
where $g_0$ and $g_2$ are given in (\ref{g_0}) and (\ref{g_2}), and $\CH^-_1$
is defined as
\begin{eqnarray}
\CH_1^-=\frac{c^2}{2}(\CH_1+c^2\CH_1^*)~.
\end{eqnarray}
Noting that
\begin{eqnarray}
M'\Gamma_{\bar B_1\cdots \bar B_5}=\Gamma_{\bar B_1\cdots\bar B_5}M~,
\end{eqnarray}
$h_7$ is expanded as
\begin{eqnarray}
Th_{7}&=&T_{\NR}\Omega^{-2}h_{7}^{\div}
+T_{\NR}h_7^{\fin}+O(\Omega^3)~
\end{eqnarray}
with
\begin{eqnarray}
h_{7}^{\div}&=&i\sqrt{s}\frac{1}{5!}\BL^{\bar A_1}_0\cdots
\BL^{\bar A_5}_0\bar L_{+0}\Gamma_{\bar A_1\cdots\bar A_5}L_{+0}
~\label{h7 div (1,5)-brane}
\end{eqnarray}
and
\begin{eqnarray}
h_{7}^{\fin}&=&
h^{(7)}_2
-\frac{c}{2}\CH_1h^{(4)}_1~,
\label{h7 fin}
\\
h^{(7)}_2&=&
\frac{i\sqrt{s}}{5!}\Big[
\BL^{\bar A_1}_0\cdots \BL^{\bar A_5}_0
\bar L_{-1}\Gamma_{\bar A_1\cdots\bar A_5}L_{-1}
+2\BL^{\bar A_1}_0\cdots \BL^{\bar A_5}_0
\bar L_{+0}\Gamma_{\bar A_1\cdots\bar A_5}L_{+2}
\nonumber\\&&
+10\BL^{\bar A_1}_0\cdots\BL^{\bar A_4}_0
\BL^{\underline A_5}_1\bar L_{+0}\Gamma_{\bar A_1\cdots\bar A_4\underline A_5}L_{-1}
+5\BL^{\bar A_1}_0\cdots\BL^{\bar A_4}_0
\BL^{\bar A_5}_2\bar L_{+0}\Gamma_{\bar A_1\cdots\bar A_5}L_{+0}
\nonumber\\&&
+20\BL^{\bar A_1}_0\cdots\BL^{\bar A_3}_0
\BL^{\underline A_4}_1 \BL^{\underline A_5}_1
\bar L_{+0}\Gamma_{\bar A_1\cdots\bar A_3\underline A_4\underline A_5}L_{+0}
\nonumber\\&&
-\delta^{(1,5)}{6\lambda}
\epsilon_{\bar a_1'\cdots\bar a_5'\underline a_6'\underline a_7'}
\BL^{\bar a_1'}_0\cdots\BL^{\bar a_5'}_0\BL^{\underline a_6'}_1\BL^{\underline a_7'}_1
\Big]~,
\\
h^{(4)}_1
&=&
c\Big[
\BL^{\bar A}_0\BL^{\bar B}_0
\bar L_{+0}\Gamma_{\bar A\bar B}L_{-1}
+\BL^{\bar A}_0\BL^{\underline B}_1
\bar L_{+0}\Gamma_{\bar A\underline B}L_{+0}
\nonumber\\&&
-\delta^{(3,3)}\frac{6\lambda}{3!}
\epsilon_{\bar a_1\bar a_2\bar a_3\underline a_4}
\BL^{\bar a_1}_0\BL^{\bar a_2}_0\BL^{\bar a_3}_0\BL^{\underline a_4}_1
\Big]
~.
\end{eqnarray}
This implies that the bosonic 6-form $C^{(6)}$ is expanded as
\begin{eqnarray}
TdC^{(6)}&=&T_{\NR}\Omega^{-2}dC^{(6)}_0+T_{\NR}dC^{(6)}_2 +O(\Omega^2)~,\\
dC^{(6)}_0&=&0~,\\
dC^{(6)}_2&=&
-\frac{6c^2}{5!}\delta^{(1,5)}
\epsilon_{\bar a_1'\cdots\bar a_5'\underline a_6'\underline a_7'}
e^{\bar a_1'}_0\cdots e^{\bar a_5'}_0
e^{\underline a_6'}_1e^{\underline a_7'}_1
+\frac{c^2}{2}\lambda\delta^{(3,3)}
\epsilon_{\bar a_1\cdots\bar a_3\underline a_4}
e^{\bar a_1}_0\cdots e^{\bar a_3}_0
e^{\underline a_4}_1
H_1~.~~~
\end{eqnarray}

Since
\begin{eqnarray}
d\CL_{\PST}^{\div}
=d(\det (\BL_0^{\bar A})_id^{6}\xi)
=-h_{7}^{\div}~,
\end{eqnarray}
the fermionic contribution of $\CL_{\PST}^{\div}+\CL_{\WZ}^{\div}$
cancels out.
The bosonic terms of $\CL_{\PST}^{\div}+\CL_{\WZ}^{\div}$,
$
\frac{1}{6!}\epsilon_{\bar A_0\cdots\bar A_5}
e^{\bar A_0}_0\cdots e^{\bar A_5}_0
+C_0^{(6)}
$,
are deleted by choosing
\begin{eqnarray}
C_0^{(6)}=-
\frac{1}{6!}\epsilon_{\bar A_0\cdots\bar A_5}
e^{\bar A_0}_0\cdots e^{\bar A_5}_0
\end{eqnarray}
which satisfies $dC_0^{(6)}=0$.
In summary, we find that
the gluing matrix $M$
leads to the consistent non-relativistic limit of
the M5-brane in AdS$_{4/7}\times$S$^{7/4}$.

The non-relativistic M5-brane action is composed of $\CL_{\PST}^{\fin}$ in
(\ref{PST fin})
and
\begin{eqnarray}
\CL_{\WZ}^{\fin}&=&c^2
\int_0^1\!\!dt\Bigg[
\Big(
\frac{2}{5!}\hat\BL_0^{\bar A_1}\cdots\hat\BL_0^{\bar A_5}(
\hat{\bar L}_{-1}\Gamma_{\bar A_1\cdots\bar A_5}\theta_-
+\hat{\bar L}_{+2}\Gamma_{\bar A_1\cdots\bar A_5}\theta_+
)
\nonumber\\&&
+\frac{2}{4!}\hat\BL_0^{\bar A_1}\cdots\hat\BL_0^{\bar A_4}\hat\BL_1^{\underline A_5}
(
\hat{\bar L}_{-1}\Gamma_{\bar A_1\cdots\bar A_4\underline A_5}\theta_+
+\hat{\bar L}_{+0}\Gamma_{\bar A_1\cdots\bar A_4\underline A_5}\theta_-
)
\nonumber\\&&
+\frac{2}{3!}\hat\BL_0^{\bar A_1}\cdots\hat\BL_0^{\bar A_3}
\hat\BL_1^{\underline A_4}\hat\BL_1^{\underline A_5}
\hat{\bar L}_{+0}\Gamma_{\bar A_1\cdots\bar A_3\underline A_4\underline A_5}\theta_+
\nonumber\\&&
+\frac{2}{4!}\hat\BL_0^{\bar A_1}\cdots\hat\BL_0^{\bar A_4}\hat\BL_2^{\bar A_5}
\hat{\bar L}_{+0}\Gamma_{\bar A_1\cdots\bar A_5}\theta_+
\Big)
\nonumber\\&&
+\frac{1}{2}\Big(
\hat\BL^{\bar A}_0\hat\BL^{\bar B}_0(
\hat{\bar L}_{+0}\Gamma_{\bar A\bar B}\theta_-
+\hat{\bar L}_{-1}\Gamma_{\bar A\bar B}\theta_+)
+2\hat\BL^{\bar A}_0\hat\BL^{\underline B}_1
\hat{\bar L}_{+0}\Gamma_{\bar A\underline B}\theta_+
\Big)\hat\CH_1
\Bigg]
\nonumber\\&&
+C^{(6)}_2~.
\end{eqnarray}
The bosonic contribution is
\begin{eqnarray}
\int_\Sigma C^{(6)}_2&=&
6c^2\lambda
\delta^{(1,5)}
\int_\Sigma
\vol_{\Sigma_5'}
\epsilon_{\underline a'\underline b'}y^{\underline a'}dy^{\underline b'}
+3c^2\lambda\delta^{(3,3)}
\int_\Sigma
\vol_{\Sigma_3}
yH_1
\nonumber\\
&=&3i\sqrt{s}\lambda
\int\!\!d^6\xi\sqrt{s\det g_0}\Big[
2\delta^{(1,5)}\epsilon_{\underline a'\underline b'}y^{\underline a'}
\partial_{\xi}y^{\underline b'}
-\delta^{(3,3)}\partial_{i'}y(^*B_1)^{i'}
\Big]
\end{eqnarray}
where $\xi$ and $i'$ represent coordinates on $\Sigma_1$ and $\Sigma_3'$
respectively,
and $y$ is the transverse direction in AdS$_4$ or S$^4$.
$*$ means the Hodge dual in $\Sigma_3'$.

Let us fix the $\kappa$-symmetry by $\theta_+=0$.
The $\theta$-dependent term in $\CH$ disappears
and so we have $\CH_1=H_1$
in this gauge.
The $t$-integration is easily done,
and the action is drastically simplified as
\begin{eqnarray}
\CL_{\PST}^{\fin}&=&
d^6\xi\sqrt{s\det g_0}\Big[
-\bar\theta_-\gamma^i\mathrm{D}_i\theta_-
+\frac{1}{2}g_0^{ij}\partial_iy^{\underline A}\partial_jy^{\underline B}
 \eta_{\underline A\underline B}
\nonumber\\&&
+\frac{1}{2}(H_1^-)_{ij}(H_1^*)^{ij}
+\frac{\epsilon^2\lambda^2}{2}(4my^2-n{y'}^2)
\Big]~,\\
\CL_{\WZ}^{\fin}&=&
\int^1_0\!\!dt\,
c^2\frac{2}{5!}e_0^{\bar A_1}\cdots e_0^{\bar A_5}\mathrm{D}(t\bar\theta_-)
\Gamma_{\bar A_1\cdots\bar A_5}\theta_-
+C^{(6)}_2\nonumber\\
&=&d^6\xi\sqrt{s\det g_0}\left[
-\bar\theta_-\gamma^i\mathrm{D}_i\theta_-
\right]+C^{(6)}_2~.
\end{eqnarray}
Combining these results,
we obtain the non-relativistic M5-brane action
\begin{eqnarray}
S_{\NR}&=&T_{\NR}\int\!\!
d^6\xi\sqrt{s\det g_0}\Big[
\frac{1}{2}g_0^{ij}\partial_iy^{\underline A}\partial_jy^{\underline B}
 \eta_{\underline A\underline B}
+\frac{\epsilon^2\lambda^2}{2}(4my^2-n{y'}^2)
\nonumber\\&&
+\frac{1}{2}(H_1^-)_{ij}(H_1^*)^{ij}
-2\bar\theta_-\gamma^i\mathrm{D}_i\theta_-
\Big]
+T_{\NR}\int_\Sigma C_2^{(6)}~.~~
\end{eqnarray}
In the flat limit $\lambda\to 0$,
it is reduced to the linearized M5-brane action considered in \cite{CKP}.

\section{Summary and Discussions}

We have derived the NH superalgebra
for AdS branes
as IW contractions of the super-AdS$\times$S
algebras in ten- and eleven-dimensions.
Requiring that the isometry on the AdS brane worldvolume
and the Lorentz symmetry in the transverse space
extend to the super-isometry,
we classified possible branes.
The NH superalgebra
contains the super-isometry as 
a super-subalgebra:
su(2$|$2)$\times$su(2$|$2),
osp(4$|$4),
osp(6$|$2)$\times$psu(2$|$1)
and variants of them 
for non-relativistic AdS branes in AdS$_5\times$S$^5$,
and
osp(4$|$2)$\times$osp(4$|$2),
osp(6$|$2)$\times$so(2$|$2),
sp(4$|$2)$\times$osp(4$|$2),
osp(8$|$2)$\times$su(2)
and variants of them
for non-relativistic AdS M-branes in AdS$_{4/7}\times$S$^{7/4}$.
The possible branes are summarized in Table 1 and 4.
These contain 1/2 BPS branes obtained by examining 
an open superstring in AdS$_5\times$S$^5$
and an open supermembrane in AdS$_{4/7}\times$S$^{7/4}$.
We applied the similar analyses to branes in IIB pp-wave and M pp-wave.
The possible branes are summarized in Table 2, 3, 5 and 6
and  we derived the NH superalgebras of these pp-wave branes.

The WZ terms of AdS branes in ten- and eleven-dimensions
 are examined by using
the CE cohomology on the super-AdS$\times$S algebras.
We find that WZ terms of the AdS branes
in AdS$_5\times$S$^5$ and AdS$_{4/7}\times$S$^{7/4}$
 are non-trivial elements 
of the CE cohomology except for those of strings in AdS$_5\times$S$^5$.

By taking the non-relativistic limit
of the relativistic brane actions obtained above,
we derived non-relativistic D$p$-brane actions
in AdS$_5\times$S$^5$
and non-relativistic M-brane actions
in AdS$_{4/7}\times$S$^{7/4}$.
We have seen that there exists the consistent non-relativistic limit
for
D$p$(even,even) for $p=1\mod 4$
and D$p$(odd,odd) for $p=3\mod 4$ in AdS$_5\times$S$^5$,
and 
M2(0,3), M2(2,1), M5(1,5) and M5(3,3) in AdS$_{4}\times$S$^{7}$
and S$^{4}\times$AdS$_{7}$.
We derived the non-relativistic actions for these branes.

In the flat limit,
the non-relativistic AdS D$p$- and M2-brane actions
are reduced to non-relativistic flat brane actions
\cite{Gomis:2005bj,Gomis:2004pw}.
The non-relativistic AdS M5-brane action
is reduced to the linearized M5-brane action \cite{CKP}.
It is interesting to examine these non-relativistic AdS brane actions further,
but is left for future investigations.

It is also interesting to examine the non-relativistic limit of
branes in the pp-wave.
It is known that the pp-wave superalgebra is an IW contraction
of the AdS superalgebra.
So, the brane actions in the pp-wave
can be derived from those in the AdS background
by expanding supercurrents with respect to the
contraction parameter $\Lambda$ as was presented in Appendix C.
Once having derived the brane action in the pp-wave
one can easily extract the non-relativistic brane actions.
These actions can be also derived from the non-relativistic actions
derived in the present paper
by expanding supercurrents with respect to the
contraction parameter $\Omega$.
We hope to report these points elsewhere in near future.

\section*{Acknowledgments}

The authors thank Machiko Hatsuda for useful discussions and Kiyoshi
Kamimura for useful comments. This work is supported in part by the
Grant-in-Aid for Scientific Research (No.~17540262 and No.~17540091)
from the Ministry of Education, Science and Culture, Japan.  The work of
K.~Y.\ is supported in part by JSPS Research Fellowships for Young
Scientists.

\appendix
\section*{Appendix}

\section{LI Cartan one forms}
\subsection{AdS$_5\times$S$^5$}\label{parametrization 10}

Supervielbeins on the AdS$_{5}\times S^{5}$ can be obtained via  
the coset construction with the coset supermanifold: 
\begin{eqnarray}
\mbox{AdS}_5\times S^5 \sim \frac{PSU(2,2|4)}{SO(1,4)\times SO(4)}~.
\end{eqnarray} 
We parametrize the group manifold as
\begin{eqnarray}
g=g_xg_\theta~,~~~
g_{\theta}=\e^{Q\theta}~,~~~Q=(Q_1,Q_2)~,~~\theta=\left(
  \begin{array}{c}
    \theta_1   \\
    \theta_2   \\
  \end{array}
\right)\,.
\end{eqnarray}
where $g_x$ is concretely specified later. The supervielbeins $\BL^{A}$
and $L^\alpha$, and super spin connection $\BL^{AB}$ are the LI Cartan
one forms defined by
\begin{eqnarray}
g^{-1}dg&=&\BL^AP_A +\frac{1}{2}\BL^{AB}J_{AB}
+Q_\alpha L^\alpha~,\\
g_x^{-1}dg_x&=&e^AP_A+\frac{1}{2}\omega^{AB}J_{AB}\,, 
\end{eqnarray}
where $e^A$ and $\omega^{AB}$ are the vielbein and the spin connection
of the AdS$_5\times$S$^5$.  After some algebra, we obtain\footnote{ The
differential $d$ acts as $d(F\wedge G)=dF\wedge G+(-1)^{f}F\wedge dG$
(where $f$ is the degree of $F$), and commutes with $\theta$.  }
\begin{eqnarray}
\BL^A&=&
e^A+2i\sum_{n=1}^\infty\bar\theta\Gamma^A\frac{\CM^{2n-2}}{(2n)!}D\theta
=
e^A+2i\bar\theta\Gamma^A
\left(
\frac{\cosh\CM-1}{\CM^2}
\right)
D\theta
~,\\
L^\alpha&=&
\sum_{n=0}^\infty\frac{\CM^{2n}}{(2n+1)!}D\theta
=
\frac{\sinh\CM}{\CM}D\theta~,\\
\BL^{AB}&=&
\omega^{AB}
-2i\lambda\bar\theta\widehat\Gamma^{AB}i\sigma_2
\sum_{n=1}^\infty\frac{\CM^{2n-2}}{(2n)!}D\theta
=\omega^{AB}
-2i\lambda\bar\theta\widehat\Gamma^{AB}i\sigma_2
\frac{\cosh\CM-1}{\CM^2}
D\theta
\end{eqnarray}
with
\begin{eqnarray}
\CM^2&=&
i\lambda\left(
\widehat\Gamma_Ai\sigma_2\theta\,
\bar\theta\Gamma^A
-\frac{1}{2}\Gamma_{AB}\theta\,
\bar\theta\widehat\Gamma^{AB}i\sigma_2
\right)~,\\
D\theta&=&
d\theta
+\frac{\lambda}{2}e^A\widehat\Gamma_Ai\sigma_2\theta
+\frac{1}{4}\omega^{AB}\Gamma_{AB}\theta~,\\
\widehat\Gamma_A&=&(-\Gamma_a\CI,\Gamma_{a'}\CJ)~,~~~
\widehat\Gamma_{AB}=(-\Gamma_{ab}\CI,\Gamma_{a'b'}\CJ)~.
\end{eqnarray}

The bosonic subalgebra is a direct product of so(2,4) and so(6), and so
we may consider these parts separately.  For an ($m,n$)-brane, it is
convenient to parametrize the group manifold on the so(2,4) algebra as
\begin{eqnarray}
g_{AdS}=g_N
\e^{y^{\underline a}P_{\underline a}}
~,~~~g_N=\e^{x^{\bar a_m}P_{\bar a_m}}\cdots \e^{x^{\bar a_1}P_{\bar a_1}}
~.
\end{eqnarray}
For this parametrization, we obtain
\begin{eqnarray}
e^{\bar a_\ell}&=&e^{\bar a_\ell}_N \cosh r_y ~,~~~\ell=1,2,...,m~,\nonumber\\
e^{\underline a}&=&\left(\frac{\sinh Y}{Y}dy\right)^{\underline a}~,\nonumber\\
\omega^{\bar a_k\bar a_\ell}&=&\omega^{\bar a_k\bar a_\ell}_N~,\nonumber\\
\omega^{\underline a\bar a_\ell}&=&
-\lambda^2y^{\underline a}
\frac{\sinh r_y}{r_y}e_N^{\bar a_\ell}~,\nonumber\\
\omega^{\underline a\underline b}&=&
-2\lambda^2y^{[\underline a}
\left(\frac{\cosh Y-1}{Y^2}dy\right)
^{\underline b]}~,
\end{eqnarray}
where $r_y^2=\lambda^2y^{\underline a}y^{\underline b}\eta_{\underline
a\underline b} =\lambda^2y^2$ and $(Y^2)^{\underline a}{}_{\underline
b}=\lambda^2(y^2\delta^{\underline a}_{\underline b} -y^{\underline
a}y_{\underline b})$.  $e_N$ and $\omega_N$ defined by
\begin{eqnarray}
g_N^{-1}dg_N=e_N^{\bar a}P_{\bar a}+\frac{1}{2}\omega^{\bar a\bar b}_NJ_{\bar a\bar b}
\end{eqnarray}
are obtained as
\begin{eqnarray}
e_N^{\bar a_\ell}&=&\cosh r_1\cdots\cosh r_{\ell-1}dx^{\bar a_\ell}~,\nonumber\\
\omega_N^{\bar a_k\bar a_\ell}&=&
-\lambda^2x^{\bar a_k}\frac{\sinh r_k}{r_k}
\cosh r_{k+1}\cdots \cosh r_{\ell-1}dx^{\bar a_\ell}~,~~~k<\ell~
\end{eqnarray}
where $r_\ell^2=\lambda^2x^{\bar a_\ell}x^{\bar a_\ell}\eta_{\bar
a_\ell\bar a_\ell}$\,.

The vielbein $e^{a'}$ and the spin connection $\omega^{a'b'}$ of S$^5$
are obtained as those of AdS$_5$ with the replacement
\begin{eqnarray}
\lambda^2\to-\lambda^2\,, \quad 
\bar a\to\bar a'\,, \quad 
\underline a\to\underline a'\,, \quad m\to n\,. 
\end{eqnarray}

Under the scaling with $\Omega$ defined in (\ref{coordinate scaling}),
the above vielbeins and spin connections are expanded as
\begin{eqnarray}
e^{\bar a_\ell}&=&
e_0^{\bar a_\ell}+\Omega^2e_2^{\bar a_\ell}
+O(\Omega^4)~,~~~
e_0^{\bar a_\ell}=e^{\bar a_\ell}_N~,~~~
e_2^{\bar a_\ell}=e^{\bar a_\ell}_N\frac{1}{2}r_y^2~,
\\
e^{\underline a}&=&
\Omega 
e^{\underline a}_1
+O(\Omega^3)~,~~~
e^{\underline a}_1=dy^{\underline a}\\
\omega^{\bar a_k\bar a_\ell}&=&\omega^{\bar a_k\bar a_\ell}_0~,~~~
\omega^{\bar a_k\bar a_\ell}_0=\omega^{\bar a_k\bar a_\ell}_N~,
\\
\omega^{\underline a\bar a_\ell}&=&
-\Omega\lambda^2y^{\underline a}
e_N^{\bar a_\ell} +O(\Omega^3)~,\\
\omega^{\underline a\underline b}&=&
-2\Omega^2\lambda^2 y^{[\underline a}
dy^{\underline b]}+O(\Omega^4)~.
\end{eqnarray}

\subsection{AdS$_{4/7}\times$S$^{7/4}$}\label{parametrization 11}

Supervielbeins on the AdS$_{4/7}\times S^{7/4}$ can be obtained via  
the coset construction with the coset supermanifolds: 
\begin{eqnarray}
\mbox{AdS}_4\times S^7 \sim \frac{OSp(8|4)}{SO(3,1)\times SO(7)}\,, \quad 
\mbox{AdS}_7\times S^4 \sim \frac{OSp(8^*|4)}{SO(6,1)\times SO(4)}\,. 
\end{eqnarray} 
Parametrizing the manifolds as
${g}(X,\theta)= g_x\e^{\theta Q}$\,, we obtain the expression
of supervielbeins: 
\begin{eqnarray}
\BL^A &=& {e}^A
-2\bar\theta\Gamma^A\sum_{n=1}\frac{\CM^{2n-2}}{(2n)!}D\theta~,\\
\BL^{AB}&=&
\omega^{AB}+2\lambda\bar\theta\widehat\Gamma^{AB}
\sum_{n=1}\frac{\CM^{2n-2}}{(2n)!}D\theta~,\\
L&=&
\sum_{n=0}\frac{\CM^{2n}}{(2n+1)!}D\theta~,
\end{eqnarray}
where we have introduced the following quantities: 
\begin{eqnarray*}
 (D\theta)^{\bar{\alpha}} 
&\equiv& d\theta
+-\frac{\lambda}{2}e^A\widehat\Gamma_A\theta
+\frac{1}{4}\omega^{AB}\Gamma_{AB}\theta
\,,\\
\mathcal{M}^2 &=& 
\lambda(
\widehat\Gamma^A\theta\,\bar\theta\Gamma_A
+\frac{1}{2}\Gamma^{AB}\theta\, \bar\theta\widehat\Gamma_{AB}
)\,,\\
\widehat\Gamma_A&=&(2\CI\Gamma_a,\CI\Gamma_{a'})~,~~~
\widehat\Gamma_{AB}=(2\CI\Gamma_{ab},\CI\Gamma_{a'b'})~.
\end{eqnarray*}
Here $e^A_M$ and $\omega^{AB}_M$ are the vielbein and the spin connection,
respectively. 

Since the bosonic subalgebra is the direct product of so(3,2) (so(5))
and so(8)(so(6,2)), we may consider these parts separately as in the
case of AdS$_5\times$S$^5$\,. 
For the former group manifold, a group element is represented by  
\begin{eqnarray}
g_4=g_Ne^{y^{\underline a}P_{\underline a}}~,~~~
g_N=\e^{x^{\bar a_m}P_{\bar a_m}}
\cdots
\e^{x^{\bar a_1}P_{\bar a_1}}~.
\end{eqnarray}
It is straightforward to derive
\begin{eqnarray}
e^{\bar a_\ell}&=&
e^{\bar a_\ell}_N\cosh r_y~,~~~~
\ell=1,\cdots,m\\
e^{\underline a}&=&\left(
\frac{\sinh Y}{Y}dy
\right)^{\underline a}~,\\
\omega^{\bar a_k\bar a_\ell}&=&
\omega^{\bar a_k\bar a_\ell}_N~,\\
\omega^{\underline a\underline b}&=&
-8\epsilon^2\lambda^2y^{\underline a}\left(
\frac{\cosh Y-1}{Y}dy
\right)^{\underline b}~,\\
\omega^{\bar a_\ell\underline b}&=&
4\epsilon^2\lambda^2e^{\bar a_\ell}_N
y^{\underline b}\frac{\sinh r_y}{r_y}~
\end{eqnarray}
with
\begin{eqnarray}
e_N^{\bar a_\ell}&=&
dx^{\bar a_\ell}\cosh r_1\cdots\cosh r_{\ell-1}~,\\
\omega^{\bar a_k\bar a_\ell}_N&=&
-4\epsilon^2\lambda^2x^{\bar a_k}dx^{\bar a_\ell}
\frac{\sinh r_k}{r_k}
\cosh r_{k+1}
\cdots
\cosh r_{\ell-1}~,~~~~
k<\ell~,
\end{eqnarray}
where $r_k^2=4\epsilon^2\lambda^2x^{\bar a_k}x^{\bar a_k}\eta_{\bar
a_k\bar a_k}$, 
$r_y^2=4\epsilon^2\lambda^2y^{\underline a}y^{\underline
b}\eta_{\underline a\underline b} =4\epsilon^2\lambda^2y^2$ and
$Y^2=4\epsilon^2\lambda^2(y^2\delta^{\underline a}_{\underline b}-
y^{\underline a}y_{\underline b})$\,. For the latter group manifold,
the vielbein and the spin connection can be obtained from those for the
former case with the replacement
\begin{eqnarray}
\epsilon^2\to-\frac{1}{4}\epsilon^2~,~~~
\bar a\to\bar a'~,~~~
\underline a\to\underline a'~,~~~
m\to n~.
\end{eqnarray}

Under the $\Omega$-scaling (\ref{scale 11}),
these scale as
\begin{eqnarray}
e^{\bar a}&=&e^{\bar a}_0
+\Omega^2e^{\bar a}_2
+O(\Omega^4)~,~~~
e^{\bar a}_0=e^{\bar a}_N
~,~~~
e^{\bar a}_2=e^{\bar a}_N\frac{r_y^2}{2}
\\
e^{\underline a}&=&
\Omega e^{\underline a}_1
+O(\Omega^3)~,~~~
e^{\underline a}_1=dy^{\underline a}~,\\
\omega^{\bar a\bar b}&=&\omega^{\bar a\bar b}_0~,~~~
\omega^{\bar a\bar b}_0=\omega^{\bar a\bar b}_N~,\\
\omega^{\underline a\underline b}&=&
O(\Omega^2)~,\\
\omega^{\bar a\underline b}&=&
\Omega 4\epsilon^2\lambda^2 e^{\bar a}_Ny^{\underline b}
+O(\Omega^3)~.
\end{eqnarray}

\section{$\kappa$-symmetry}\label{appendix:kappa}
\subsection{D-branes in AdS$_5\times$S$^5$} 

Here we recall the $\kappa$-variation of the action (\ref{S 10}) by
following \cite{D:curved,F1:AdS,D3:AdS}. Here we consider both
Lorentzian branes and Euclidean branes.

Following (\ref{MC AdSxS in 10-dim}), one can derive 
a variation of the supercurrents by using the homotopy formula as
follows:
\begin{eqnarray}
\delta\BL^A&=&
d\delta x^A
+\eta_{BC}\BL^B\delta x^{CA}
+\eta_{BC}\BL^{AB}\delta x^C
-2i\bar L\Gamma^A\delta\theta~,\nonumber\\
\delta L&=&
d\delta\theta
-\frac{\lambda}{2}\delta x^A\widehat\Gamma_Ai\sigma_2 L
+\frac{\lambda}{2}\BL^A\widehat\Gamma_Ai\sigma\delta\theta~,\nonumber\\
\delta\BL^{ab}&=&
d\delta x^{ab}
-2\lambda^2\BL^{a}\delta x^{b}
+2\eta_{cd}\BL^{ac}\delta x^{db}
+2i\lambda\bar L\widehat\Gamma^{ab}i\sigma_2\delta\theta~,\nonumber\\
\delta\BL^{a'b'}&=&
d\delta x^{a'b'}
+2\lambda^2\BL^{a'}\delta x^{b'}
+2\eta_{c'd'}\BL^{a'c'}\delta x^{d'b'}
+2i\lambda\bar L\widehat\Gamma^{a'b'}i\sigma_2\delta\theta\,,
\label{variation}
\end{eqnarray}
where
\begin{eqnarray}
\delta x^A=\delta Z^{\hat M}\BL_{\hat M}^A~,~~~
\delta x^{AB}=\delta Z^{\hat M}\BL_{\hat M}^{AB}~,~~~
\delta \theta^\alpha=\delta Z^{\hat M}L_{\hat M}^{\alpha}~.
\label{delta x definition}
\end{eqnarray}
A universal feature of the  $\kappa$-variation is
\begin{eqnarray}
\delta_\kappa x^A=0~.
\end{eqnarray}
By using (\ref{variation}), one can find that 
\begin{eqnarray}
\delta_\kappa g_{ij}&=&-4i\BL^A_{(i}\bar L_{j)}\Gamma_A\delta\theta~
\end{eqnarray}
and
\begin{eqnarray}
\delta_\kappa d\CF=-2id(\BL^A\bar L\Gamma_A\sigma\delta\theta)
~~~\to~~~
\delta_\kappa \CF=-2i\BL^A\bar L\Gamma_A\sigma\delta\theta\,, 
\label{CF kappa}
\end{eqnarray}
where the exact term is deleted by $\delta_\kappa A$\,. By using
(\ref{variation}) and (\ref{CF kappa}), we find that
\begin{eqnarray}
\delta_\kappa h_{p+2}&=&d[\CC_\kappa\wedge e^{\CF}]_{p+1}~,~~~
\CC_\kappa=\bigoplus_{\ell=\mathrm{even}}\CC_\kappa^{(\ell)}~,
\nonumber\\
\CC_\kappa^{(2n)}&=&
\frac{2\sqrt{s}}{(2n-1)!}
\BL^{A_1}\cdots\BL^{A_{2n-1}}\bar L\Gamma_{A_1\cdots A_{2n-1}}(\sigma)^{n}i\sigma_2
\delta_\kappa\theta~
\end{eqnarray}
so that
\begin{eqnarray}
\delta_\kappa\CL_{WZ}=[\CC_\kappa\wedge e^{\CF}]_{p+1}\,, 
\end{eqnarray}
where $[\bullet]_{p+1}$ represents the $(p+1)$-form part of $\bullet$\,. 

By using these expressions, one finds that the action (\ref{S 10}) is
invariant under
\begin{eqnarray}
\delta_\kappa\theta&=&(1+\Gamma)\kappa~,\\
\Gamma&=&\frac{s\sqrt{-s}}{\sqrt{s\det (g+\CF)}}
\sum_{n=0}\frac{1}{2^nn!}
\gamma^{j_1k_1\cdots j_nk_n}\CF_{j_1k_1}\cdots \CF_{j_nk_n}
\nonumber\\&&
\times(-1)^n
(\sigma)^{n-\frac{p-3}{2}}i\sigma_2
\frac{1}{(p+1)!}\epsilon^{i_1\cdots i_{p+1}}
\gamma_{i_1\cdots i_{p+1}}\,, 
\end{eqnarray}
where $\gamma_i=\BL_i^A\Gamma_A$\,. 

Under the $\Omega$-scaling,
$\Gamma$ is expanded as
\begin{eqnarray}
\Gamma&=&\Gamma_0+O(\Omega)~,\\
\Gamma_0&=&
\frac{s\sqrt{-s}}{\sqrt{s\det g_0}}
\frac{1}{(p+1)!}\epsilon^{i_1\cdots i_{p+1}}
(\BL^{\bar A_0}_0)_{i_1}\cdots (\BL^{\bar A_{p}}_0)_{i_{p+1}}
\Gamma_{\bar A_0\cdots \bar A_{p}}
(\sigma)^{-\frac{p-3}{2}}i\sigma_2
\nonumber\\
&=&M~.
\end{eqnarray}
Expanding $\kappa$ as
\begin{eqnarray}
\kappa=\kappa_++\Omega\kappa_-~,~~~ 
\kappa_\pm=\CP_{\pm}\kappa_{\pm}
\label{expansion kappa}
\end{eqnarray}
leads to
\begin{eqnarray}
\delta_\kappa\theta_+
= (1+\Gamma_0) \kappa_+
=2\kappa_+
~.
\end{eqnarray}
This implies that the $\kappa$-symmetry can 
be gauge fixed by choosing $\theta_+=0$
since $\delta_\kappa\theta_+|_{\theta_+=0}=2\kappa_+$.

For an F-string, we obtain
the similar expression with $\sigma=-\sigma_1$ and $\CF=0$. Hence the
action is $\kappa$-invariant, and the $\kappa$-gauge symmetry is fixed
by $\theta_+=0$\,.

\subsection{M-brane in AdS$_{4/7}\times$S$^{7/4}$}

Following \cite{M2:curved,PST,M5:AdS}, we recall the $\kappa$-symmetry
of the M-brane actions. Here we shall consider Euclidean branes as well
as Lorentzian branes.

A variation of the supercurrents 
is derived from (\ref{MC 11}) as
\begin{eqnarray}
\delta\BL^{A}&=&
d\delta x^A-\eta_{BC}\delta x^{AB}\BL^C
+\eta_{BC}\BL^{AB}\delta x^{C}
+2\bar L\Gamma^A\delta\theta~,
\nonumber\\
\delta\BL^{ab}&=&
d\delta x^{ab}
+8\epsilon^2\lambda^2\BL^{a}\delta x^{b}
+2\eta_{cd}\BL^{ca}\delta x^{bd}
-2\lambda\bar L\widehat\Gamma^{ab}\delta\theta~,
\nonumber\\
\delta\BL^{a'b'}&=&
d\delta x^{a'b'}
-2\epsilon^2\lambda^2\BL^{a'}\delta x^{b'}
+2\eta_{c'd'}\BL^{c'a'}\delta x^{b'd'}
-2\lambda\bar L\widehat\Gamma^{a'b'}\delta\theta~,
\nonumber\\
\delta L&=&
d\delta\theta
+\frac{1}{2}\delta x^A\widehat\Gamma_A L
-\frac{\lambda}{2}\BL^A\widehat\Gamma_A\delta\theta
-\frac{1}{4}\delta x^{AB}\Gamma_{AB}L
+\frac{1}{4}\BL^{AB}\Gamma_{AB}\delta\theta\,, 
\label{variation 11}
\end{eqnarray} 
where $\delta x^A$, $\delta x^{AB}$ and $\delta \theta$ have been
defined in (\ref{delta x definition}). For the $\kappa$-variation, we
require that $\delta_\kappa x^A=0$.

\subsubsection{M2-brane}

Let us first consider the case of an M2-brane. From (\ref{variation
11}), one can obtain
\begin{eqnarray}
\delta_\kappa g_{ij}&=&
4\BL^A_{(i}\bar L_{j)}\Gamma_A\delta_\kappa\theta~,
\end{eqnarray}
and
\begin{eqnarray}
\delta_\kappa h_4&=&
d(-c\CL^A\BL^B\bar L\Gamma_{AB}\delta_\kappa\theta)
~~~\to~~~
\delta_\kappa\CL_{\WZ}^{M2}=-c\CL^A\BL^B\bar L\Gamma_{AB}\delta_\kappa\theta~.
\end{eqnarray}
By using them
one can see that the action (\ref{action 11}) is invariant under
\begin{eqnarray}
\delta_\kappa\theta=(1+\Gamma)\kappa~,~~~
\Gamma=\frac{i\sqrt{s}}{\sqrt{s\det g}}\frac{1}{3!}\epsilon^{ijk}\gamma_{ijk}~.
\end{eqnarray}

Under the $\Omega$-scaling, $\Gamma$ is expanded as
\begin{eqnarray}
\Gamma=\Gamma_0+O(\Omega^2)~,~~~
\Gamma_0=i\sqrt{s}\Gamma_{\bar A_0\bar A_1\bar A_2}=M~.
\end{eqnarray}
By expanding $\kappa$ as (\ref{expansion kappa}), we derive
\begin{eqnarray}
\delta_\kappa\theta_+=(1+\Gamma_0)\kappa_+=2\kappa_+
\end{eqnarray}
which implies that the $\kappa$-gauge symmetry is fixed by $\theta_+=0$.

\subsubsection{M5-brane}

Next, we consider the case of an M5-brane. A variation of $\CH$ is
taken with (\ref{variation 11}) as follows: 
\begin{eqnarray}
-c\delta_\kappa d\CH
=\delta_\kappa h_4
=d(-c\BL^A\BL^B\bar L\Gamma_{AB}\delta_\kappa\theta)
~~~\to~~~
\delta_\kappa \CH=\BL^A\BL^B\bar L\Gamma_{AB}\delta_\kappa\theta\,, 
\end{eqnarray}
where the exact term is deleted by $\delta_\kappa B$\,. Noting that
\begin{eqnarray}
\delta_\kappa h_7=c^2d\Big[
\frac{2}{5!}\BL^{A_1}\cdots\BL^{A_5}\bar L\Gamma_{A_1\cdots A_5}\delta_\kappa\theta
+\frac{1}{2}\BL^A\BL^B\bar L\Gamma_{AB}\delta_\kappa\theta\,\CH
\Big]\,, 
\end{eqnarray}
we see that
\begin{eqnarray}
\delta_\kappa\CL_{\WZ}=
c^2\Big[
\frac{2}{5!}\BL^{A_1}\cdots\BL^{A_5}\bar L\Gamma_{A_1\cdots A_5}\delta_\kappa\theta
+\frac{1}{2}\BL^A\BL^B\bar L\Gamma_{AB}\delta_\kappa\theta\,\CH
\Big]\,, 
\end{eqnarray}
where $c^2=i\sqrt{s}$\,. The $\kappa$-invariance of the action is shown
by following \cite{M5:AdS} (see also \cite{CKP}). By using the
expressions derived above and the following useful relations
\begin{eqnarray}
&&
\bar\gamma=\frac{s\sqrt{-s}}{6!\sqrt{s\det g}}
\epsilon^{i_1\cdots i_6}\gamma_{i_1\cdots i_6}~,~~~
\bar\gamma^2=1~,
\nonumber\\
&&
\epsilon^{i_1\cdots i_{6-n}j_1\cdots j_n}\gamma_{j_1\cdots j_n}
=(-1)^{[\frac{6-n}{2}]}n!
\frac{s\sqrt{-s}}{\sqrt{s\det g}}\gamma^{i_1\cdots i_{6-n}}\bar \gamma
~,
\nonumber\\
&&
\epsilon^{i_1\cdots i_{6-n}k_1\cdots k_n}
\epsilon_{j_1\cdots j_{6-n}k_1\cdots k_n}=
sn!(6-n)!\delta^{i_1}_{[j_1}\cdots\delta^{i_{6-n}}_{j_{6-n}]}~,
\nonumber\\
&&
\CH_{ijk}=
3\CH_{[ij}v_{k]}
-\frac{s}{2}
\sqrt{s\det g}\,
\epsilon_{ijklmn}\CH^{*lm}v^n~,
\end{eqnarray}
it is shown that the M5-brane action is invariant under the
$\kappa$-variation
\begin{eqnarray}
\delta_\kappa\theta&=&(1+\Gamma)\kappa~,~~~~\delta_\kappa a=0~,\\
\Gamma&=&\frac{\sqrt{s\det g}}{\sqrt{s\det (g-ic^2\CH^*)}}
\left(
\bar\gamma
-\frac{c^2}{2}\CH^*_{ij}v_k\gamma^{ijk}
-\frac{sc^2}{16\sqrt{s\det g}}\epsilon^{i_1\cdots i_6}
\CH^*_{i_1i_2}\CH^*_{i_3i_4}\gamma_{i_5i_6}
\right).~~~
\end{eqnarray}
Under the $\Omega$-scaling, $\Gamma$ is expanded as
\begin{eqnarray}
\Gamma&=&\Gamma_0+O(\Omega)~,\\
\Gamma_0&=&\frac{s\sqrt{-s}}{\sqrt{s\det g_0}}
\frac{1}{6!}\epsilon^{i_1\cdots i_6}(\BL_0^{\bar A_1})_{i_1}
\cdots
(\BL_0^{\bar A_6})_{i_6}
\Gamma_{\bar A_1\cdots\bar A_6}
=s\sqrt{-s}\Gamma_{\bar A_0\cdots\bar A_5}=M\,, 
\end{eqnarray}
which implies that
\begin{eqnarray}
\delta_\kappa\theta_+=(1+\Gamma_0)\kappa_+=2\kappa_+~.
\end{eqnarray}
Thus the $\kappa$-symmetry is gauge fixed by $\theta_+=0$\,.

\section{Penrose limit of Brane Actions}

Here we will construct the action of D-branes and M-branes on pp-wave
backgrounds via the Penrose limit, instead of non-relativistic limit.
This is a natural application of our procedure. 
The Penrose limit of an alternative action of an AdS superstring
has been discussed in \cite{Hatsuda:2002iu}.
The result includes Metsaev's results for F-string \cite{Metsaev:2001bj} 
and D3-brane \cite{D3-pp} on the
maximally supersymmetric pp-wave. 

\subsection{Branes in IIB pp-wave}

We derive the D$p$-brane action in the IIB pp-wave from the D$p$-brane
action in AdS$_5\times$S$^5$
\begin{eqnarray}
S&=&T\int\CL_{DBI}+\CL_{WZ}~,~~~\nonumber\\
\CL_{\DBI}&=&\sqrt{s\det (g+\CF)}d^{p+1}\xi~,~~~
d\CL_{\WZ}=h_{p+2}=\sum_{n=0}h^{(p+2-2n)}\CF^n~,
\nonumber\\
h^{(2n+1)}&=&\frac{c}{(2n-1)!}\Big[
\BL^{A_1}\cdots\BL^{A_{2n-1}}\bar L\Gamma_{A_1\cdots A_{2n-1}}
\sigma^{n+1}\varrho L
\nonumber\\&&~~~~~~
+\delta_{n,2}\frac{i}{5}\lambda\bigl(
 \epsilon_{a_1\cdots a_5}\BL^{a_1}\cdots \BL^{a_5}
 -\epsilon_{a'_1\cdots a'_5}\BL^{a'_1}\cdots \BL^{a'_5}\bigr)
\Big]
\end{eqnarray}
where
$\sigma=\sigma_3$ and $\varrho=\sigma_1$.
 $c=\sqrt{s}$ is required by the $\kappa$-invariance of the action.

The Penrose limit considered in section 3
 is equivalent to 
scaling the coordinates as
\begin{eqnarray}
X^+\to\Lambda^2 X^+~,~~~X^\ih\to\Lambda X^\ih~,~~~
\theta_+\to\Lambda \theta_+
\label{coordinate scaling 10}
\end{eqnarray}
and taking the limit $\Lambda\to 0$.
Under the scaling,
 LI Cartan one-forms are expanded as
\begin{eqnarray}
&&
\BL^+=\sum_{n=0}\Lambda^{2n+2}\BL^{+}_{2n+2}~,~~~
\BL^{\ih}=\sum_{n=0}\Lambda^{2n+1}\BL^\ih_{2n+1}~,~~~
\BL^{\ih}_*=\sum_{n=0}\Lambda^{2n+1}\BL^\ih_{*2n+1}~,~~~
\nonumber\\&&
L_+=\sum_{n=0}\Lambda^{2n+1} L_{+2n+1}
\label{expansion 10}
\end{eqnarray}
where
\begin{eqnarray}
\BL^\pm=\frac{1}{\sqrt{2}}(\BL^9\pm\BL^0)~,~~~
L_\pm=\ell_\pm L_\pm~,~~~
\BL^\ih_*=(\BL^{0i},\BL^{9i'})~.
\end{eqnarray}

Under the expansion (\ref{expansion 10}), we derive
\begin{eqnarray}
&&
g_{ij}=\Lambda^2g_{ij}^{(\pp)}+O(\Lambda^4)~,~~~
g_{ij}^{(\pp)}=2(\BL^+_2)_i(\BL^-_0)_j+(\BL^\ih_1)_i(\BL^\jh_1)_j\eta_{\ih\jh}~.
\end{eqnarray}
It follows from\footnote{$\BL^+$, $\BL^-$, $\BL^\ih$, $L_+$ and $L_-$
are understood as $\BL^+_2$, $\BL^-_0$, $\BL^\ih_1$, $L_{+1}$ and $L_{-0}$
respectively below.}
\begin{eqnarray}
d\CF=\Lambda^2[
-i\BL^+\bar L_-\Gamma_+\sigma L_-
-i\BL^-\bar L_+\Gamma_-\sigma L_+
-2i\BL^\ih\bar L_+\Gamma_\ih\sigma L_-
]
\end{eqnarray}
that
\begin{eqnarray}
\CF=\Lambda^2\CF_{\pp}+O(\Lambda^4)
\end{eqnarray}
where we assume that $F=\Lambda^2F_\pp$.
These imply that
\begin{eqnarray}
\CL_{\DBI}=\Lambda^{p+1}\CL_{\DBI}^\pp+O(\Lambda^{p+3})~,~~~
\CL_{\DBI}^\pp=\sqrt{s\det (g_\pp+\CF_\pp)}d^{p+1}\xi~.
\label{L DBI pp 10}
\end{eqnarray}
The factor $\Lambda^{p+1}$ is absorbed into the definition of the tension as
\begin{eqnarray}
T=\Lambda^{-(p+1)}T_\pp~.
\end{eqnarray}

One finds that the fermionic part of $h^{(2n+1)}$
is scaled as
\begin{eqnarray}
h^{(2n+1)}|_{\mathrm{fermionic}}&=&
\Lambda^{2n}h_\pp^{(2n+1)}|_{\mathrm{fermionic}}+O(\Lambda^{2n+2})~,\\
h_\pp^{(2n+1)}|_{\mathrm{fermionic}}&=&\frac{c}{(2n-1)!}
\BL^{A_1}\cdots\BL^{A_{2n-1}}\bar L\Gamma_{A_1\cdots A_{2n-1}}\sigma_3^{n+1}\sigma_1 L
\nonumber\\
&=&
\frac{c}{(2n-1)!}\BL^{\ih_1}\cdots\BL^{\ih_{2n-1}}
(\bar L_+\Gamma_{\ih_1\cdots \ih_{2n-1}}\sigma_3^{n+1}\sigma_1 L_-
\nonumber\\&&\hspace{50mm}
+\bar L_-\Gamma_{\ih_1\cdots \ih_{2n-1}}\sigma_3^{n+1}\sigma_1 L_+)
\nonumber\\&&
+\frac{c}{(2n-2)!}
\BL^{+}\BL^{\ih_1}\cdots\BL^{\ih_{2n-2}}
 \bar L_-\Gamma_{+\ih_1\cdots \ih_{2n-2}}\sigma_3^{n+1}\sigma_1 L_-
\nonumber\\&&
+\frac{c}{(2n-2)!}\BL^{-}\BL^{\ih_1}\cdots\BL^{\ih_{2n-2}}
 \bar L_+\Gamma_{-\ih_1\cdots \ih_{2n-2}}\sigma_3^{n+1}\sigma_1 L_+
\nonumber\\&&
+\frac{c}{(2n-3)!}\BL^{+}\BL^{-}\BL^{\ih_1}\cdots\BL^{\ih_{2n-3}}(
\bar L_-\Gamma_{\ih_1\cdots \ih_{2n-3}}\sigma_3^{n+1}\sigma_1 L_+
\nonumber\\&&\hspace{50mm}
-\bar L_+\Gamma_{\ih_1\cdots \ih_{2n-3}}\sigma_3^{n+1}\sigma_1 L_-)~.
\end{eqnarray}
For the bosonic part, we derive
\begin{eqnarray}
h^{(5)}|_{\mathrm{bosonic}}&=&\Lambda^4h^{(5)}_\pp|_{\mathrm{bosonic}}
+O(\Lambda^6)~,\\
h^{(5)}_\pp|_{\mathrm{bosonic}}&=&-ic\frac{4\lambda}{\sqrt{2}}
\BL^-(\BL^1\BL^2\BL^3\BL^4+\BL^5\BL^6\BL^7\BL^8)~.
\end{eqnarray}
The $O(\Lambda^6)$-term which contains $\BL^+$
disappears in the limit.

To summarize the pp-wave D$p$-brane action is given as
\begin{eqnarray}
S_\pp&=&T_\pp\int \CL_{\DBI}^\pp+\CL_{\WZ}^\pp~
\end{eqnarray}
with (\ref{L DBI pp 10}) and
\begin{eqnarray}
h_{p+2}^\pp&=&d\CL_{WZ}^\pp=\sum_{n=0}h^{(p+2-2n)}_{\pp}\CF^n_\pp~,\\
h_\pp^{(2n+1)}&=&\frac{c}{(2n-1)!}
\BL^{A_1}\cdots\BL^{A_{2n-1}}\bar L\Gamma_{A_1\cdots A_{2n-1}}\sigma^{n+1}\varrho L
\nonumber\\&&
-ic\delta_{n,2}\frac{4\lambda}{\sqrt{2}}
\BL^-(\BL^1\BL^2\BL^3\BL^4+\BL^5\BL^6\BL^7\BL^8)~.
\end{eqnarray}
This reproduces the pp-wave D3-brane action 
given in \cite{D3-pp} as the $p=3$ case.
Let $\varrho=\sigma_3$ and $\sigma=-\sigma_1$
and replace $\CL_{\DBI}$ with $\CL_{\NG}$
or with the Polyakov action, then it is reduced to the pp-wave F-string
action constructed in \cite{Metsaev:2001bj}.

The $(p+2)$-form $h_{p+2}^\pp$ can be shown to be a non-trivial element of
the CE cohomology except for $h_3$ by following the procedure
explained in section 4.1.
It is easy to obtain the $(p+1)$-dimensional form of the WZ term
as was done in section 4.2.

\subsection{Branes in M pp-wave}

The Penrose limit considered in section 7 is equivalent to scaling the
coordinates as (\ref{coordinate scaling 10}) and taking the limit
$\Lambda\to 0$.  Under the scaling, LI Cartan one-forms are expanded as
(\ref{expansion 10}) where we define Cartan one-forms as
\begin{eqnarray}
\BL^\pm=\frac{1}{\sqrt{2}}(\BL^\natural\pm\BL^0)~,~~~
\BL_*^\ih=\left\{
  \begin{array}{ll}
   (\BL^{i0},\BL^{i'\natural})    &\text{for AdS$_4\times$S$^7$}    \\
   (\BL^{i\natural},\BL^{i'0})    &\text{for AdS$_7\times$S$^4$}    \\
  \end{array}
\right.
~,~~~
L_\pm=\ell_\pm L~.
\end{eqnarray}

\subsubsection{Penrose limit of M2 brane action}
We consider the Penrose limit of the M2-brane action
\begin{eqnarray}
S&=&T\int\CL_{\NG}+\CL_{\WZ}~,\\
\CL_{\NG}&=&\sqrt{s\det g}~,\\
d\CL_{\WZ}&=&h_4=c\Big[\frac{1}{2}\BL^A \BL^B\bar L\Gamma_{AB}L
 -\frac{6\lambda}{4!}\epsilon_{a_1\cdots a_4}\BL^{a_1}\cdots \BL^{a_4}
 \Big]
\end{eqnarray}
where $c=i\sqrt{s}$ is required by the $\kappa$-invariance of the action.
Under the expansion (\ref{expansion 10}), $\CL_{\NG}$ is expanded as
\begin{eqnarray}
\CL_{\NG}=\Lambda^3\CL_{\NG}^\pp+O(\Lambda^5)~,~~~
\CL_{\NG}^\pp=\sqrt{s\det g_\pp}~
\label{NG M2 pp}
\end{eqnarray}
while $d\CL_{\WZ}=h_4$ is expanded as
\begin{eqnarray}
h_4&=&\Lambda^3h_4^\pp+O(\Lambda^5)~,\\
h_4^\pp&=&c\Big[\frac{1}{2}\BL^A \BL^B\bar L\Gamma_{AB}L
 +\frac{6\lambda}{\sqrt{2}}\BL^{-}
 \BL^1\BL^2\BL^3\BL^4
 \Big]~.
 \label{h M2 pp}
\end{eqnarray}
The $\Lambda^3$ factor is absorbed into the definition of the tension as
$T=\Lambda^{-3}T_\pp$.  The pp-wave M2-brane action
is given as
\begin{eqnarray}
S=T_\pp\int \CL_{\NG}^\pp+\CL_{WZ}^\pp
\end{eqnarray}
with (\ref{NG M2 pp}) and $h_4^\pp=d\CL_{WZ}^\pp$ with (\ref{h M2 pp}).

\subsubsection{Penrose limit of M5-brane action}

Next we will consider the M5-brane action
\begin{eqnarray}
S&=&T\int\CL_{\PST}+\CL_{\WZ}~,
\end{eqnarray}
with
\begin{eqnarray}
\CL_{\PST}&=&\sqrt{s\det(g_{ij}-ic^2\CH^*_{ij})}
+c^2\frac{\sqrt{s\det g}}{4}\CH^{*ij}\CH_{ij}\,,\\ 
\CH_{ij}&=&\CH_{ijk}v^k\,,
\quad 
\CH^{*ij}=\CH^{*ijk}v_k\,,\quad
v_i=\frac{\partial_ia}{\sqrt{g^{jk}\partial_ja\partial_ka}}\,, \nonumber \\
\CH &=& H+\CC_3\,, \quad
\CH^{*ijk}=\frac{1}{3!\sqrt{s\det g}}\epsilon^{ijklmn}\CH_{lmn}\,, \quad
H=dB\, \nonumber
\end{eqnarray}
and
\begin{eqnarray}
d\CL_{\WZ}&=&h_7~,~~~h_7=h^{(7)}-\frac{c}{2}h^{(4)}\CH~,\\
h^{(4)}&=&
 c\Big[\frac{1}{2}\BL^A \BL^B\bar L\Gamma_{AB}L
 -\frac{6\lambda}{4!}\epsilon_{a_1\cdots a_4}\BL^{a_1}\cdots \BL^{a_4}
 \Big]~,\\
h^{(7)}&=&
c^2\Big[\frac{1}{5!}\BL^{A_1}\cdots \BL^{A_5}
 \bar L\Gamma_{A_1\cdots A_5}L
-\frac{6\lambda}{7!}\epsilon_{a_1'\cdots a_7'}\BL^{a_1'}\cdots \BL^{a_7'}
\Big]~,\\
-cd\CH&=&h^{(4)}
\end{eqnarray}
with $c^2=i\sqrt{s}$ for the $\kappa$-invariance of the action.

Observe that under the expansion (\ref{expansion 10}),
\begin{eqnarray}
&&
a=a_\pp~,~~~
v_i=\Lambda v_i^\pp+O(\Lambda^3)~,~~~
\CH_{ijk}=\Lambda^3\CH_{ijk}^\pp+O(\Lambda^5)~,~~~
\CH_{ij}=\Lambda^2\CH_{ij}^\pp+O(\Lambda^4)~,
\nonumber\\&&
\CH^{*ijk}=\Lambda^{-3}\CH^{*ijk}_\pp+O(\Lambda^{-1})~,~~~
\CH^*_{ij}=\Lambda^{2}\CH^{*\pp}_{ij}+O(\Lambda^{4})~
\end{eqnarray}
where we assume that $H=\Lambda^3 H_\pp$.
These imply that
\begin{eqnarray}
\CL_{\PST}&=&\Lambda^{6}\CL_{\PST}^\pp+O(\Lambda^8)~,~~~
\nonumber\\
\CL_{\PST}^\pp&=&\sqrt{s\det(g_{ij}^\pp-ic^2\CH^{*\pp}_{ij})}
+c^2\frac{\sqrt{s\det g_\pp}}{4}\CH^{*ij}_\pp\CH_{ij}^\pp\,.
\label{PST pp}
\end{eqnarray}
The WZ term $h_7$ is expanded as
\begin{eqnarray}
h_7&=&\Lambda^6h_{7}^\pp~,~~~h_7^\pp=h^{(7)}_\pp-\frac{c}{2}h^{(4)}_\pp\CH_\pp~,
\label{h M5 pp}\\
h^{(7)}_\pp&=&c^2\Big[
\frac{1}{5!}\BL^{A_1}\cdots\BL^{A_5}\bar L\Gamma_{A_1\cdots A_5}L
-\frac{6\lambda}{\sqrt{2}} \BL^-\BL^4\cdots\BL^9
\Big]
~,\\
h^{(4)}_\pp&=&c\Big[
\frac{1}{2}\BL^{A}\BL^B\bar L\Gamma_{AB}L
 +\frac{6\lambda}{\sqrt{2}}\BL^{-}
 \BL^1\BL^2\BL^3\BL^4
 \Big]~.
\end{eqnarray}
The $\Lambda^6$ factor is absorbed into the definition of the tension as
$T=\Lambda^{-6}T_\pp$.
The pp-wave M5-brane action is given  as
\begin{eqnarray}
S=T_\pp\int \CL_{\PST}^\pp+\CL_{WZ}^\pp
\end{eqnarray}
with (\ref{PST pp}) and $h_7^\pp=d\CL_{WZ}^\pp$ with (\ref{h M5 pp}).

\medskip
Following the procedure
explained in section 8.1,
one can show that
the $(p+2)$-form $h_{p+2}^\pp$ is a non-trivial element of
the CE cohomology.
The $(p+1)$-dimensional form of the WZ term
can be obtained easily
as was done in section 8.2.


\end{document}